\documentclass[preprint,floats,aps,epsfig,nofootinbib,amssymb]{JHEP3}
\usepackage{graphicx}

\def\gev{\rm GeV}
\def\mev{\rm MeV}
\def\ev{\rm eV}
\def\s{\rm s}
\def\nul{\nu_L^{}}

\def\dl{\Delta L}
\def\pslash{\not{\hbox{\kern-4pt $p$}}}
\def\qslash{\not{\hbox{\kern-4pt $q$}}}
\def\lv{LV}
\def\ptmiss{p\!\!\!\slash_T^{}}
\def\etmiss{E\!\!\!\slash_T^{}}
\def\beq{\begin{equation}}
\def\eeq{\end{equation}}
\def\bea{\begin{eqnarray}}
\def\eea{\end{eqnarray}}
\def\bef{\begin{figure}}
\def\eef{\end{figure}}
\def\bet{\begin{table}}
\def\eet{\end{table}}

\def\nuh{\mbox{$N_4$}}

\def\lsim{\mathrel{\raise.3ex\hbox{$<$\kern-.75em\lower1ex\hbox{$\sim$}}}}
\def\gsim{\mathrel{\raise.3ex\hbox{$>$\kern-.75em\lower1ex\hbox{$\sim$}}}}
\def\ifmath#1{\relax\ifmmode #1\else $#1$\fi}

\preprint{FERMILAB-PUB-08-086-T,~NSF-KITP-08-54,~MADPH--06--1466,~DCPT/07/198,~IPPP/07/99}

\title{The Search for Heavy Majorana Neutrinos}

\author{Anupama Atre$^{1,2}$, Tao Han$^{2,3,4}$, Silvia Pascoli$^{5}$,
Bin Zhang$^{4}$\thanks{avatre@fnal.gov,~ than@hep.wisc.edu,
~silvia.pascoli@durham.ac.uk,
~zb@mail.tsinghua.edu.cn (Communication author) }\\
$^1${\it Fermi National Accelerator Laboratory,
MS106, P.O.Box 500, IL 60510, U.S.A.}\\
$^2${\it Kavli Institute of Theoretical Physics, University of California, Santa Barbara, CA 93107, U.S.A.}\\
$^3${\it Department of Physics, University of Wisconsin, 1150 University Ave, Madison, WI 53706, U.S.A.}\\
$^4${\it Center for High Energy Physics, Department of Physics, Tsinghua University, Beijing 100084, P.R. China}\\
$^5${\it Institute for Particle Physics Phenomenology, Department
of Physics, Durham University, Durham DH1 3LE, United Kingdom }}

\abstract{ The Majorana nature of neutrinos can be experimentally
verified only via {\it lepton-number} violating processes
involving charged leptons. We study $36$ lepton-number violating
($\lv$) processes from the decays of tau leptons and pseudoscalar
mesons. These decays are absent in the Standard Model but,
in presence of Majorana neutrinos in the mass range $\sim 100
\mbox{ } \mev$ to $5 \mbox{ } \gev$, the
rates for these processes would be enhanced due to their resonant
contribution. We calculate the transition
rates and branching fractions and compare them to the current
bounds from direct experimental searches for $\dl=2$ tau and rare
meson decays. The experimental non-observation of such $\lv$
processes places stringent bounds on the Majorana neutrino mass
and mixing and we summarize the existing limits. We also extend
the search to hadron collider experiments. We find that, at the
Tevatron with $8\ \mbox{fb}^{-1}$ integrated luminosity,
 there could be $2\sigma$ ($5\sigma$)
 sensitivity for resonant production of a Majorana neutrino in the $\mu^\pm \mu^\pm$
 modes in the mass range of $\sim 10 - 180\ \mbox{\gev} \ (10 - 120\ \mbox{\gev})$. This reach
can be extended  to $\sim 10 - 375\ \mbox{\gev}\ (10 - 250\
\mbox{\gev})$ at the LHC of 14 TeV with $100\ \mbox{fb}^{-1}$.
The production cross section at the LHC of 10 TeV is also presented for comparison.
We study the $\mu^\pm e^\pm$ modes as well and find that the signal
could be large enough even taking into account the current bound from
neutrinoless double-beta decay.
The signal from the gauge boson fusion channel $W^+ W^+
\rightarrow \ell^+_1 \ell^+_2$ at the LHC is found to be very weak
given the rather small mixing parameters. We comment on the search
strategy when a $\tau$ lepton is involved in the final state. }

\begin{document}
\section{Introduction}

 In the Standard Model (SM) of strong and electroweak interactions, neutrinos
are strictly massless due to the absence of right-handed chiral
states ($N_R$) and the requirement of $SU(2)_L$ gauge invariance
and renormalizability. Recent neutrino oscillation experiments
have conclusively shown that  neutrinos are massive
\cite{BargerReview}.  This discovery presents a pressing need to
consider physics beyond the SM. It is straightforward to obtain a
Dirac mass term $m_D (\overline{\nul} N_R + \mathrm{h.c.})$ for a
neutrino by including the right-handed state, just like the
treatment for all other fermions via Yukawa couplings to the Higgs
doublet in the SM. However,  a profound question arises:  Since
$N_R$ is a SM gauge singlet, why should a gauge-invariant Majorana
mass term ${1\over 2}M N_R N_R $ not exist in the  theory? In
fact, there is strong theoretical motivation for the Majorana mass
term to exist since it could naturally explain the smallness of
the observed neutrino masses via the so-called ``see-saw"
mechanism \cite{seesaw}
\beq m_\nu \approx {m^2_D \over M}. \label{seesaw}
\eeq
From a model-building point of view, there are many scenarios that could incorporate the Majorana mass. Examples include Left-Right symmetric gauge theories \cite{LRModels};  $SO(10)$ Supersymmetric (SUSY) grand unification \cite{SO10SUSYGUT} and other grand unified theories \cite{MGUT};  models  with exotic Higgs representations \cite{ZeeModel,MaModels}; R-parity violating interactions ($\dl=1$) in Supersymmetry (SUSY) \cite{RparitySUSY} and theories with extra dimensions \cite{ExtraDim}. There are other proposals to generate Majorana masses for neutrinos at a higher scale $M$ without relying on the right-handed state $N_R$ \cite{TypeII, TypeIII}. According to the scheme in generating the mass scale $M$ in Eq.~(\ref{seesaw}), it has been customary to call them Type I  \cite{seesaw}, Type II  \cite{TypeII} or Type III  \cite{TypeIII}.

Within the context of the SM, there is only one gauge-invariant
operator  \cite{dim6} that is relevant to the neutrino mass,
\beq {\kappa \over \Lambda} l_L^{} H\  l_L^{} H, \label{Wen} \eeq
where $l_L^{}$ and $H$ are the SM lepton and Higgs doublets,
respectively. The constant $\kappa$
is a model-dependent effective coupling and $\Lambda$
is the new physics cut-off scale.
It is a dimension-5 non-renormalizable operator, and
leads to Majorana neutrino masses of the order $\kappa
v^2/\Lambda$,  after the Higgs field acquires a vacuum expectation
value $v$, in accordance with the see-saw scheme. Higher
dimensional operators that give rise to Majorana neutrino masses
have also been constructed in a model-independent  manner
\cite{babuleung}. The challenging task is to look for experimental
evidence to probe the new physics scale $\Lambda$ and to
distinguish the underlying theoretical models mentioned above.

\begin{figure}[tb]
\center
\includegraphics[width=2.5in]{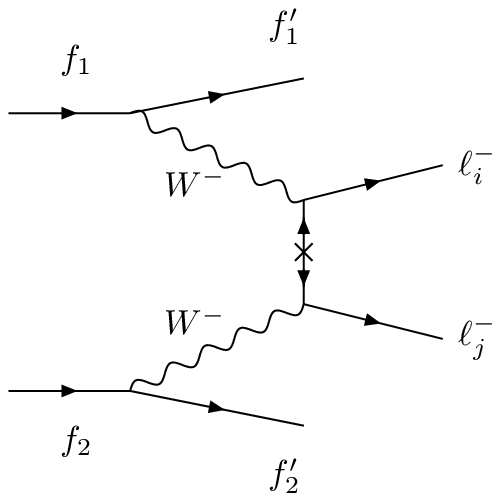}
\caption{A generic diagram for $\dl=2$ processes via Majorana
neutrino exchange.} \label{wwll}
\end{figure}

In the neutrino sector, besides the rich phenomena of neutrino flavor oscillations and
the possible existence of new sources of CP-violation,
lepton number violation by two units ($\dl=2$), as implied by
a Majorana mass term, plays a crucial role. Not only may it result in important
consequences in particle physics, nuclear physics and cosmology
but it would also guide us in understanding the fundamental
symmetries of physics beyond the SM.
Although the prevailing theoretical prejudice prefers Majorana
neutrinos, experimentally testing the nature of neutrinos and
lepton-number violation ($\lv$) in general, is of fundamental
importance. In accelerator-based experiments, neutrinos in the
final state are undetectable by the detectors, leading to the
so-called ``missing energy" and therefore missing lepton numbers
as well. One is thus  forced to look for charged leptons in the
final state. The basic process with $\dl=2$ can be generically
expressed by
\bea W^-W^- \rightarrow \ell^-_1 \ell^-_2, \label{ll} \eea
where $W^-$ is a virtual SM weak boson and $\ell_{1,2}=e,\mu,\tau$. By coupling fermion currents to the $W$ bosons as
depicted in Fig.~\ref{wwll}, and arranging the initial and final
states properly, one finds various physical processes that can be
experimentally searched for. The best known example is
neutrinoless double-beta decay ($0\nu\beta\beta$)
 \cite{nuless,DoiKotani,ElliottEngel}, which proceeds via the parton-level
subprocess $dd \to uu\ W^{-*} W^{-*} \to uu\ e^- e^-$. Other
interesting classes of $\lv$ processes involve tau decays such as
$\tau^- \to \ell^+ M_1^- M_2^-$ \cite{TauDecay, Atre} where the
light mesons $M_1,M_2$ are $\pi, K$, rare meson decays such as
$M^+_1 \to \ell^+_1 \ell^+_2 M^-_2$ \cite{Ng:1978ij,AliBorisov,
Atre} and hyperon decays such as $\Sigma^- \to \Sigma^+ e^- e^-$,
$\Xi^- \to p \mu^- \mu^-$ etc. \cite{Barbero}. One could also
explore additional processes like $e^- \to \mu^+$ \cite{CSLim},
$\mu^- \to e^+$ \cite{Atre,muep} and $\mu^- \to \mu^+$ conversion
\cite{Atre,mumup}. One may also consider searching for signals at
accelerator and collider experiments via $e^-e^- \to W^-W^-$
\cite{TRizzo}, $e^+e^- \to Z^0 \to N+X$ \cite{Dittmar:1989yg},
$e^\pm p \to \nu_e(\overline{\nu_e})\ell_1^\pm \ell_2^\pm X$
\cite{Rodejohann}, neutrino nucleon scattering
$\nu_\ell(\overline{\nu_\ell}) {\cal N} \to \ell^\mp \ell_1^\pm
\ell_2^\pm X$ \cite{Flanz}, $pp \to  \ell^+_1 \ell^+_2 X$
\cite{goran,NCollider,HZ,more,delAguila:2007em,NRCMS}, top-quark
decays $t \to b \ell^+_1 \ell^+_2 W^-$ \cite{Jing}, charged-Higgs
production $e^\pm e^\pm \rightarrow H^\pm H^\pm$ \cite{Soni}, and
in the decay $N\to \ell^\pm H^\mp$ \cite{BarShalom:2008gt}.

The dynamics for $\dl=2$ processes as in Eq.~(\ref{ll}) is
dictated by the properties of the exchanged neutrinos. For a
Majorana neutrino that is light compared to the energy scale in
the process, the transition rates for $\lv$ processes are
proportional to the product of two flavor mixing matrix elements
among the light neutrinos and a $\lv$ mass insertion
\bea \left<m\right>_{\ell_1\ell_2}^2 = \biggl | \sum_{m=1}^3
U_{\ell_1 m }U_{\ell_2 m} m_{\nu_m}^{} \biggr |^2 , \eea
where $\left<m\right>_{\ell_1\ell_2}$ are the  ``effective
neutrino masses". If the neutrinos are heavy compared to the
energy scale involved, then the contribution scales as
\beq \left| \sum_{m'=4}^{3+n}  \frac {V_{\ell_1 m'} V_{\ell_2
m'}}{ m_{N_{m'}}} \right|^2, \eeq
where $V$ is the mixing matrix between the light flavor and heavy
neutrinos. Unfortunately, both situations encounter a severe
suppression either due to the small neutrino mass like
$m_{\nu_m}^2/M_W^2$, or due to the small mixing $\left|V_{\ell_1
m'} V_{\ell_2 m'}  \right|^2$. An important observation is that
when the heavy neutrino mass is kinematically accessible, a
process may undergo resonant production of the heavy neutrino. The
transition  rate can be substantially enhanced and goes like
\beq {\Gamma(N_{m'} \to i)\ \Gamma(N_{m'}\to f) \over
m_{N_{m'}}^{}\Gamma_{N_{m'}}}, \eeq
where $i,f$ refer to the initial and final state during the
transition.

The possible existence of sterile neutrinos in the mass range
relevant for resonant enhancement of $\Delta L=2$ processes
studied in this paper is motivated in several scenarios. Models
which implement the see-saw mechanism at low energies have been
recently considered ~\cite{Bajc:2007zf,degouvea}: the neutrino
masses generated are accidentally small and active-sterile mixing
can be as large as few percent. See-saw models at the electroweak
scale  can explain neutrino masses if appropriate symmetries are
imposed and at the same time provide an appealing mechanism for
baryon asymmetry generation via resonant
leptogenesis~\cite{Pilaftsis}. In theories with dynamical
electroweak symmetry breaking, sterile neutrinos with masses in
the $100$s of MeV to GeV range are invoked to explain light
neutrino masses~\cite{ETC}. Sterile neutrinos can also play a role
in understanding the flavour problem in the leptonic sector. It
has been shown that mixing with sterile neutrinos can be at the
origin of the large angles in the neutrino sector
\cite{largeangle}.

Heavy, mostly-sterile neutrinos have been investigated for their
role in cosmology and astrophysics, in particular in Big Bang
Nucleosynthesis, Large Scale Structure
formation~\cite{dmstructure}, cosmic microwave background, diffuse
extragalactic background radiation, supernovae~\cite{kusenkoSN} and as dark matter
candidates~\cite{dodelson,fullerpatel,gelmini} (for a review on
MeV sterile neutrinos, see  Ref.~\cite{dolgov}).
A keV sterile
neutrino is a viable dark matter candidate
\cite{dodelson,gelmini}, which can also explain the origin of
pulsar kicks \cite{kusenko}. Decays of heavy,  mostly-sterile
neutrinos have been proposed to explain the early ionization of
the Universe \cite{hansen}. Due to mixing, dark matter sterile
neutrinos would decay radiatively contributing to the Diffuse
Extragalactic Background Radiation and inducing x-ray emission
from galaxy clusters~\cite{fullerxray,nuSMxray}. A large coupling
between sterile neutrinos and light dark matter scalars can be at
the origin of neutrino masses and of the observed dark matter
abundance~\cite{Boehm}. A model with sterile neutrinos in the
keV-GeV mass range has been proposed to explain the dark matter of
the Universe as well as baryogenesis \cite{nuSM,Asaka:2006ek}. Its
phenomenological and astrophysical signatures have been considered
in detail in Refs.~\cite{nuSMconstraints,nuSMxray}. This model
assumes the existence of one sterile neutrino with keV mass for
dark matter and two heavier neutrinos with quasi-degenerate GeV
masses for successful baryogenesis. The required mixing $|V_{\ell
m'}|^2$ of the latter neutrinos with the active ones is mass
dependent and lies in the range $10^{-11} - 10^{-8}$, for a mass
of $1$~GeV. Additional constraints on the heavy neutrino mass and
mixing angles can be derived from astrophysical observations.
Sterile neutrinos mixed with active ones would be efficiently
produced in supernovae cores, escaping from it and depleting
substantially the supernova core energy, and, therefore, might
modify the supernova evolution.
Recently, it was shown that sterile neutrinos in the mass range $\sim 0.2 \ \mathrm{GeV}$
and small mixing angle with $\mu$ and $\tau$ neutrinos could enhance
the energy transport from the core to the stalled shock and favor
the supernova explosion~\cite{kusenkoSN}.
They could also explain the high
velocity of pulsars if the momentum carried away by heavy sterile
neutrinos is emitted asymmetrically ~\cite{kusenko}. Detailed
reviews and discussions of heavy neutrinos in the Early Universe
and their present bounds can be found, e.g., in
Refs.~\cite{dolgov,Smirnov:2006bu,Kusenko:2007wv}.

Cosmological and astrophysical constraints on sterile neutrinos
are typically very strong but are not as robust as the ones from
laboratory searches as they typically depend on the production
mechanism of sterile neutrinos in the Early Universe and on the
cosmological evolution. For example, they can be significantly
weakened or evaded if the reheating temperature is low
~\cite{gelmini,EGPP}, if their density in the Early Universe is
diluted by entropy injected at late times~\cite{Asaka:2006ek} or
if they have non-standard interactions. In these cases, much
larger mixing angles with active neutrinos are allowed by
cosmological observations and can be tested in terrestrial
experiments. Therefore, it is important to perform experimental
searches of heavy sterile neutrinos with increased sensitivity
and, specifically for Majorana neutrinos, to consider $\dl=2$
processes. If a positive signal is found and is incompatible with
the cosmological and/or astrophysical observations, one would need
to consider modifications to the standard cosmological scenario
and/or would gain new insight on the evolution of astrophysical
objects.

In this paper, we study  resonant contributions of heavy Majorana
neutrinos to $\dl=2$ processes involving two charged leptons in
accelerator-based experiments. We establish our conventions  and
discuss the current constraints on the mass and mixing of heavy
neutrinos in Sec.~\ref{hvynus}. In Section \ref{lvps} we lay out
the general expressions for the heavy neutrino contributions to
low energy $\lv$ decays and study  two classes of $\dl=2$
processes,
\begin{itemize}
\item[(a)] tau decays $\tau^- \rightarrow \ell^+ M_1^- M_2^-$,
\item[(b)] rare meson decays $K^+,\ D^+,\ D^+_s,\ B^+  \rightarrow
\ell_1^+\ \ell_2^+\ M_2^-$.
\end{itemize}
We calculate the enhanced transition rates and branching fractions
and compare them to the bounds set by direct experimental
searches. A non-observation of such $\dl = 2$ processes places
stringent constraints on the mass and mixing of Majorana neutrinos
which are also presented in this section. The resonant production
of Majorana neutrinos at hadron colliders, namely the Tevatron and
LHC are studied and updated in Sec.~\ref{colsig}. We draw our
conclusions in Sec.~\ref{concld}. We discuss in detail the
formalism, the decay modes and the total decay width of heavy
Majorana neutrinos and the transition rates of $\lv$ processes in
the Appendices.

\section{Majorana neutrinos in extension of the standard model}
\label{hvynus}

To set up our notation and convention, we first discuss the
formalism for the simplest extension of the SM which includes
right handed singlets.  Also in this section, we present the
current constraints on the mass and mixing of a heavy neutrino
from various direct detection experiments, accelerator searches
and electroweak precision constraints.

\subsection{Formalism for Heavy Neutrino Mixing}
\label{formalism}

The leptonic content in our simplest extension of the SM includes
three generations of left-handed SM $SU(2)_L$ doublets and $n$
right-handed SM singlets
\beq L_{aL}=\left(
\begin{array}{c} \nu_a\\
l_a
\end{array}
\right)_L ,\quad   N_{b R}, \eeq
where $a=1,2,3$ and $b=1,2, \cdots, n$. The gauge-invariant Yukawa
interactions lead to Dirac masses for the charged leptons and
neutrinos after the Higgs field develops a vacuum expectation
value $v$. It  is also possible for the singlet neutrinos to have
a heavy Majorana mass term. The full neutrino mass terms  as well
as the diagonalized eigenvalues can be expressed as
\bea -{\cal L}_{m}^\nu &=&\frac{1}{2} \left(\ \sum_{a=1}^{3}
\sum_{b=1}^{n} \ ( \overline{\nu_{aL}}\ m^\nu_{a b} \ N_{b R} +
\overline{N^c_{bL}}\ m^{\nu *}_{ ba} \ \nu^c_{a R} ) +
\sum_{b,b'=1}^{n} \  \overline{N^c_{b L}}\ B_{b b'}\ N_{b' R}
\right)+\mathrm{h.c.} \nonumber \\
&=&
 {1\over 2}
\left(\sum_{m=1}^3 m_{\nu_m}\ \overline{\nu_{m L}}\  \nu^c_{m R} +
\sum_{m'=4}^{3+n} M_{N_{m'}}\ \overline{N^c_{m'L}}\ N_{m'R} \right)
+ \mathrm{h.c.} \label{numass} \eea
with the mixing relations  between the gauge and mass eigenstates
\bea
&& \nu_{aL} = \sum_{m=1}^3 U_{a m}\nu_{m L}+\sum_{m'=4}^{3+n} V_{am'} N^c_{m'L}, \\
&& U U^\dag + V V^\dag = I. \label{numix} \eea

In terms of the mass eigenstates, the gauge interaction Lagrangian
can be written as
\bea \nonumber \label{CCc} -{\cal L} &=& \frac{g}{\sqrt{2}}
W^+_\mu \left( \sum_{\ell=e}^\tau \sum_{m=1}^3 U^*_{\ell m}\
\overline{\nu_m} \gamma^\mu P_L \ell + \sum_{\ell=e}^\tau
\sum_{m'=4}^{3+n}
V^{*}_{ \ell m'}\ \overline{N^c_{m'}} \gamma^\mu P_L \ell  \right)+ \mathrm{h.c.} \\
&+& \frac{g}{2\cos \theta_W}Z_\mu \left( \sum_{\ell=e}^\tau
\sum_{m=1}^3 U^{*}_{\ell m}\  \overline{\nu_m} \gamma^\mu P_L\
\nu_\ell + \sum_{\ell=e}^\tau \sum_{m'=4}^{3+n} V^{*}_{ \ell m'}\
\overline{N^c_{m'}} \gamma^\mu P_L\ \nu_\ell  \right) + \mathrm{h.c.}
\label{NCc} \eea
Further details about the mixing formalism are given  in Appendix
\ref{appmix}.

A few important  remarks are in order before the detailed
considerations. First of all, parameterically, the light neutrino
masses $m_{\nu, \ diag}$ are of the order of magnitude
$(m^\nu_D)^2/B$, while the heavy neutrino masses are
$M_{N, \ diag}\simeq B$. Secondly, the mixing parameters would
typically scale as  $U^\dag U\approx I$ and  $V^\dag V\approx
m_\nu/M_N$. Thirdly, the Majorana mass term for the flavor states $\nu_{a L}$,
absent in Eq.~(\ref{numass}) and
corresponding to the null entry $0_{3\times 3}$ in
Eq.~(\ref{A-numass}), may receive non-zero
contributions as  Majorana masses  for the light active neutrinos,
for instance from higher dimensional $\dl=2$ operators or in
theories with a triplet Higgs field. The general formalism
presented here remains the same. In this paper, we will take a
phenomenological approach toward these parameters. We will simply
take the masses and mixing elements of the heavy neutrino as free
parameters, only subject to some constraints from experimental
observations. The assumption that the masses and mixing elements
are not rigorously related by the see-saw relations is feasible
from a model-building point of view, since some fine-tuning or
some ansatz of the neutrino mass matrix can always alter the
general relations. Several scenarios where it is possible to have
rather low mass of the heavy neutrino were mentioned in the
previous section. Here and henceforth, we consider the case when
only one heavy Majorana neutrino is kinematically accessible and
denote it by $N_4$, with the corresponding mass $m_4$ and mixing
with charged lepton flavors $V_{\ell 4}$. If we stick with this
simple parameterization, the SM Higgs boson will couple to the
heavy neutrinos as well. We present the couplings  in Appendix
\ref{appmix}. When appropriate, we will include this effect. As
noted above, some fine-tuning \cite{Chao:2008mq} would be needed
to avoid excessive contributions to the light neutrino mass.

\subsection{Current Constraints on $\nuh$ Masses And Mixing}
\label{m4mix}

In laboratory searches, no positive evidence of sterile neutrinos
has been found so far,\footnote{Indications of the existence of a
neutrino with 17 keV mass were subsequently shown to be non valid.
For a review, see Ref.~\cite{Wietfeldt:1995ja}. Studying
interactions of neutrinos from $\pi$ and $\mu$ decays, an anomaly
in time distribution was found~\cite{Armbruster:1995nr}. It could
be interpreted as the existence of a neutrino emitted in pion
decays with mass of 33.9 MeV. Searches for this neutral fermion
have not given any positive signature and have allowed to
constrain the mixing to be $|V_{\mu 4}|^2 < 9.2 \times 10^{-8}$ at
95\% C.L. \cite{Daum}.}  in the mass range of interest,
100~eV--100~GeV.\footnote{For sterile neutrinos with smaller
masses a rather complete analysis of the bounds can be found in
Ref.~\cite{Cirelli:2004cz}. See also the implications of the
recent results from the MiniBooNE
collaboration~\cite{Aguilar-Arevalo:2007it,Maltoni:2007zf}.}

A very powerful probe  of the mixing of heavy neutrinos with both
$\nu_e$ and $\nu_\mu$ are peak searches in leptonic decays of
pions and kaons ~\cite{Shrock:1980vy}. If a heavy neutrino is
produced in such decays, the lepton spectrum would show a
monochromatic  line at
\beq E_\ell = \frac{m_M^2 + m_\ell^2 - m_4^2}{2 m_M}, \eeq
where $E_\ell$ and $m_\ell$ are respectively the lepton energy and
mass, $m_M$ is the meson mass. The mixing angle controls the
branching ratio of this process as:
\beq \frac{\Gamma \big( M^+ \rightarrow \ell^+ N_4\big) }{\Gamma
\big( M^+ \rightarrow \ell^+ \nu_\ell \big)} =
\frac{|V_{\ell4}|^2} {\sum^3_{m=1}|U_{\ell m}|^2} \rho \approx
|V_{\ell4}|^2 \rho~, \label{rho} \eeq
where $\rho$ is a kinematical factor~\cite{Shrock:1980vy}:
\beq \rho = \frac{\sqrt{1 + \mu^2_\ell + \mu^2_4 - 2
\big(\mu_\ell+ \mu_4 + \mu_\ell \mu_4 \big) } \Big(\mu_\ell+ \mu_4
- \big( \mu_\ell- \mu_4 \big)^2\Big) }{ \mu_\ell \big( 1 -
\mu_\ell \big)^2}, \eeq
with $\mu_i=m_i^2/m_M^2$. For large $m_4$, the helicity
suppression of the $\pi, K \rightarrow \ell \nu_\ell$ decays
weakens and there is an enhancement for $M^+ \rightarrow \ell^+
N_4$ by a relative factor $m_4^2/m_\ell^2$, reaching up to
$10^4-10^5$ compared to that of $\pi \rightarrow e \nu_e$ and $K
\rightarrow e \nu_e$ in the SM, respectively. These bounds are
very robust because they rely only on the assumption that a heavy
neutrino exists and mixes with $\nu_e$ and/or $\nu_\mu$.

Another strategy to constrain heavy neutrinos mixed with $\nu_e$,
$\nu_\mu$ and $\nu_\tau$, is via searches of the products of their
decays. If kinematically allowed, $\nuh$ would be produced in
every process in which active neutrinos are emitted with a
branching fraction proportional to the mixing parameter
$|V_{\ell4}|^2$. They would subsequently decay via Charged Current
(CC) and Neutral Current (NC) interactions into neutrinos and
other ``visible" particles, such as electrons, muons and pions.
Searches for these ``visible" decay-products were performed and
were used to constrain the mixing parameters. In beam dump
experiments, $\nuh$ would be produced in meson decays, with the
detector located far away from the production site. The
suppression of the flux of $\nuh$ needs to be taken into account
if the decay length is very short and, therefore, typically both
an upper and a lower bound on the mixing angle can be set.
Otherwise, the production can happen in the detector itself, as
for the limits obtained from a reanalysis of LEP data, using the
possible decays of the $Z^0$~\cite{Dittmar:1989yg} into heavy
neutrinos. In this case, large values of the mixing angle can be
excluded. These bounds are less robust than the ones previously
discussed. In fact, if the heavy neutrinos have other dominant
decay channels into invisible particles, these bounds would be
weakened, if not completely evaded. For example, a coupling of the
type $g N \nu \phi$ (see Ref.~\cite{gNnuphi,Boehm}), with $\phi$ a
scalar, can induce very fast decays,  which might dominate over
the ones induced by CC and NC interactions. In this case, if the
decay length is very short due to these strong interactions, the
flux of $\nuh$ might be suppressed at the far detector and the
bound would not apply. If the production happens in the detector
itself, the bounds would need to be recomputed considering the
branching fraction into ``visible" channels. Notice that we do not
report here the bounds from Ref.~\cite{decayswrong} as they do not
apply to the heavy neutrinos under consideration. In these
analyses it was assumed that heavy neutrinos were produced via
$Z^0 \rightarrow \nuh \bar{\nuh}$ with the same strength as an
active neutrino. In our scheme, this would correspond to having
mixing angle equal to 1. Then, the search for $\nuh$ decays in the
detector was used to constrain the heavy neutrino parameters.
These data should be reanalyzed considering that the production of
$\nuh$ is suppressed by $|V_{\ell4}|^2$. Comparing the expected
number of events with the backgrounds we estimate that typically
bounds of order $|V_{\ell4}|^2 < {\rm few} \ \times
10^{-3}$--$10^{-2}$ could be deduced. However, a detailed analysis
should be performed and we do not report these limits in our
figures.

For masses above the production threshold, additional constraints
can be obtained from lepton universality as the decay rates for
muons, pions, taus as well as the invisible decay width for the
$Z^0$ boson are modified with respect to the SM
predictions~\cite{Nardi:1994iv,bergmann,delAguila:2008pw}. Flavour changing neutral
current processes such as $\mu \rightarrow e \gamma$, $\mu
\rightarrow e e^+ e^-$ and $\mu$--$e$ conversion in nuclei are
affected by the existence of heavy sterile neutrinos and strong
limits can be obtained on the mixing with active neutrinos
~\cite{Ma:1980gm,Langacker:1988up,Tommasini:1995ii}. These bounds
are reported in Section~\ref{EWlepto}.

Finally, in Section~\ref{ewpt} we discuss the very strong
constraints on $|V_{e4}|^2$ which can be obtained from the
non-observation of neutrinoless double beta decay. It should be
noticed that in the presence of more than one sterile neutrino,
possible cancellations between the contributions to the decay rate
can be achieved and the bounds would be consequently much weaker.

Next, we review the laboratory constraints on the mixing between
heavy and active neutrinos, depending on flavour and the mass of
sterile neutrinos.

\subsubsection{\bf Mixing with $\nu_e$}

The mixing parameter $V_{e4}$ can be tested in searches of kinks
in the $\beta$-decay spectrum, of peaks in the spectrum of
electrons in meson decays and, finally, of $\nuh$ decays in
reactor and accelerator neutrino experiments.

\begin{figure}
\center
\includegraphics*[width=0.95\textwidth] {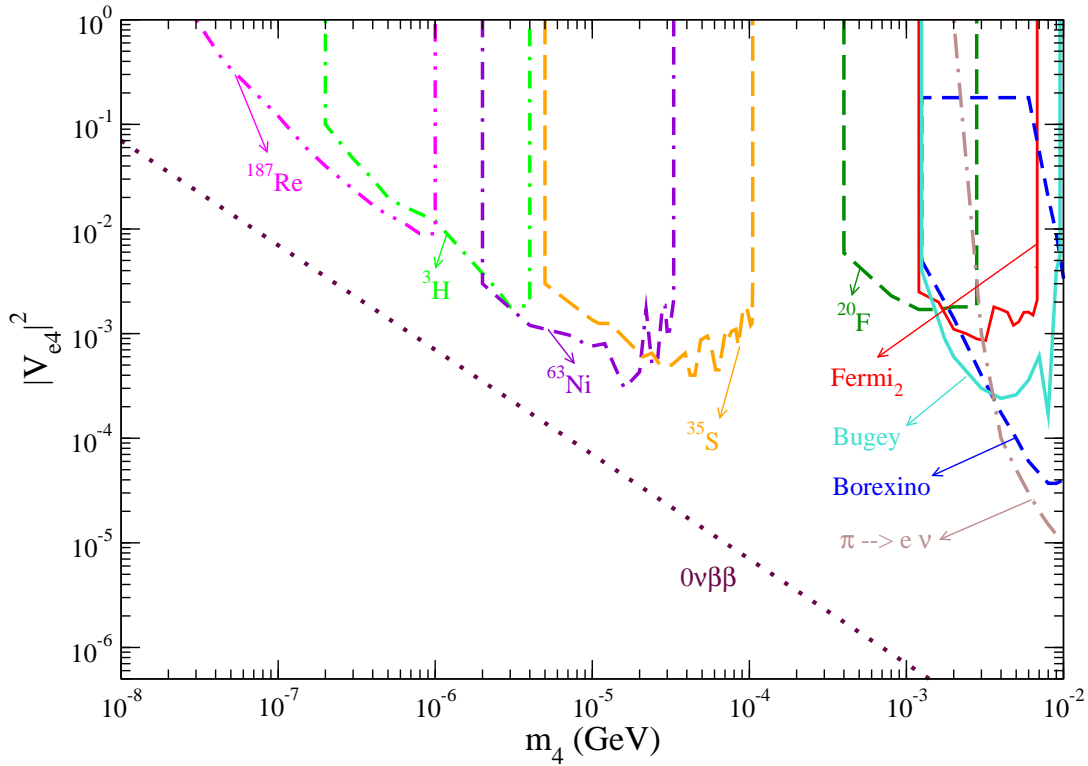}
\caption{Bounds on $|V_{e4}|^2$ versus $m_4$ in the mass range
10~eV--10~MeV. The excluded regions with contours labeled
$^{187}$Re \protect~\cite{Galeazzi:2001py}, $^3$H
\protect\cite{Hiddemann:1995ce} , $^{63}$Ni
\protect\cite{Holzschuh:1999vy} , $^{35}$S
\protect\cite{Holzschuh:2000nj} , $^{20}$F and Fermi$_2$
\protect\cite{Deutsch:1990ut} refer to the bounds from kink
searches. All the limits are given at 95\%~C.L. except for the
ones from Ref.~\protect\cite{Deutsch:1990ut} which are at
90\%~C.L.. The areas delimited by short dashed (blue) contour
labeled Borexino and solid (cyan) contour labeled Bugey are
excluded at 90\%~C.L. by searches of \protect$\nuh$ decays from
the Borexino Counting Test facility~\protect\cite{Back:2003ae} and
Ref.~\protect\cite{Hagner:1995bn} respectively. The region with
long-dash-dotted (grey) contour, labelled $\pi \rightarrow e \nu$,
is excluded by peak searches~\cite{Britton:1992pg}. The dotted
(maroon) line labeled $0\nu\beta\beta$ indicates the bound from
searches of neutrinoless double beta-decay
\protect\cite{Benes:2005hn}. } \label{fig:Uekinks}
\end{figure}

For masses $30 \ {\rm eV} \simeq m_4 \simeq 1$~MeV, the most
sensitive probe is the search for kinks in the $\beta$-decay
spectra~\cite{Shrock:1980vy}. In the presence of heavy neutrinos
mixed with $\nu_e$, the Kurie plot would be given by the
contributions of the decays into light neutrinos as well as into
heavy ones. This induces a kink in the Kurie plot at the end point
electron energy $E_e$
\beq E_e = \frac{M_i^2 + m_e^2 - (M_f + m_4)^2}{2 M_i}, \eeq
where $M_{i, f}$ are the mass of the initial and final nuclei,
respectively, and $m_e$ is the electron mass. In
Fig.~\ref{fig:Uekinks} we report the most stringent limits,
obtained by using different nuclei
~\cite{Galeazzi:2001py,Hiddemann:1995ce,Holzschuh:1999vy,Holzschuh:2000nj,Deutsch:1990ut}.
In reactors and in the Sun only low mass, $m_4 <$~few~MeV, heavy
neutrinos can be produced. The constraints obtained by looking for
their decays into electron-positron pairs \cite{Back:2003ae,Hagner:1995bn} are reported in
Fig.~\ref{fig:Uekinks} with solid (cyan) contour labeled Bugey and
short dashed (blue) contour labeled Borexino. The region with long
dash dotted (grey) contour, labelled $\pi \rightarrow e \nu$, is
excluded by peak searches~\cite{Britton:1992pg}. 

\begin{figure}
\center
\includegraphics[width=0.95\textwidth]{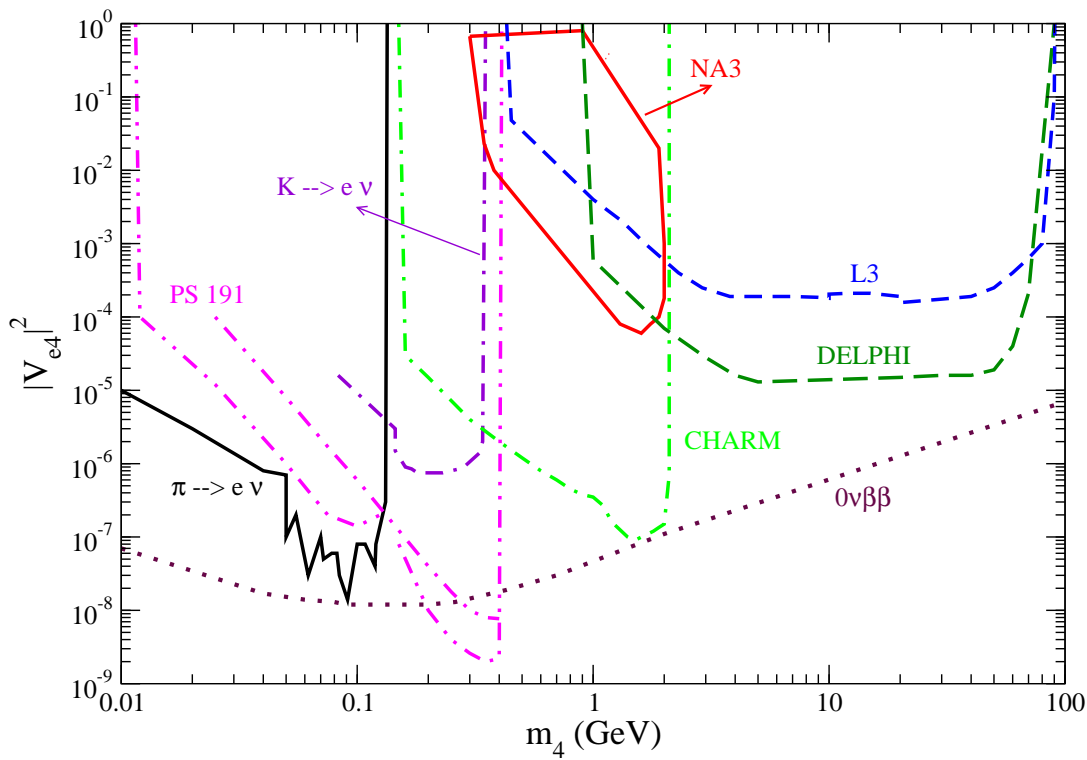}
\caption{ Bounds on $|V_{e4}|^2$ versus $m_4$ in the mass range
10~MeV--100~GeV. The areas with solid (black) contour labeled $\pi
\rightarrow e \nu$ and double dash dotted (purple) contour labeled
$K  \rightarrow e \nu$ are excluded by peak
searches~\protect\cite{Britton:1992pg,knupeak}. Limits at
90\%~C.L. from beam-dump experiments are taken from
Ref.~\protect\cite{Bernardi:1987ek} (PS191),
Ref.~\protect\cite{Badier:1986xz} (NA3) and
Ref.~\protect\cite{Bergsma:1985is} (CHARM).
The limits from contours labeled DELPHI and L3 are at 95\%~C.L.
and are  taken from Refs.~\protect\cite{Abreu:1996pa} and
\protect\cite{Adriani:1992pq} respectively. The excluded region
with dotted (maroon) contour is derived from a reanalysis of
neutrinoless double beta decay experimental data
\protect\cite{Benes:2005hn}. } \label{fig:Uepeak}
\end{figure}

For heavier masses peak searches give the most stringent bounds,
shown in Fig.~\ref{fig:Uepeak}. Notice that, due to the weakened
helicity suppression of the $\pi$ decay,  the sensitivity on
$V_{e4}$  increases with $m_4$ till phase space becomes relevant
at $m_4 > 80$~MeV, for $\pi \rightarrow e \nu_e$. The excluded
region, at 90\%~C.L., from Ref.~\cite{Britton:1992pg}, is
indicated with the solid (black) line labeled $\pi \rightarrow e
\nu$. For heavier masses, stringent bounds are obtained by looking
at the electron spectrum in $K$ decays~\cite{knupeak} and are
indicated by the double dash dotted (purple) line labeled $K
\rightarrow e \nu$ in Fig.~\ref{fig:Uepeak}. Assuming that only CC
and NC interactions are at play, stringent bounds have been
obtained on $|V_{e4}|^2$ and are reported in Fig.~\ref{fig:Uepeak}
by the rest of the contours (except dotted (maroon) line labeled
$0\nu\beta\beta$). In particular, the limits at 90\%~C.L. from
Refs.~\cite{Bernardi:1987ek,Badier:1986xz,Bergsma:1985is}, assume
the production of $\nuh$ in meson decays and look for visible
channels in a detector located some distance from the source. The
limits at 95\%~C.L. in Refs.~\cite{Abreu:1996pa,Adriani:1992pq}
analyse the data from DELPHI and L3 detectors, looking for
$\nuh$ from $Z^0$-decays. In Fig.~\ref{fig:Uepeak} we also report
the excluded region from neutrinoless double beta-decay
experiments~\cite{sub,Benes:2005hn}, bounded by dotted (maroon) line,
valid if the heavy neutrinos are Majorana particles (see further).

\subsubsection{\bf Mixing with $\nu_\mu$}

The bounds on $|V_{\mu 4}|^2$ come from searches of peaks in the
spectrum of muons in pion and kaon decays and of the decays of
$\nuh$ produced in neutrino beams and $e^+ e^-$ collisions.

\begin{figure}
\includegraphics[width=0.95\textwidth]{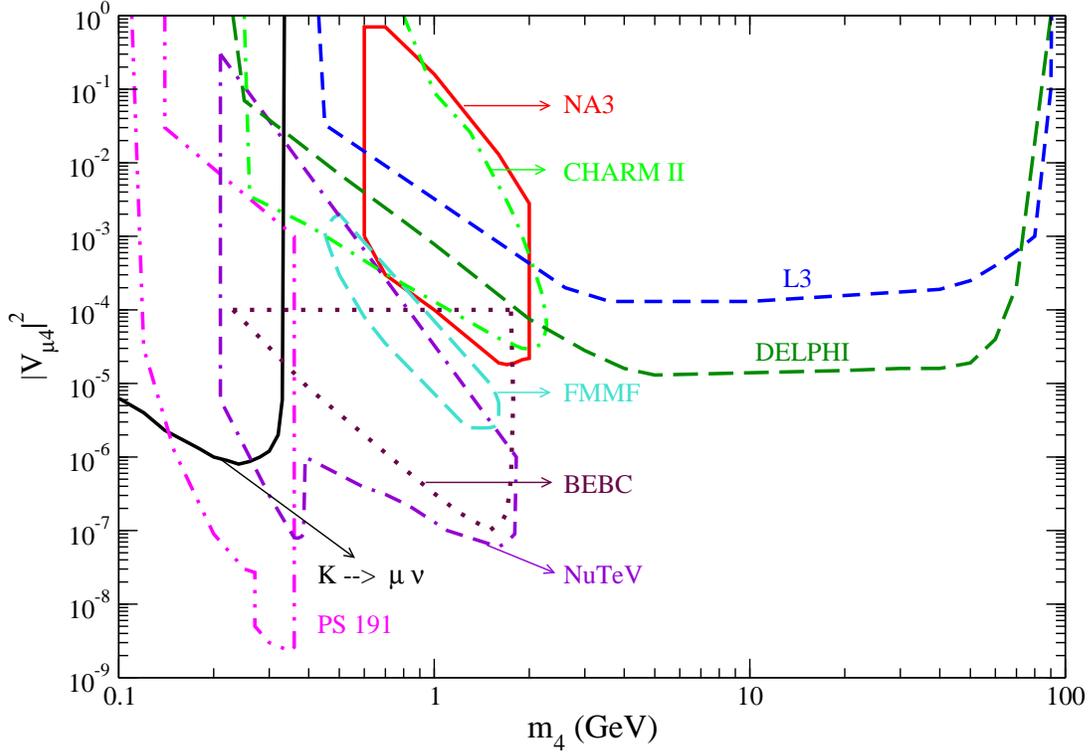}
\caption{ Limits on $|V_{\mu 4}|^2$ versus $m_4$ in the mass range
100~MeV--100~GeV come from peak searches and from $\nuh$ decays. The
area with solid (black) contour labeled $K \rightarrow \mu \nu$
\protect\cite{Kusenko:2004qc} is excluded by peak searches. The
bounds indicated by contours labeled by
PS191~\protect\cite{Bernardi:1987ek}, NA3
\protect\cite{Badier:1986xz}, BEBC \protect\cite{BEBC}, FMMF
\protect\cite{Gallas:1994xp}, NuTeV \protect\cite{Vaitaitis:1999wq}
and CHARMII \protect\cite{Vilain:1994vg} are at 90\%~C.L.,  while
DELPHI \protect\cite{Abreu:1996pa} and L3
\protect\cite{Adriani:1992pq} are at 95\%~C.L. and are deduced from
searches of visible products in \protect$\nuh$ decays. For the beam
dump experiments, NA3, PS191, BEBC, FMMF and NuTeV we give an
estimate of the upper limit for the excluded values of the mixing
angle.} \label{fig:Umu}
\end{figure}

As already discussed in the case of mixing with $\nu_e$, peak
searches provide very robust and stringent bounds, by looking at
pion decays for masses up to 34~MeV, and at kaon decays for higher
masses. A detailed review is given in Figs.~1 and 2 in
Ref.~\cite{Kusenko:2004qc} and for masses larger than 100~MeV the
limits are reported in Fig.~\ref{fig:Umu}.
The other limits on $|V_{\mu 4}|^2$ are found in decay searches and
are also shown in Fig.~\ref{fig:Umu}. They come from beam dump
experiments
\cite{Badier:1986xz,Bernardi:1987ek,BEBC,Gallas:1994xp,Vaitaitis:1999wq}
and from direct $N_4$ production in the detectors
DELPHI~\cite{Abreu:1996pa}, L3~\cite{Adriani:1992pq} and
CHARM~\cite{Vilain:1994vg}.

\subsubsection{\bf Mixing with $\nu_\tau$}

\begin{figure}
\includegraphics[width=0.95\textwidth]{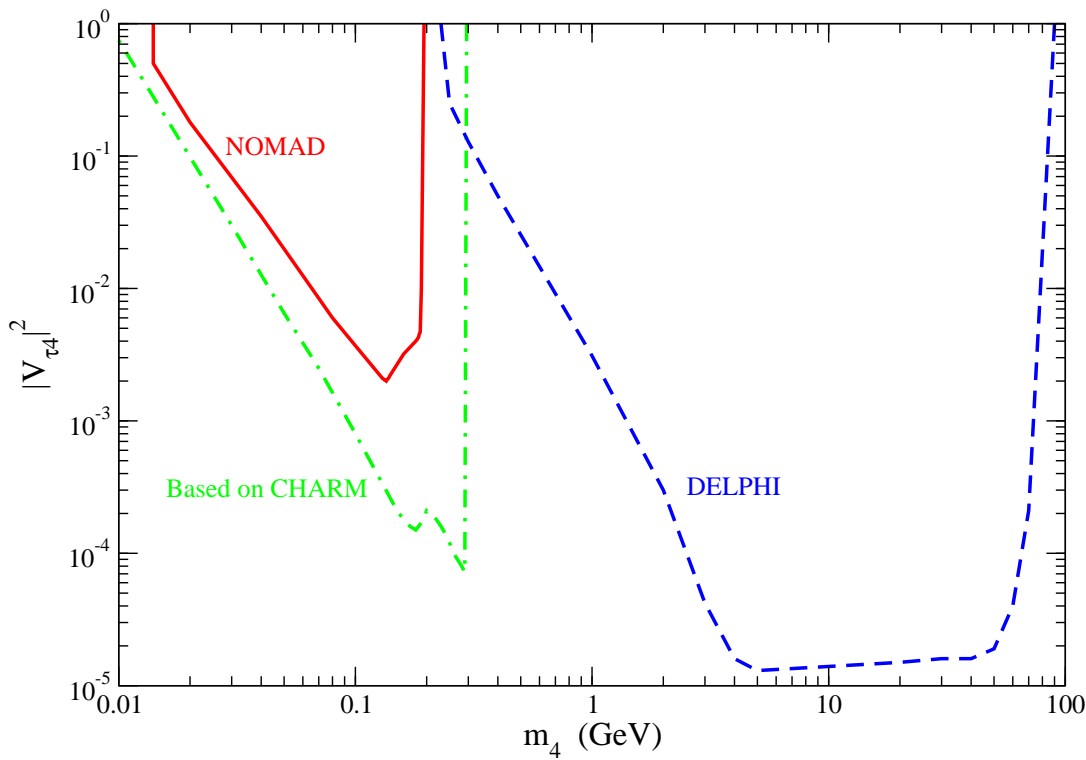}
\caption{ Bounds on $|V_{\tau 4}|^2$ versus $m_4$ from searches of
decays of heavy neutrinos, given in
Ref.~\protect\cite{Orloff:2002de} (CHARM) and in
Ref.~\protect\cite{Astier:2001ck} (NOMAD) at 90\%~C.L., and in
Ref.~\protect\cite{Abreu:1996pa} (DELPHI) at 95\%~C.L. }
\label{fig:Utau}
\end{figure}

Heavy neutrinos mixed with $\tau$ neutrinos can be produced either
via CC interactions if a $\tau$ is produced or in NC interactions.
The only limits come from searches of $\nuh$ decays and are
reported in Fig.~\ref{fig:Utau}. The bounds at 90\%~C.L.  from
CHARM~\cite{Orloff:2002de} and NOMAD ~\cite{Astier:2001ck} assume
production via $D$ and $\tau$ decays. The DELPHI bound at
95\%~C.L. \cite{Abreu:1996pa} assumes $\nuh$ production in $Z^0$
decays and with respect to the bound on $|V_{e4}|^2$ and
$|V_{\mu4}|^2$ there is $\tau$-production kinematical suppression
for low masses which weakens the constraint for masses in the
range $m_4 \sim 2$--3~GeV.

\subsubsection{\bf Electroweak Precision Tests}
\label{EWlepto}

The presence of heavy neutral fermions affects processes below
their mass threshold due to their mixing with standard
neutrinos~\cite{Nardi:1994iv} and significant bounds can be set by
precision electroweak data. The effective $\mu$-decay constant
$G_\mu$, measured in muon decays, is modified with respect to the
SM value and can be related to the fundamental coupling $G_F$ as:
\beq G_\mu = G_F \sqrt{(1 - |V_{e 4}|^2)(1 - |V_{\mu 4}|^2)}~.
\label{gfermi} \eeq
The $\mu-e$ universality test, done by comparing the decay rate of
pions into $e \bar{\nu}$ and $\mu \bar{\nu}$,  can be used to
constrain the ratio
\beq \frac{1 - |V_{e 4}|^2}{1 - |V_{\mu 4}|^2}, \eeq
for $m_4 > m_\pi$~\cite{Nardi:1994iv,bergmann}. The analysis of
experimental data leads to $\frac{1 - |V_{\mu 4}|^2}{1 - |V_{e
4}|^2} = 1.0012\pm 0.0016$ \cite{bergmann}, which implies $|V_{e
4}|^2 < 0.004$ at $2\sigma$ for the least conservative case of
$|V_{\mu 4}|^2=0$. For $m_4 > m_\tau$, the $\mu-\tau$ universality
sets limits on:
\beq \frac{1 - |V_{\tau 4}|^2}{1 - |V_{\mu 4}|^2}, \eeq
and can be tested by looking at the $\tau$ leptonic and hadronic
decays which give $|V_{\tau 4}|^2 - |V_{\mu 4}|^2 = 0.0057\pm
0.0065$~\cite{bergmann} and $|V_{\tau 4}|^2 - |V_{e 4}|^2 =
0.0054\pm 0.0064$~\cite{bergmann}. The most constraining bound on
$|V_{\tau 4}|^2$ is obtained for $|V_{e 4}|^2, |V_{\mu 4}|^2=0$
and reads $|V_{\tau 4}|^2 < 0.018$ at $2\sigma$. The unitarity
constraint on the first row of the CKM matrix~\cite{PDG} reads
\beq \sum_{i=1,2,3} |V^{\rm CKM}_{ui}|^2 = \frac{1}{1 - |V_{\mu
4}|^2}=0.9992 \pm 0.0011, \eeq
and  translates into a very strong bound on $|V_{\mu 4}|^2$,
$|V_{\mu 4}|^2 < 0.0003\ (0.0014)$, at $1\  (2) \sigma$, which
holds for sterile neutrinos heavier than the $\Lambda$ baryon.

In the presence of heavy singlet neutrinos heavier than half the
$Z^0$ mass, the invisible decay rate of $Z^0$ would be reduced
with respect to the SM one, $\Gamma_{Z \rightarrow
\mathrm{inv}}^{\rm SM}$, as:
\beq \frac{\Gamma_{Z \rightarrow \mathrm{inv}}}{\Gamma_{Z
\rightarrow \mathrm{inv}}^{\rm SM}} \simeq ( 1- \frac{1}{6} |V_{e
4}|^2 -\frac{1}{6} |V_{\mu 4}|^2 - \frac{2}{3} |V_{\tau 4}|^2).
\label{zwid} \eeq
 By a standard model fit to LEP data, the effective number of
neutrinos is now determined to be $N_\nu = 2.984 \pm
0.008$~\cite{PDG} and provides a bound on $|V_{\ell 4}|^2$ similar
to but somewhat weaker than the ones obtained by
lepton-universality.

A combined analysis of an old set of unitarity
bounds~\cite{bergmann}, which does not include the one from the
CKM matrix determination, leads to the following limits at
90\%~C.L. $ |V_{e4}|^2 < 0.012, |V_{\mu 4}|^2<  0.0096\
\mbox{and}\ |V_{\tau 4}|^2 <0.016. $ If the CKM matrix constraint
is included and partial cancellations between the contributions of
different flavors are taken into account, a previous combined
study~\cite{Nardi:1994iv} then gives the more robust limits at
$90\%$~C.L., $|V_{e4}|^2 < 0.0066,  |V_{\mu 4}|^2<  0.0060$
 and $ |V_{\tau 4}|^2 <0.018$.
A very recent analysis \cite{delAguila:2008pw} has updated these results
using the latest electroweak precision data, except for the CKM observables.
They find at 90\%~C.L.
\beq
|V_{e4}|^2 < 0.003, \qquad |V_{\mu 4}|^2 < 0.003, \qquad |V_{\tau 4}|^2 < 0.006 ~.
\eeq
If the constraints from CKM observables are included, we expect the bounds to become somewhat stronger, given by $|V_{e4}|^2 < 0.002, |V_{\mu 4}|^2 < 4 \times 10^{-5}, |V_{\tau 4}|^2 < 0.006 $~\cite{delaguilaprivate}. In the following, we take the bound
$|V_{\mu4}|^2 < 0.0060$ as a conservative reference limit on the mixing for comparison with the results of our study.

Indirect limits on the parameters characterizing heavy sterile
neutrinos can be obtained from searches for flavour changing
neutral current processes such as $\mu \rightarrow e \gamma$, $\mu
\rightarrow e e^+ e^-$ and $\mu-e$ conversion in
nuclei~\cite{Tommasini:1995ii}. The branching fraction for
$\mu\rightarrow e \gamma$ induced by the mixing with heavy singlet
neutrinos is given
by~\cite{Ma:1980gm,Langacker:1988up,Tommasini:1995ii}:
\beq {\rm Br }(\mu \rightarrow e \gamma) = \frac{3 \alpha}{8 \pi}
\left| \sum_{m'} V_{em'} V_{\mu m'}^\ast\ g
\left(\frac{m_{N_{m'}}^2}{m^2_W}\right) \right|^2 ~, \eeq
where $m'$ indicates the heavy sterile neutrinos with mass
$m_{N_{m'}}$, and $m_W$ is the mass of the $W$ boson. The function
$g(x)$ is given by
\beq g(x) = \frac{x ( 1 - 6 x + 3 x^2 + 2 x^3 -
6 x^2 \ln(x) ) }{2 (1-x)^4},
\eeq
where $g(x)$ goes from 0 to 1 as $x$ varies from 0 to infinity. At
present, the branching fraction is constrained to be ${\rm Br }(\mu
\rightarrow e \gamma)< 1.2 \times 10^{-11}$~\cite{Brooks:1999pu} at
90\%~C.L. implying that, for one extra sterile neutrino, $|V_{e4}
V_{\mu 4}^\ast| < 0.015 \ (3.5 \times 10^{-4}) \ [ 1.2 \times
10^{-4}]$ for $m_4 = 10 \ {\rm GeV} \ (100 \ {\rm GeV}) \ [1000 \
{\rm GeV}]$. Similar constraints are imposed by searches for the
processes $\mu-e$ conversion in nuclei and $ \mu \rightarrow e e^+
e^-$~\cite{Tommasini:1995ii}.  The current strongest bound comes
from the search for $\mu-e$ conversion in $\mathrm{Ti}$ for which
the branching ratio with respect to the total nuclear muon capture
rate is constrained to be ${\rm Br }(\mu \, {\rm Ti} \rightarrow e
\, {\rm Ti}) < 4.3 \times 10^{-12}$ at
90\%~C.L.~\cite{Dohmen:1993mp}. For one sterile neutrino, this
translates into a bound on the following quartic combination of
mixing angles $\left| V_{e4} V_{\mu 4}^\ast \sum_\ell |V_{\ell 4}|^2
\right| < 1.3 \times 10^{-3} \left( 100 \, {\rm GeV}/m_4\right)^2$,
which is weaker than the bounds from $ \mu \rightarrow e \gamma $
searches but becomes important at very high values of the masses,
$m_4 \gsim 10 \ {\rm TeV}$. In the presence of more than one sterile
neutrino, partial cancellations between their contributions are
possible, with a consequent weakening of the bounds. Future more
sensitive searches will further improve these limits.

The most stringent EW precision constraints are compiled in
Table~\ref{tabewpt} and include the bounds reported above from
universality tests and lepton flavor changing processes. These
bounds are obtained barring cancellations between mixing angles
and therefore could be weakened if some parameters are of the same
order.
\bet[tb] \caption{Most stringent model-independent constraints on
the mixing elements of the heavy neutrino from precision
electro-weak measurements. The bounds on $|V_{\ell 4}|^2$,
$\ell=e, \mu, \tau$ and on $|V_{e
4}V_{\mu 4}|$ at 90\%~C.L.. See text for details. }
\begin{center}
\begin{tabular}
{|c|c|c|} \hline
Mixing element & Range of $m_4$ & EW Measurement  \\
\hline
 $|V_{e 4}|^2$ & $m_4 \gsim {\cal O}(m_\pi)$ & $<0.003  ~\protect\cite{delAguila:2008pw}$ \\
 \hline
 $|V_{\mu 4}|^2$ & $m_4 \gsim {\cal O}(m_\Lambda)$ & $<0.003 ~\protect\cite{delAguila:2008pw} $ \\
 \hline
 $|V_{\tau 4}|^2$ & $m_4 \gsim {\cal O}(m_\tau)$ & $<0.006 ~\protect\cite{delAguila:2008pw}$ \\
 \hline
 $|V_{e 4}V_{\mu 4}|$ &  10~GeV \  (100~GeV) \ [1000~GeV] \ & $<0.015 \ (3.5 \times 10^{-4}) \ [1.2 \times 10^{-4}] \ $ \\
\hline
\end{tabular}
\end{center}
\label{tabewpt} \eet

\subsubsection{\bf Neutrinoless Double Beta Decay ($0 \nu \beta \beta$) }
\label{ewpt}

The most well studied among $\dl = 2$ processes is neutrinoless
double beta decay ($0 \nu \beta \beta$) and the constraints from
it deserve special attention. The constraints on $|V_{e 4}|^2$ for
a wide range of heavy neutrino masses ($10\ \mev \le m_4 \le 100\
\gev$) are shown in Figs.~\ref{fig:Uekinks} and \ref{fig:Uepeak}.
For heavy neutrinos with mass, $m_{N_{m'}} \gg 1~{\rm GeV}$, the
bound is \cite{sub,Benes:2005hn}
 \bea
\sum_{m'} \frac {\left| V_{e m'}\right| ^{2}}{m_{N_{m'}}}%
<5\times 10^{-5}~{\rm TeV}^{-1}. \label{0nubb} \eea
The constraint above is very strong and makes it impossible to
observe at colliders the like-sign dilepton signature with
electrons (see Sec.~\ref{colsig}).

\section{Lepton-Number Violating Decays}
\label{lvps}

The key point for the search of lepton-number violating  processes
in this paper is to consider the substantial enhancement via
resonant neutrino production. One thus needs to evaluate the decay
widths of $N_4$ to various channels. We consider the decay width
of the heavy Majorana neutrino in two regimes: when the mass is
much smaller than that of the $W$ boson and when the mass is
larger than the mass of the $W$ boson. Based on this, we then
compute the $\dl=2$ decay branching fractions for $\tau$ lepton
and $K, D, D_s$ and $B$ mesons.

\subsection{Decay Modes of Heavy Majorana Neutrino}
\label{decaymodes}

\subsubsection{Decay Modes of Heavy Majorana Neutrino with mass $m_4 \ll m_W$}
\label{smlm4}

For the $\lv$ low energy tau decays and rare meson decays the
resonant contribution is from a heavy Majorana neutrino with mass
of order $\mev$ to $\gev$. In this section we discuss the decay
modes of a Majorana neutrino which is lighter than the $W$ boson,
so that $m_4 \ll m_W$. The heavy neutrino decays via charged and
neutral current interactions to the modes listed below.
The partial decay widths of the heavy Majorana neutrino with the
leading terms in mixing and in the massless limit of the final
state particles are given below. The full detailed expressions for
the same are given in Appendix \ref{decmod}.
\bea \label{stnudecs} \label{2bodyCC}
\Gamma^{\ell P} &\equiv& \Gamma(N_4 \rightarrow \ell^- P^+)= \frac {G^2_F}{16 \pi} f^2_P\ |V_{q \bar q'}|^2\ |V_{\ell 4}|^2\ m^3_4,\\
\label{2bodyNC}
\Gamma^{\nu_\ell P} &\equiv& \Gamma(N_4 \rightarrow \nu_\ell P^0)= \frac {G^2_F}{64 \pi} f^2_P\ |V_{\ell 4}|^2\ m^3_4,\\
\label{2bodyCCV}
\Gamma^{\ell V} &\equiv& \Gamma(N_4 \rightarrow \ell^- V^+)= \frac {G^2_F}{16 \pi} f^2_V\ |V_{q \bar q'}|^2\  |V_{\ell 4}|^2\ m^3_4 ,\\
\label{2bodyNCV}
\Gamma^{\nu_\ell V} &\equiv& \Gamma(N_4 \rightarrow \nu_\ell V^0) = {\frac {G^2_F}{2 \pi}} {\kappa^2_V}\ {f^2_V}\ {{|V_{\ell 4}|}^2}\ {m^3_4} ,\\
\label{3bodycc}
\Gamma^{\ell_1 \ell_2 \nu_{\ell_2}} &\equiv& \Gamma(N_4 \rightarrow \ell^-_1 \ell^+_2 \nu_{\ell_2}) = \frac{G^2_F}{192 \pi^3}\ {|V_{\ell_1 4}|}^2\ m^5_4,\\
\label{3bodyNC1}
\Gamma^{\nu_{\ell_1} \ell_2 \ell_2}&\equiv& \Gamma(N_4 \rightarrow \nu_{\ell_1} \ell^-_2 \ell^+_2 )= \frac {G^2_F}{96 \pi^3}\ {|V_{\ell_1 4}|}^2\ m^5_4\ [\alpha_1 + \delta_{\ell_1 \ell_2} \alpha_2 ],\\
\label{3bodyNC2} \Gamma^{\nu_{\ell_1} \nu \nu} &\equiv& \sum_{\ell_2=e}^\tau
\Gamma(N_4 \rightarrow \nu_{\ell_1} \nu_{\ell_2} \overline{\nu_{\ell_2}} )=
\frac {G^2_F}{96\pi^3}\ |V_{\ell_14}|^2\ m^5_4,
\eea
where $P^{+(0)}$ and $V^{+(0)}$ are charged (neutral) pseudoscalar
and vector mesons, $f_{M}$ are the meson decay constants and $V_{q
\bar q'}$ are the CKM matrix elements.

All the decay modes listed above contribute to the total decay
width of the heavy Majorana neutrino which is given by:
\bea \label{totwid_small} \nonumber
\Gamma_{N_4}& =& \sum_{\ell, P}{ \Gamma^{\nu_\ell P}} + \sum_{\ell, V} {\Gamma^{\nu_\ell V}} + \sum_{\ell,P} {2 \Gamma^{\ell P}} +  \sum_{\ell,V} {2 \Gamma^{\ell V}} \\
&+&  \sum_{\ell_1,\ell_2(\ell_1 \ne \ell_2)}{2\Gamma^{\ell_1
\ell_2 \nu_{\ell_2}}} + \sum_{\ell_1, \ell_2} {\Gamma^{\nu_{\ell_1} \ell_2 \ell_2}} +  \sum_{\ell_1}
{\Gamma^{\nu_{\ell_1} \nu \nu}},
 \eea
where $\ell, \ell_1, \ell_2 = e, \mu, \tau$. For a
Majorana neutrino, the $\Delta L = 0$ process $N_4 \rightarrow
\ell^- P^+$ as well as its charge conjugate $\Delta L = 2$ process
$N_4 \rightarrow \ell^+ P^-$ are possible and have the same width
$\Gamma^{\ell P}$. Hence the factor of 2 associated with the decay
width of this mode in Eq.~(\ref{totwid_small}). Similarly, the
$\Delta L = 0$ and its charge conjugate $\Delta L = 2$ process are
possible for the decay modes $N_4 \rightarrow \ell^- V^+$ and $N_4
\rightarrow \ell^-_1 \ell^+_2 \nu_{\ell_2}$ and hence have a factor of 2
associated with their width in Eq.~(\ref{totwid_small}).

For the low energy $\lv$ tau decays and rare meson decays we
consider, the mass of the heavy neutrino is in the range $140\
{\mev} \lsim m_4 \lsim 5278\  \mev$. For this mass range we list
all the possible decay channels for $N_4$ in Table
\ref{newchannel} in Appendix \ref{decmod}. The mass and decay
constants of pseudoscalar and vector mesons used in the
calculation of partial widths given in
Eqs.~(\ref{2bodyCC})$-$(\ref{3bodyNC2}) are listed in Table
\ref{constants} in Appendix \ref{apprmd}.

Next, we get a rough numerical estimate of the total width of the
heavy neutrino with the decay modes discussed in
Eqs.~(\ref{2bodyCC})$-$(\ref{3bodyNC2}). To do this, we consider
the massless limit of the decay products of the heavy neutrino,
include only leading terms in mixing of ${\cal O}({|V_{\ell
4}|^2})$ and ignore small factors like $\pi$ and $|V^{CKM}|^2$ in
calculating the partial decay widths. We can only get a rough
estimate of the width in this approximation, but it is sufficient
to see that it warrants the use of narrow width approximation. The
two body decays of the heavy neutrino have a general form
\beq \label{2bodyapprox} \Gamma^{2 body} \sim \frac {G^2_F f^2_M
m^3_4}{10 \pi}{|V_{\ell 4}|^2} \sim \frac {G^2_F f^2_M
m^3_4}{10}{|V_{\ell 4}|^2} \sim (10^{-13} \mbox { } {|V_{\ell
4}|^2}) \mbox { }\gev, \eeq
where typical values of $m_4 \sim 1\ \gev$, $f_M \sim 0.1\  \gev$
and $G_F \sim  10^{-5}\ \gev^{-2}$ have been used. The three body
decays of the heavy neutrino have a general form
\beq \label{3bodyapprox} \Gamma^{3 body} \sim \frac {G^2_F
m^5_4}{100 \pi^3}{|V_{\ell 4}|^2} \sim \frac {G^2_F
m^5_4}{1000}{|V_{\ell 4}|^2} \sim (10^{-13}\mbox { }{|V_{\ell
4}|^2}) \mbox { }\gev , \eeq
where typical values of $m_4 \sim 1\ \gev$ and $G_F \sim  10^{-5}\
\gev^{-2}$ have been used. The total width of the heavy neutrino
is then given by
\bea \label{totwidapprox} \nonumber \Gamma_{N_4} &=& \mbox
{(number of decay modes)} \times (\Gamma^{2 body} +
\Gamma^{3body}) \\ &\sim & 50 \times (10^{-13}\mbox { }\gev +
10^{-13} \mbox { } \gev) \mbox { }{|V_{\ell 4}|^2} \sim  (10^{-11}
\mbox { } {|V_{\ell 4}|^2}) \mbox { }\gev. \eea
As shown above, the width of the heavy neutrino $\sim {\cal
O}(10^{-11} \mbox { } {|V_{\ell 4}|^2}) \mbox { } \gev$ is much
smaller than the mass of the heavy neutrino $\sim {\cal O}(1\
\gev)$ and we can use the narrow width approximation to an
excellent approximation.

Now we look at the lifetime of the heavy Majorana neutrino to
determine the decay length. The lifetime is given by
\bea \nonumber
\tau_{N_4} &=& \frac{1}{\Gamma_{N_4}} \sim \frac{1}{10^{-11}\ {|V_{\ell 4}|^2}\ \gev}\ ,\\
&\sim& 10^{11}\ |V_{\ell 4}|^{-2}\ \gev^{-1} \sim 6.58 \times
10^{-14}\ |V_{\ell 4}|^{-2}\ s, \eea
which gives a typical decay length $c \tau_{N_4} \sim 1 \times
10^{-5}\ |V_{\ell 4}|^{-2}\ \mathrm{m}$. Note that for a very small mixing,
$|V_{\ell 4}|^{2} < {\cal O}(10^{-5}),$ the $N_4$ may escape from
the detector if it is not much heavier than a GeV. We will take
this effect into account in the following studies.

\subsubsection{Decay Modes of Heavy Majorana Neutrino with mass $m_4 > m_W$}
\label{bigm4}

In this section we discuss the decay modes of the Majorana
neutrino which is heavier than the $W$ gauge boson, so that $m_4 >
m_W$. The decay modes of the heavy Majorana neutrino are to a $W$
or a $Z$ gauge boson plus the corresponding SM lepton.  The
partial decay widths for longitudinal and transverse gauge bosons
$W^{\pm},Z^0$ in static heavy neutrino frame are
\bea
\Gamma^{\ell W_L} &\equiv& \Gamma(N_4\rightarrow \ell^- W^+_L) = \Gamma(N_4\rightarrow \ell^+ W^-_L) = \frac{g^2}{64\pi M^2_W} \left|V_{\ell 4} \right|^{2}\ m^3_4\ (1-\mu_W)^2,\\
\Gamma^{\ell W_T} &\equiv& \Gamma(N_4\rightarrow \ell^- W^+_T) = \Gamma(N_4\rightarrow \ell^+ W^-_T) = \frac{g^2}{32\pi}\left|V_{\ell 4}\right| ^{2}\ m_4\ (1-\mu_W)^2, \\
\label{nuZL}
\Gamma^{\nu_\ell Z_L} &\equiv& \Gamma(N_4\rightarrow \nu_\ell Z_L)= \frac{g^2}{64\pi M^2_W} {{|V_{\ell 4}|}^2}\ m^3_4\ (1-\mu_Z)^2,\\
\label{nuZT} \Gamma^{\nu_\ell Z_T} &\equiv&
\Gamma(N_4\rightarrow \nu_\ell
Z_T)=\frac{g^2}{32\pi\cos^2_W} {{|V_{\ell 4}|}^2}\ m_4\ (1- \mu_Z )^2, \eea
where $\mu_i$ are the masses of the gauge bosons scaled by the
mass of the heavy neutrino and are given by $\mu_i = m^2_i/m^2_4$.
To obtain the total decay width for $N_4$, we sum over the charged
leptons $\ell$ and as discussed earlier include the $\Delta L = 0$
process $N_4 \rightarrow \ell^- W_{L,T}^+$ as well as the charge
conjugate $\Delta L = 2$ process $N_4 \rightarrow \ell^+
W_{L,T}^-$. Hence the factor of $2$ associated with the decay
width of these modes in the expression for the total width below.
\beq \label{totwid_big} \nonumber \Gamma_{N_4}= \sum_\ell{
\Bigl ( 2\Gamma^{\ell W_L}} + {2 \Gamma^{\ell W_T}} + {\Gamma^{\nu_\ell Z_L}} +  {\Gamma^{\nu_\ell Z_T}} \Bigr ).
 \eeq
In Eqs.~(\ref{nuZL})$-$(\ref{totwid_big}), we have used the relation
(see Appendix for details)
\beq \sum_{m=1}^3 \left|U^{\nu N}_{m 4} \right|^{2}\ = \Bigl [
\sum_{\ell = e}^{\tau} {{|V_{\ell 4}|}^2} \Bigl (1 -
\sum_{\ell_1 = e}^{\tau}{|V_{\ell_1 4}|}^2 \Bigr )\Bigr ], \ \ {\rm since}\ \ UU^\dag + VV^\dag = I. \eeq
Ignoring terms of order $\left|V_{\ell 4} \right|^{4}$ we have
\bea \sum_m \left| U^{\nu N}_{m 4} \right|^2 \approx \sum_\ell
\left|V_{\ell 4} \right| ^{2}. \eea
In this approximation, the total width of a heavy Majorana
neutrino can be written as
\beq \Gamma_{N_4} \left\{
\begin{array}{ll} \displaystyle
\approx \sum_\ell \left|V_{\ell 4}\right| ^{2} \frac{3G_F m^3_4
}{8\pi\sqrt 2}\quad
 & {\rm for}\ \ m_4 > m_W,\\ [6mm]
  \displaystyle
\propto \sum_\ell \left|V_{\ell 4}\right| ^{2} G_F^2  m^3_4 (f^2_M
+ m^2_4) \quad & {\rm for}\ \ m_4 \ll m_W ,
\end{array}
\right. \label{approxwid} \eeq
where the expression when $m_4 \ll m_W$ is obtained from
Eq.~(\ref{totwid_small}) and $f_M$ are the meson decay constants.
We note that the approximate form of the total width as given in
Eq.~(\ref{approxwid}) is only for intuitive purposes to infer the
general behaviour of the total width as a function of mass. The
precise expressions for the total width of the heavy Majorana
neutrino as given in Eqs.~(\ref{totwid_small}), (\ref{totwid_big})
and (\ref{apptotwid}) have been used in the numerical analysis.

\begin{figure}[tb]
\center
\includegraphics[width=10.5truecm,clip=true]{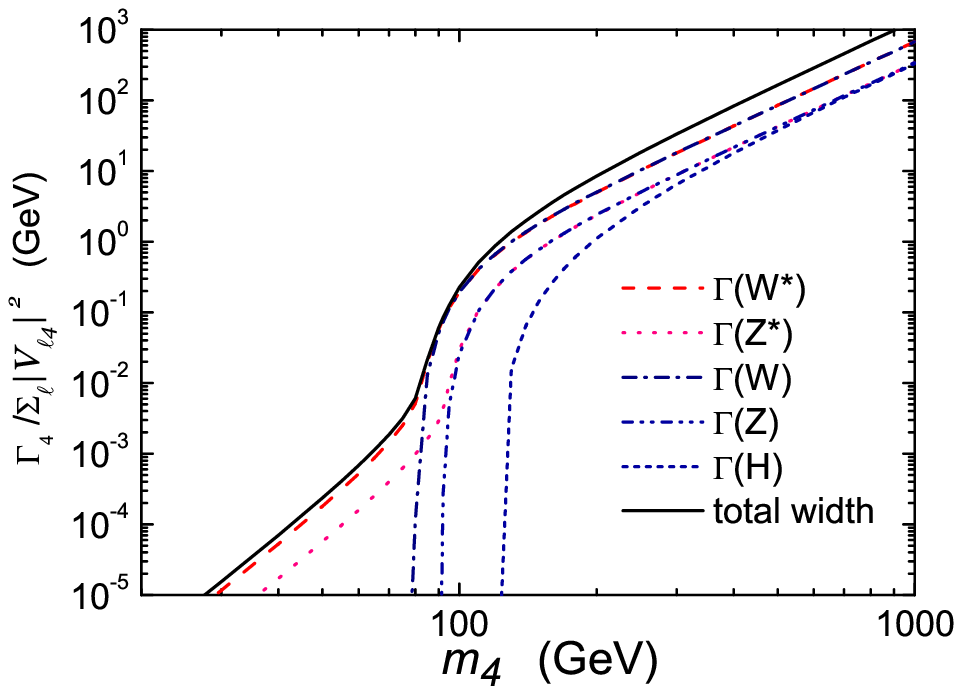}
\includegraphics[width=10.5truecm,clip=true]{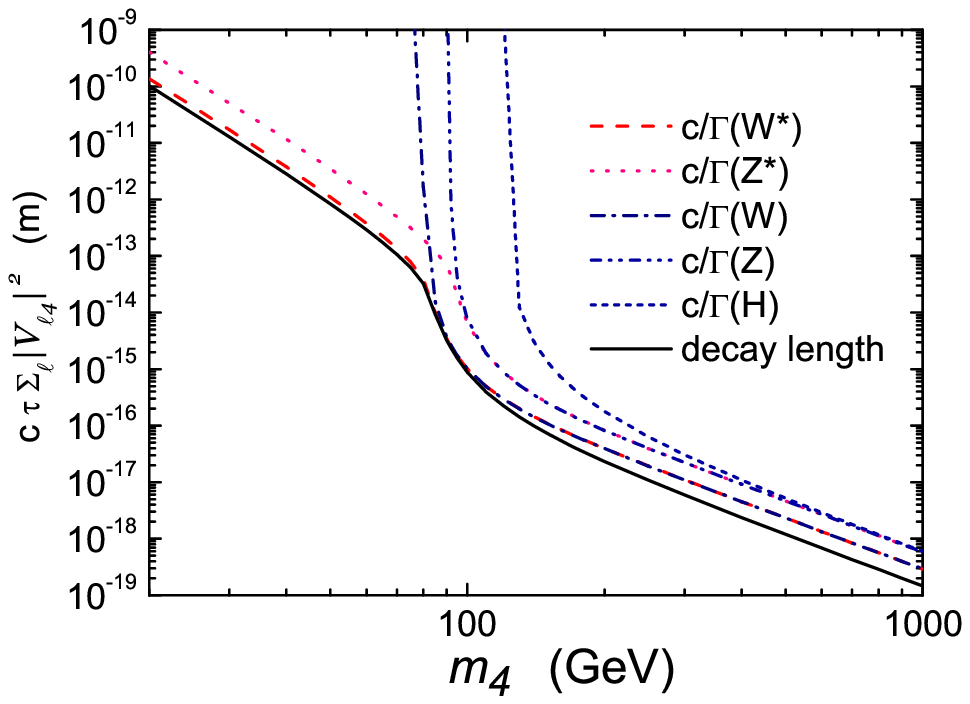}
 \caption{ (a) Top: decay width  and (b) bottom: decay length
 (normalized by $\sum_\ell \left|V_{\ell 4}\right| ^{2}$) versus mass of heavy Majorana neutrino for real and virtual weak bosons with the inclusion of Higgs decay channel for $m_H=120$ GeV . }
\label{Fig:Ndecaywidth}
\end{figure}
\begin{figure}[tb]
\includegraphics[width=8.3truecm,clip=true]{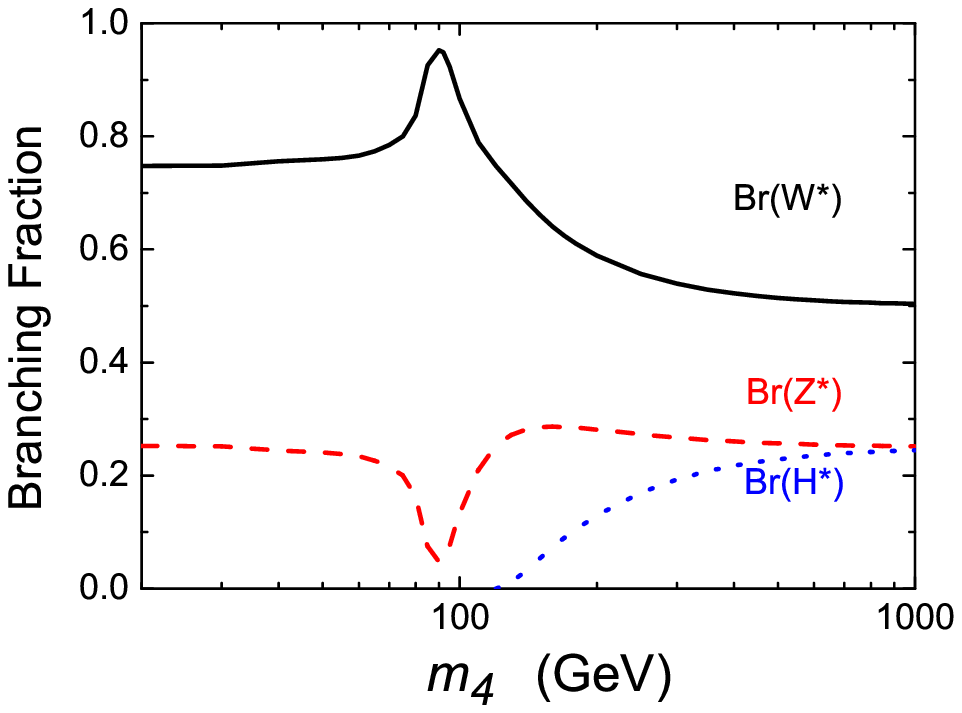}
\includegraphics[width=8truecm,clip=true]{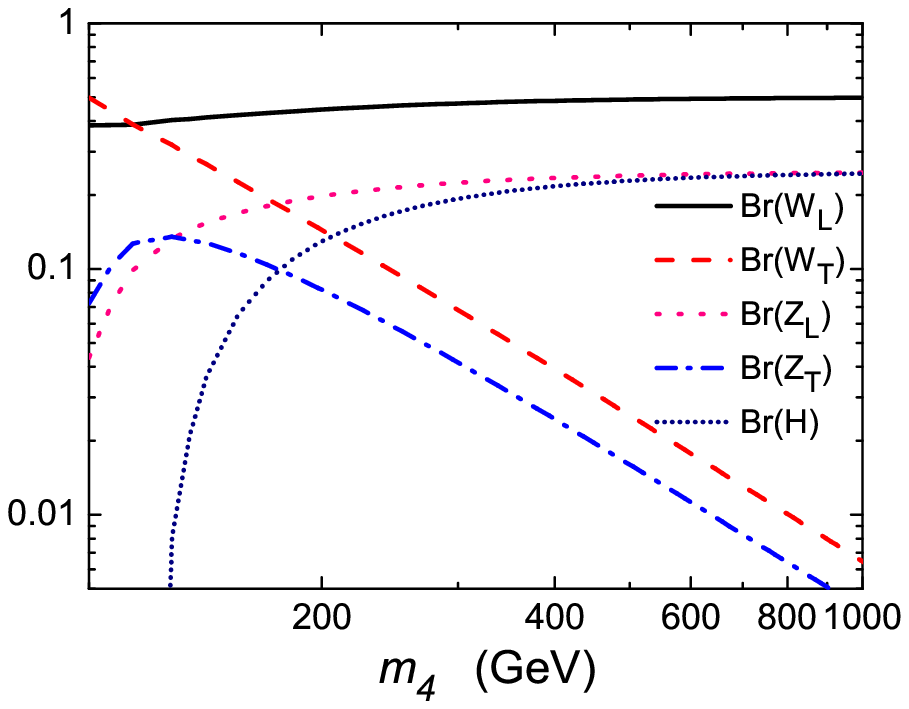}
 \caption{ (a) Left: branching fractions for decay of heavy Majorana neutrino into $W^*$ and $Z^*$ bosons with varying heavy neutrino mass; (b) right: branching fractions for decay of heavy Majorana neutrino into longitudinal and transverse gauge bosons in static heavy neutrino frame with the inclusion of Higgs decay channel for $m_H=120$ GeV . }
\label{Fig:NdecayBr}
\end{figure}
It should be noted that in the SM, if $N_4$ is heavier than the
Higgs boson, then the decay to a Higgs will be present and the
partial width is given by
\bea \Gamma^{\nu H} \equiv \Gamma(N_4 \rightarrow \nu_\ell H) =
{g^2 \over 64 \pi m_W^2}\  \left|V_{\ell  4} \right|^{2}\ m^3_4\
(1-\mu_H^{})^2. \eea
In Fig.~\ref{Fig:Ndecaywidth} we plot the decay width of the heavy
Majorana neutrino versus its mass normalized by the common factor
$\sum_\ell \left|V_{\ell 4}\right| ^{2}$. We can see in
Fig.~\ref{Fig:Ndecaywidth}(a) that for a heavy neutrino with mass
$m_4 > m_W$, the decay width increases as $G_F m^3_4$ as given in
Eq.~(\ref{approxwid}). Given the rather small mixing parameter,
the width remains narrow even for $m_4\sim {\cal O}$(1 TeV). For a
lighter neutrino with $m_4 \ll m_W$, the width can be very small.
The proper decay length is presented in
Fig.~\ref{Fig:Ndecaywidth}(b). We see from this that for $m_4
\lsim 20$ GeV and $|V_{\mu4}|^2 \lsim 10^{-4}$ from
Fig.~\ref{fig:Umu}, we have $c\tau \sim 1\ \mu \mathrm{m}$.

In Fig.~\ref{Fig:NdecayBr}(a) we plot the branching fractions of
the heavy Majorana neutrino decay to $W \ell$ and $Z\nu$ versus
varying heavy neutrino mass $m_4$. In Fig.~\ref{Fig:NdecayBr}(b)
we plot the branching fractions for the decays into longitudinal
and transverse gauge bosons in static heavy neutrino frame. When
the neutrino mass is large, it mainly decays to longitudinal gauge
bosons and $\mathrm{Br}(N_4 \rightarrow W^+ \ell^-) \simeq
\mathrm{Br}(N_4 \rightarrow Z\nu) = \mathrm{Br}(N_4 \rightarrow
H\nu)=25\%.$ In terms of the search at hadron colliders, we prefer
to adopt the $W\ell$ mode since we wish to reconstruct the full
event including the lepton number.

\subsection{Lepton-Number Violating Tau Decays}
\label{taud}

In this section we examine tau decays into an anti-lepton and two
mesons
\beq \tau^-(p_1) \rightarrow \ell^+(p_2)\  M_1^-(q_1)\  M_2^-(q_2)
\label{taudec} \eeq
which is a process with $\Delta L=-2$. The decay amplitude for the
above process is given by
\beq {i\cal M} = 2G_F^2 V^{CKM}_{M_1} V^{CKM}_{M_2} f_{M_1}
f_{M_2} {V_{\tau 4}^*}{V^*_{\ell 4}}\   m_4  \Biggl [\frac
{\overline{v_\tau} \qslash_1 \qslash_2 P_R v_\ell}{(p_1 - q_1)^2 -
m_4^2 +i\Gamma_{N_4}m_4}\Biggr ] + (q_1 \leftrightarrow q_2), \eeq
where $V^{CKM}_{M_i}$  and  $f_{M_i}$ are the quark flavor mixing
element and the decay constant for the meson $M_i$
respectively.  From this decay amplitude, we can calculate the transition rate
$\Gamma^{\tau}_{\lv}$ and the branching fraction normalised by the
tau decay width. In Appendix \ref{apptd}, we give the calculations
and the full expressions for the decay branching fraction of the
process (\ref{taudec}) in terms of the mass of heavy neutrino,
$m_4$, and the mixing $|V_{\tau 4} V_{\ell 4}|^2$. To understand
the physical picture, we can express the branching fraction in an
intuitive form, in the massless limit of the final state
particles, as
\bea \nonumber
\mathrm{Br} &=& \frac {\Gamma^{\tau}_{\!\!\! \mbox{}_{\lv}}}{\Gamma_{\tau}} = \Gamma^{\tau}_{\!\!\!\mbox{}_{\lv}} \Bigl(\frac {192 \pi^3} {G^2_F m^5_\tau} \Bigr ),\\
\nonumber
& \sim & \frac{3}{2}\pi(1- \frac{1}{2} \delta_{M_1 M_2})G^2_F f^2_{M_1}f^2_{M_2}|V^{CKM}_{M_1}V^{CKM}_{M_2}|^2 \mbox { } |V_{\tau 4} V_{\ell 4}|^2 \Bigl (1-\frac{m^2_4}{m^2_\tau}\Bigr ) \Bigl (\frac {m_4}{\Gamma_{N_4}} \Bigr),\\
&\sim& 10^{-3} \mbox { } |V^{CKM}_{M_1}V^{CKM}_{M_2}|^2 \mbox { }
|V_{\tau 4} V_{\ell 4}|, \eea
where we have used typical values of $m_4 \sim 1\ \gev$, $f_{M_i}
\sim 0.1\ \gev$, $G_F \sim 1 \times 10^{-5}\ \gev^{-2}$ and
$\Gamma_{N_4} \sim 10^{-11}\ |V_{\ell 4}|^2\ \gev$. From the
simple expression given above one can easily make a rough estimate
of the required sensitivity and hence the feasibility of
observation in terms of the mixing parameters for a given model.

A direct search for $\lv$ tau decays has been made at the BaBar
detector and the limits on the branching fractions were reported
in Ref.~\cite{aubert}. The experimental limits for various decay
modes are typically of the order of $10^{-7}$, as given in Table
\ref{mixingtd}. From the non-observation of the $\lv$ tau decay
modes one can determine bounds on the mixing parameters ${|V_{\ell
4}V_{\tau 4}|}^2$ as a function of the heavy neutrino mass $m_4$.
To do this in a comprehensive manner, we carry out a Monte Carlo
sampling of the mixing parameters and the mass of the heavy
neutrino. For simplicity, the mixing elements $V_{e4}, V_{\mu 4}$
and $V_{\tau 4}$ are allowed to vary in the range from 0 to 1. The
ranges of mass sampled for the heavy neutrino are listed in Table
\ref{mixingtd} for the various tau decay modes. We only sample the
range of masses that lead to a resonant enhancement of the width
as the other mass regions have very small transition rates as
discussed earlier.  We then calculate the transition rates and
branching fractions over the entire range of mixing and mass of
the heavy neutrino and the results of the Monte Carlo sampling are
discussed next.

\bet[tb] \caption{Mass and mixing elements of heavy neutrino and
the decay mode constraining them with the corresponding
experimental bounds on branching fractions. Bounds for $\dl=2$ tau
decays are from Ref. \cite{aubert}}
\begin{center}
\begin{tabular}
{|c|c|l|l|} \hline
Mixing element & Range of $m_4\ (\mev)$ & Decay mode & $B_{exp}$ \\
\hline
 & 140 - 1637 &$\tau^- \rightarrow e^+ \pi^- \pi^-$  & $2.7 \times 10^{-7}$   \\
 $|V_{e4}V_{\tau 4}|$ & 140 - 1637 & $\tau^- \rightarrow e^+ \pi^- K^-$  & $1.8 \times 10^{-7}$     \\
 & 494 - 1283 & $\tau^- \rightarrow e^+ K^- K^-$ & $1.5 \times 10^{-7}$     \\
\hline
 & 245 - 1637 &$\tau^- \rightarrow \mu^+ \pi^- \pi^-$  & $0.7 \times 10^{-7}$   \\
$|V_{\mu 4}V_{\tau 4}|$ & 245 - 1637 & $\tau^- \rightarrow \mu^+ \pi^- K^-$  & $2.2 \times 10^{-7}$     \\
 & 599 - 1283 & $\tau^- \rightarrow \mu^+ K^- K^-$ & $4.8 \times 10^{-7}$     \\
\hline
\end{tabular}
\end{center}
\label{mixingtd} \eet

\begin{figure}
\center
\begin{tabular}{cc}
\includegraphics[width=3.0in]{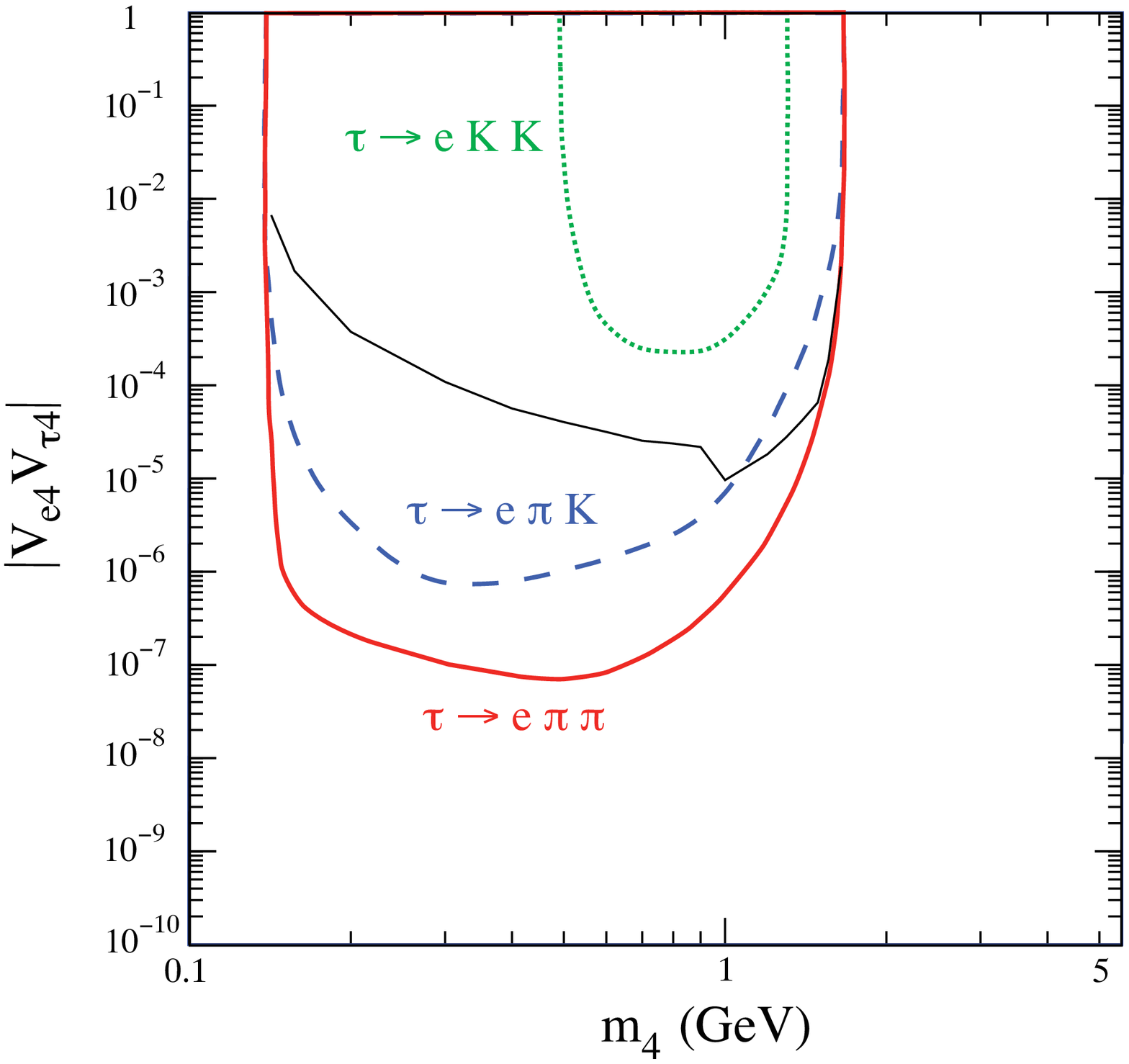}&
\includegraphics[width=3.0in]{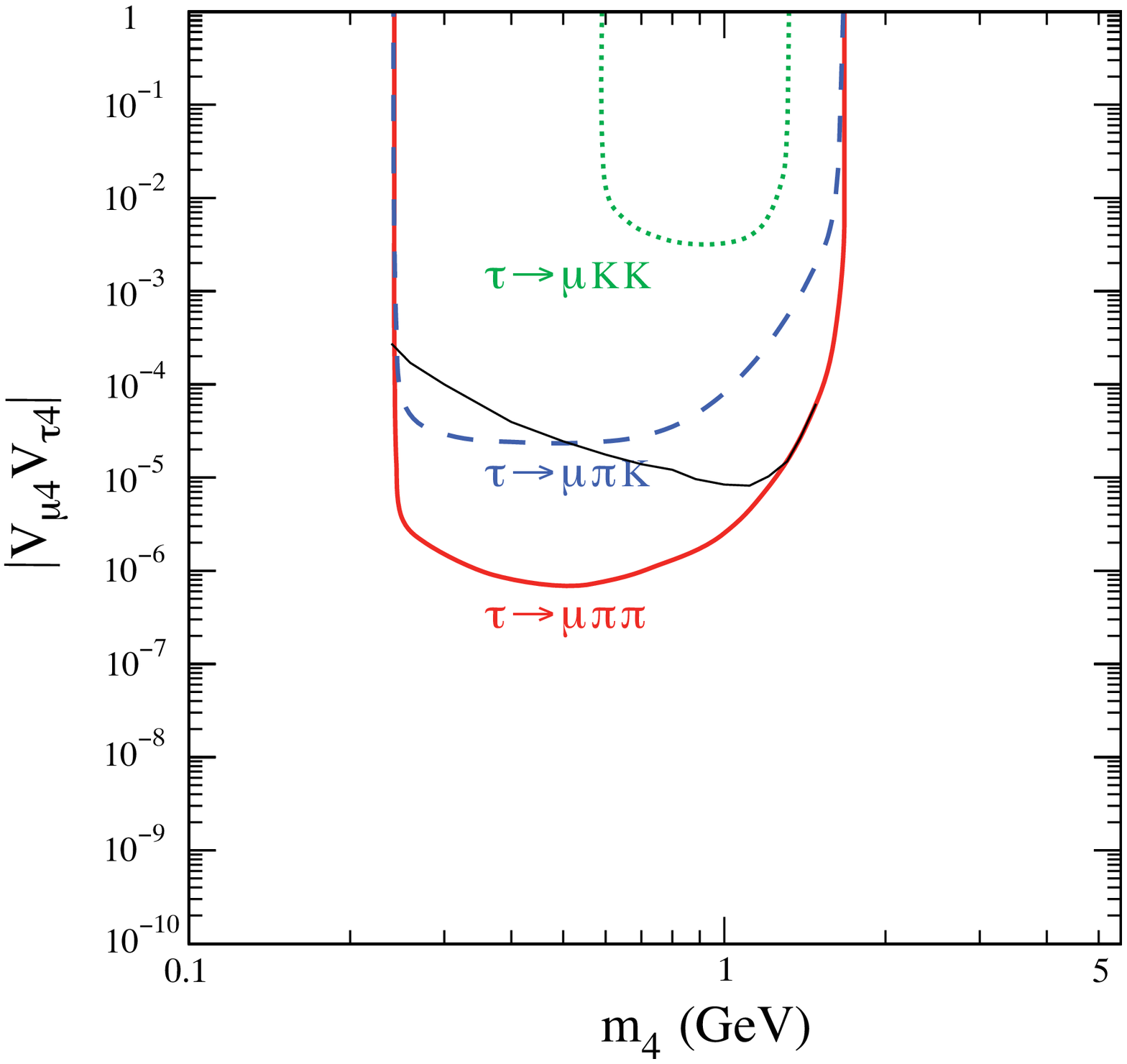}
\end{tabular}
\caption{(a) Left: excluded regions above the curves for $|V_{e
4}V_{\tau 4}|$ versus $m_4$; (b) right: same as (a) but for
$|V_{\mu 4}V_{\tau 4}|$. The thin black lines correspond to the
estimate of the bound (for $\tau \rightarrow e \pi \pi$ and $\tau
\rightarrow \mu \pi \pi$) once the probability of $N_4$ decay in
the detector is taken into account. }\label{taudfig}
\end{figure}

The relevant mixing parameters ${|V_{e 4}V_{\tau 4}|}$ and
${|V_{\mu 4}V_{\tau 4}|}$ are probed as a function of the heavy
neutrino mass $m_4$ and are shown in Fig.~\ref{taudfig}(a) and
Fig.~\ref{taudfig}(b), respectively. Under the assumption that the
detector was able to reconstruct all the signal events, the region
above the curves is excluded by the current direct experimental
search for $\lv$ tau decays. The most stringent bound on ${|V_{e
4}V_{\tau 4}|}$ is of ${\cal O}(10^{-6})$ and comes from $\tau^-
\rightarrow e^+ \pi^- \pi^-$.  The most stringent bound on
${|V_{\mu 4}V_{\tau 4}|}$ is also of ${\cal O}(10^{-6})$ and comes
from $\tau^- \rightarrow \mu^+ \pi^- \pi^-$. This is three orders
of magnitude more sensitive than the limits from precision
electroweak data which constrain the square of the mixing
${|V_{\ell 4}|}^2$ to be less than few times $10^{-3}$. In the
absence of detection of $\lv$ processes the constraints on mixing
from peak searches, accelerator experiments, reactor experiments
and others (collectively called laboratory constraints here and
henceforth) described in Fig.~\ref{fig:Uekinks}  $-$
Fig.~\ref{fig:Utau} are also applicable here. In the mass region
probed by $\lv$ tau decays the most stringent current constraints
are $|V_{e4}|^2 < 10^{-7} - 10^{-8}$,  $|V_{\mu4}|^2 < 10^{-6} -
10^{-8}$ and $|V_{\tau4}|^2 < 10^{-1} - 10^{-4}$. This would
roughly translate into constraints on $|V_{e4}V_{\tau4}| < 10^{-4}
- 10^{-6}$ and $|V_{\mu4}V_{\tau4}| < 10^{-4} - 10^{-6}$ which are
comparable to the limits from $\lv$ tau decay modes. We explore
more combinations of mixing elements and also provide better
constraints on mixing in some mass regions. To summarize, the
constraints on mixing from $\lv$ tau decays are always competitive
with or better than precision EW constraints and laboratory
constraints in the corresponding mass region. The experimental
bounds can improve in future and an order of magnitude
improvement in the experimental branching fraction will give
approximately an order of magnitude improvement in the constraints
for the mixing parameters ${|V_{\ell 4}V_{\tau 4}|}$. More
importantly, a detection in one of the laboratory experiments
implies the existence of a sterile neutrino while a detection in
one of the modes studied in our analysis would imply $\lv$ and
hence the existence of a Majorana neutrino.

It should be noted that the intermediate heavy Majorana neutrino
is treated as a real particle
 which propagates before decaying. If it exits the experimental apparatus
prior to decaying, then the signal corresponding to the $\Delta
L=2$ process cannot be reconstructed and no bound could be deduced
 from the non-observation of such events.
 In Figs.~\ref{taudfig}(a) and ~\ref{taudfig}(b),
 we provide an estimate of the bound on
the mixing parameters which takes into
 account the probability of the heavy Majorana neutrino to decay within the
detector of size $L_{\mathrm{exp}}$. This probability is given by
\beq P = 1-\exp(-L_{\mathrm{exp}} \Gamma_N) \eeq
 and for small masses and/or small mixing parameters
and consequently long decay lengths, it can be approximated with
$P \simeq L_{\mathrm{exp}} \Gamma_N$. We take
$L_{\mathrm{exp}}=10$~m, the typical size of the detectors used in
the experiments under consideration.
 For simplicity, we take $N_4$ to be relativistic but we keep its gamma
factor $\gamma=1$, as
 a more precise value requires a full understanding of the
experimental setup.
 We assume $|V_{e4}|=|V_{\mu4}|=|V_{\tau4}|$.
An estimate of the realistic bound on the mixing parameter
$|V_{e4} V_{\tau 4}|$ is then given by \beq |V_{e4} V_{\tau 4}| (=
|V_{e4}|^2) = \sqrt{|V_{e4}|^2_{\infty} / (L_{\mathrm{exp}}
\Gamma_{N0})}, \eeq where $|V_{e4}|^2_{\infty}$ is the bound
obtained assuming that all the $N_4$ decay in the detector and
discussed above, and $ \Gamma_{N0}$ is the decay rate for a fully
active heavy Majorana neutrino, i.e. when the mixing parameter $|V_{\ell 4}| = 1$ for $\ell = e, \mu, \tau$. The bounds remain unchanged for
large values of the mixing angle and /or large values of $m_4$, as
the decay length in these cases is very short.
 However, the most sensitive limit on $|V_{e4} V_{\tau 4}| (= |V_{e4}|^2)$ coming from $\tau \rightarrow e \pi \pi$ searches gets weakened to $\sim 4 \times 10^{-4} \ (4 \times 10^{-5}) \ (1 \times 10^{-5})$
for $m_4 = 0.2 \ (0.5) \ (1.0)$~GeV.
Similarly, the searches for $\tau \rightarrow \mu \pi \pi$ allows
to set a bound on $|V_{\mu4} V_{\tau 4}| (= |V_{\mu4}|^2)$ which
weakens to $|V_{\mu4} V_{\tau 4}| <  1 \times 10^{-4} \ (2 \times
10^{-5}) \ (1 \times 10^{-5})$ for $m_4 = 300 \ (600) \
(900)$~MeV. A detailed analysis taking into account the
experimental setup should be performed in order to obtain more
precise bounds.

\subsection{Lepton-Number Violating Rare Meson Decays}
\label{mesond}

\bet[tb] \caption{Same as Table \protect\ref{mixingtd} but for
$\dl=2$ rare meson decays. The experimental bounds are from Ref.
\cite{PDG}, the bounds for $D^+ \rightarrow e^+ e^+ \pi^- (K^-)$
are from Ref.~\cite{he}.}  
\begin{center}
\begin{tabular}
{|c|c|l|l|} \hline
Mixing element & Range of $m_4\ (\mev)$ & Decay mode & $B_{exp}$ \\
\hline
 & 140 - 493 & $K^+ \rightarrow e^+ e^+ \pi^-$ & $6.4 \times 10^{-10}$ \\
 & 140 - 1868 & $D^+ \rightarrow e+ e^+ \pi^-$&$3.6 \times 10^{-6}$ \\
 & 494 - 1868 & $D^+ \rightarrow e^+ e^+ K^-$&$4.5 \times 10^{-6}$ \\
 & 140 - 1967 & $D^+_s \rightarrow e^+ e^+ \pi^-$&$6.9 \times 10^{-4}$ \\
$|V_{e4}|^2$& 494 - 1967 & $D^+_s \rightarrow e^+ e^+ K^-$&$6.3 \times 10^{-4}$ \\
 & 140 - 5278 & $B^+ \rightarrow e^+ e^+ \pi^-$&$1.6 \times 10^{-6}$ \\
 & 494 - 5278 & $B^+ \rightarrow e^+ e^+ K^-$&$1.0 \times 10^{-6}$ \\
 & 776 - 5278 & $B^+ \rightarrow e^+ e^+ \rho^-$&$2.6 \times 10^{-6}$ \\
 & 892 - 5278 & $B^+ \rightarrow e^+ e^+ K^{*-}$&$2.8 \times 10^{-6}$ \\
\hline
 & 245 - 388 & $K^+ \rightarrow \mu^+ \mu^+ \pi^-$& $3.0 \times 10^{-9}$ \\
 & 245 - 1763 & $D^+ \rightarrow \mu^+ \mu^+ \pi^-$&$4.8 \times 10^{-6}$ \\
 & 599 - 1763 & $D^+ \rightarrow \mu^+ \mu^+ K^-$&$1.3 \times 10^{-5}$ \\
 & 881 - 1763 & $D^+ \rightarrow \mu^+ \mu^+ \rho^-$&$5.6 \times 10^{-4}$ \\
 & 997 - 1763 & $D^+ \rightarrow \mu^+ \mu^+ K^{*-}$&$8.5 \times 10^{-4}$ \\
$|V_{\mu 4}|^2$& 245 - 1862 & $D^+_s \rightarrow \mu^+ \mu^+ \pi^-$&$2.9 \times 10^{-5}$ \\
 & 599 - 1862 & $D^+_s \rightarrow \mu^+ \mu^+ K^-$&$1.3 \times 10^{-5}$ \\
 & 997 - 1862 & $D^+_s \rightarrow \mu^+ \mu^+ K^{*-}$&$1.4 \times 10^{-3}$ \\
 & 245 - 5173 & $B^+ \rightarrow \mu^+ \mu^+ \pi^-$&$1.4 \times 10^{-6}$ \\
 & 599 - 5173 & $B^+ \rightarrow \mu^+ \mu^+ K^-$&$1.8 \times 10^{-6}$ \\
 & 881 - 5173 & $B^+ \rightarrow \mu^+ \mu^+ \rho^-$&$5.0 \times 10^{-6}$ \\
 & 997 - 5173 & $B^+ \rightarrow \mu^+ \mu^+ K^{*-}$&$8.3 \times 10^{-6}$ \\
\hline
 & 140 - 493 & $K^+ \rightarrow e^+ \mu^+ \pi^-$& $5.5 \times 10^{-10}$ \\
 & 140 - 1868 & $D^+ \rightarrow e^+ \mu^+ \pi^-$&$5.0 \times 10^{-5}$ \\
 & 494 - 1868 & $D^+ \rightarrow e^+ \mu^+ K^-$&$1.3 \times 10^{-4}$ \\
 & 140 - 1862 & $D^+_s \rightarrow e^+ \mu^+ \pi^-$&$7.3 \times 10^{-4}$ \\
$|V_{e4}V_{\mu 4}|$ & 494 - 1967 & $D^+_s \rightarrow e^+ \mu^+ K^-$&$6.8 \times 10^{-4}$ \\
 & 140 - 5278 & $B^+ \rightarrow e^+ \mu^+ \pi^-$&$1.3 \times 10^{-6}$ \\
 & 494 - 5278 & $B^+ \rightarrow e^+ \mu^+ K^-$&$2.0 \times 10^{-6}$ \\
 & 776 - 5278 & $B^+ \rightarrow e^+ \mu^+ \rho^-$&$3.3 \times 10^{-6}$ \\
 & 892 - 5278 & $B^+ \rightarrow e^+ \mu^+ K^{*-}$&$4.4 \times 10^{-6}$ \\
\hline
\end{tabular}
\end{center}
\label{mixingrmd} \eet

We now investigate the $\lv$ processes in which a meson decays
into two like-sign leptons and another meson
\beq M_1^+(q_1) \rightarrow \ell^+(p_1)\  \ell^+(p_2)\
M_2^-(q_2). \label{mesdec} \eeq
These decays are  similar to the tau decay modes described in the
previous section. The decay amplitude for the above process is
given by
\bea \nonumber
 i{\cal M}^P &=&  2G_F^2 V^{CKM}_{M_1} V^{CKM}_{M_2} f_{M_1} f_{M_2}{V_{\ell_1 4}}{V_{\ell_2 4}}\  m_4 \\
 &\times& \Biggl [ \frac {\overline{u_{\ell_1}} \qslash_1 \qslash_2 P_R v_{\ell_2}}{(q_1-p_1)^2-{m_4}^2 + i\Gamma_{N_4} m_4}\Biggr ] + (p_1 \leftrightarrow p_2),\\
\nonumber
i{\cal M}^V & = &  2G_F^2 V^{CKM}_{M_1} V^{CKM}_{M_2} f_{M_1} f_{M_2}{V_{\ell_1 4}}{V_{\ell_2 4}}\ m_4\ m_{M_2} \\
& \times& \Biggl [ \frac {\overline{u_{\ell_1}} \qslash_1 \not
\epsilon^\lambda (q_2) P_R v_{\ell_2}}{(q_1-p_1)^2-{m_4}^2 +
i\Gamma_{N_4} m_4}\Biggr ] + (p_1 \leftrightarrow p_2), \eea
where $i{\cal M}^P$ and $i{\cal M}^V$ are the decay amplitudes
when the meson $M_2$ is a pseudoscalar or vector meson
respectively and $V^{CKM}_{M_i}$  and  $f_{M_i}$ are the quark
flavor mixing element and the decay constant for the meson $M_i$
respectively. From this decay amplitude, we can calculate the
transition rate $\Gamma^{M_1}_{\!\!\! \mbox{}_{\lv}}$ and the
branching fraction normalised by the decay width of the meson
$M_1$. In Appendix \ref{apprmd}, we give the calculations for the
decay branching fraction of the process (\ref{mesdec}) in terms of
the mass of heavy neutrino, $m_4$, and the mixing $|V_{\ell_1 4}
V_{\ell_2 4}|$. We can express the branching fraction in an
intuitive form, in the massless limit of the final state
particles, as
\bea \nonumber
\mathrm{Br} &=& \frac {\Gamma^{M_1}_{\!\!\! \mbox{}_{\lv}}}{\Gamma_{M_1}} = \Gamma^{M_1}_{\!\!\! \mbox{}_{\lv}} \mbox { } \tau_{M_1},\\
\nonumber
& \sim & \frac{1}{64 \pi^2}(1- \frac{1}{2} \delta_{\ell_1 \ell_2})G^4_F f^2_{M_1}f^2_{M_2}|V^{CKM}_{M_1}V^{CKM}_{M_2}|^2 \mbox { } |V_{\ell_1 4} V_{\ell_2 4}|^2 \Bigl (1-\frac{m^2_4}{m^2_\tau}\Bigr ) m^5_{M_1} \tau_{M_1} \Bigl (\frac {m_4}{\Gamma_{N_4}} \Bigr),\\
&\sim& (10^{-16}\ \gev)\ \tau_{M_1} \mbox { }
|V^{CKM}_{M_1}V^{CKM}_{M_2}|^2 \mbox { } |V_{\ell_1 4} V_{\ell_2
4}|, \eea
where we have used typical values of $m_4 \sim 1\ \gev$, $f_{M_i}
\sim 0.1\ \gev$, $G_F \sim 1 \times 10^{-5}\ \gev^{-2}$,
$\Gamma_{N_4} \sim 10^{-11}\ |V_{\ell 4}|^2\ \gev$ and
$\tau_{M_1}$ is in seconds. Using the values for the lifetimes of
the mesons in Appendix \ref{apprmd}, the branching fractions for
the various mesons are given by
\bea
\mathrm{Br} (K) &\sim&  |V^{CKM}_{M_1}V^{CKM}_{M_2}|^2 \mbox { } |V_{\ell_1 4} V_{\ell_2 4}|, \\
\mathrm{Br} (D,\ B) &\sim& 10^{-4}\  \mbox { } |V^{CKM}_{M_1}V^{CKM}_{M_2}|^2 \mbox { } |V_{\ell_1 4} V_{\ell_2 4}|,\\
\mathrm{Br} (D_s) &\sim& 10^{-5} \mbox { }
|V^{CKM}_{M_1}V^{CKM}_{M_2}|^2 \mbox { } |V_{\ell_1 4} V_{\ell_2
4}|. \eea
As mentioned earlier, with the simple expressions above one can
easily make a rough estimate of the required sensitivity and hence
the feasibility of observation in terms of the mixing parameters
for a given model.

Searches for rare meson decay modes have been made in numerous
experiments. Table \ref{mixingrmd} summarizes the current
experimental limits on branching fractions given by
Refs.~\cite{PDG, he}. From the non-observation of these $\lv$ rare
meson decay modes one can  determine constraints on mixing
parameters ${|V_{\ell_1 4} V_{\ell_2 4}|}$ as a function of the
heavy neutrino mass $m_4$. To do this in a comprehensive manner,
we carry out a Monte Carlo sampling of the mixing parameters and
the mass of the heavy neutrino similar to tau decay. The mixing
elements $V_{e4}, V_{\mu 4}$ and $V_{\tau 4}$ are allowed to range
from 0 to 1 for simplicity. Only the range of mass that leads to a
resonant enhancement of the width is sampled for the heavy
neutrino and listed in Table \ref{mixingrmd} for the various meson
decay modes. The transition rates and branching fractions are then
calculated over the entire range of mixing and mass of the heavy
neutrino and the results of the Monte Carlo sampling are discussed
next.

\begin{figure}[tb]
\center
\includegraphics[width=0.6\textwidth]{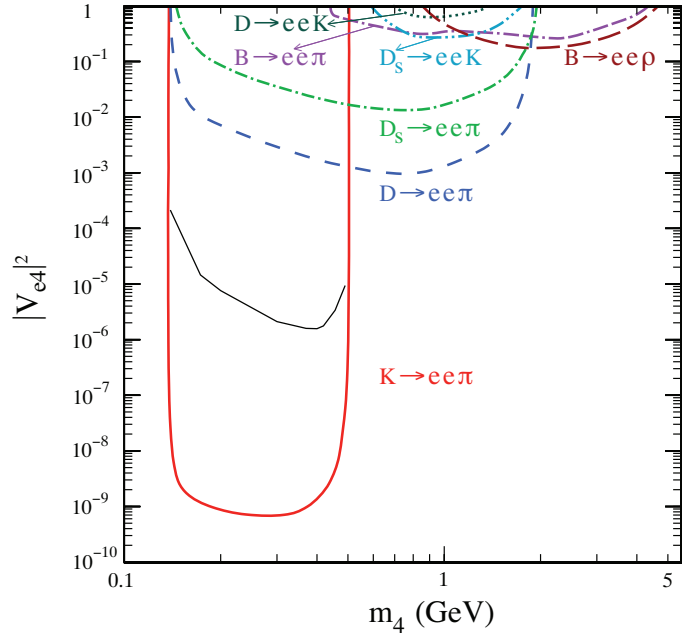}
\caption{Excluded regions above the curves for ${|V_{e 4}|}^2$
versus $m_4$ from $M_1^+ \rightarrow e^+ e^+ M_2^-$ searches. The
thin black line corresponds to an estimate of the bound from $K^+
\rightarrow e^+ e^+ \pi^-$ once the probability of decay of $N_4$
in the detector is taken into account. } \label{ve4fig}
\end{figure}

\begin{figure}[tb]
\center
\includegraphics[width=0.6\textwidth]{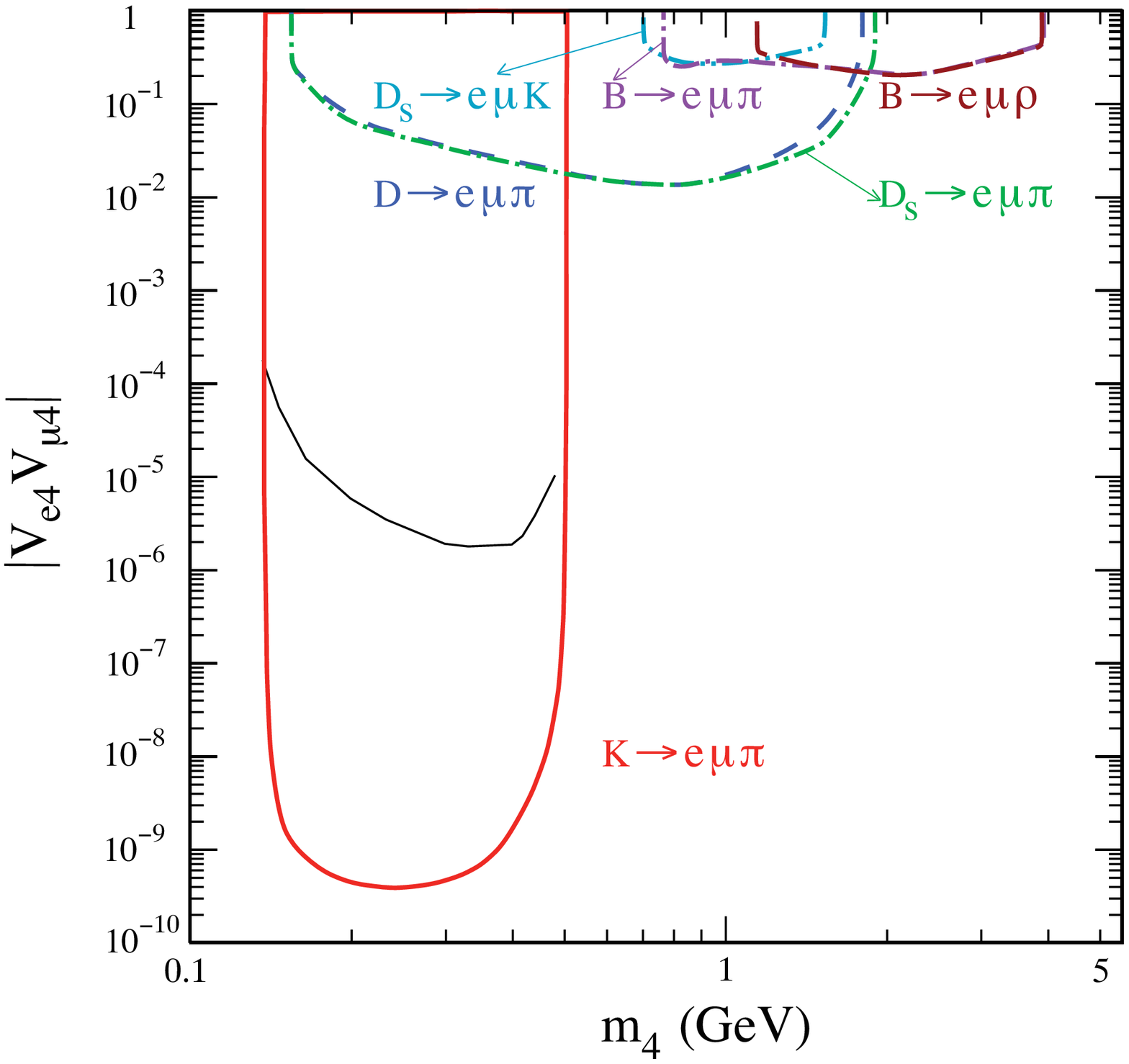}
\caption{Same as Fig.~\protect\ref{ve4fig} but for $|V_{e 4}V_{\mu
4}|$
from  $M_1^+ \rightarrow e^+ \mu^+ M_2^-$ searches.} 
\label{ve4mu4fig}
\end{figure}

\begin{figure}[tb]
\center
\includegraphics[width=0.6\textwidth]{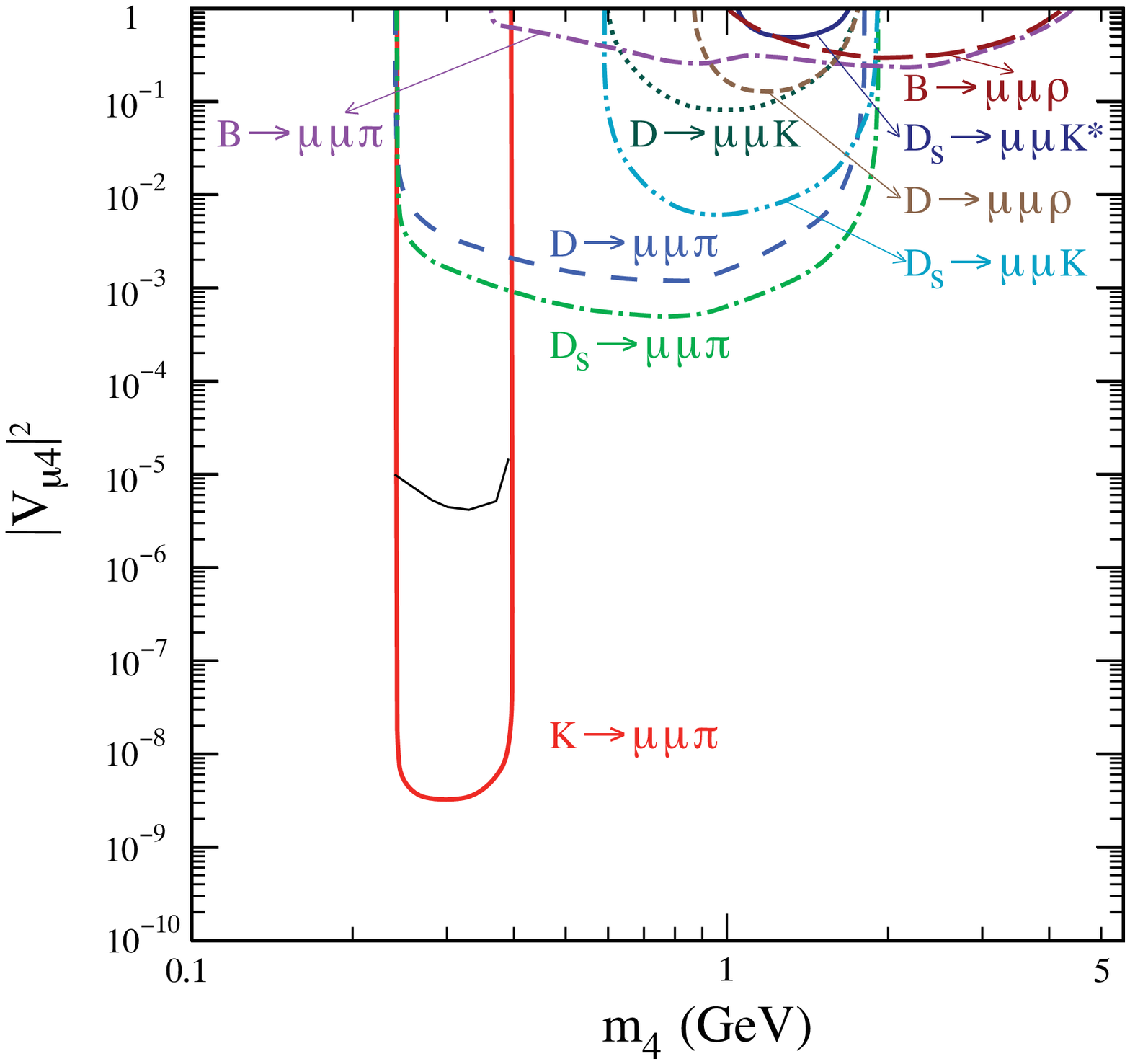}
\caption{Same as Fig.~\protect\ref{ve4fig} but for ${|V_{\mu
4}|}^2$ from
$M_1^+ \rightarrow \mu^+ \mu^+ M_2^-$ searches.} \label{vmu4fig} 
\end{figure}

For the various decay modes, the mixing parameters probed are
${|V_{e4}|}^2$, ${|V_{e4}V_{\mu 4}|}$ and ${|V_{\mu 4}|}^2$
depending on the final state leptons. Again, we plot the excluded
region of the mixing parameters as a function of neutrino mass,
as shown in Fig.~\ref{ve4fig} for  ${|V_{e4}|}^2$,
Fig.~\ref{ve4mu4fig} for  ${|V_{e4}V_{\mu 4}|}$ and in
Fig.~\ref{vmu4fig} for $ {|V_{\mu 4}|}^2$. The regions above the
curves are excluded by the current direct experimental searches
for $\lv$ meson decays. First we plot the limits which can be
derived assuming that all $N_4$ decay in the detector and give a
positive signature. The most stringent constraints are from the
$K^+ \rightarrow \ell^+_1 \ell^+_2 \pi^-$ mode with mixings of
${\cal O}(10^{-9})$ excluded for ${|V_{e4}|}^2$, ${|V_{e4}V_{\mu
4}|}$ and ${|V_{\mu 4}|}^2$. This is six orders of magnitude more
sensitive than the limits from precision electroweak data which
constrains the square of the mixing ${|V_{\ell 4}|}^2$ to be less
than few times $10^{-3}$. Next in sensitivity are the $D$ and
$D_s$ decay modes with constraints of order few times $10^{-3}$
which are similar to the constraints from precision electroweak
data. The bounds for the same mixing elements are much weaker in
the mass range above $2\ \gev$. Even though the limits are weak in
this region, it is important not to neglect the experimental study
of these processes. It only implies that there is a large
parameter space available for the mass and mixing of heavy
neutrinos.

As discussed for the $\dl = 2$ tau decays, in the absence of
detection of $\lv$ processes the laboratory constraints on mixing
described in Fig.~\ref{fig:Uekinks} $-$ Fig.~\ref{fig:Utau} are
also applicable here. In the mass region probed by $\lv$ meson
decays the most stringent laboratory bounds are $|V_{e4}|^2 <
10^{-7} - 10^{-8}$,  $|V_{\mu4}|^2 < 10^{-6} - 10^{-8}$ for $m_4 <
2$ GeV and $|V_{\mu4}|^2 < 10^{-4}$ for $m_4 > 2$ GeV. This would
roughly translate into constraints on $|V_{e4}V_{\mu4}| < 10^{-6}
- 10^{-8}$ for $m_4 < 2$ GeV and $|V_{e4}V_{\mu4}| < 10^{-5} -
10^{-6}$ for $m_4 > 2$ GeV. It should be noted that, if these
experiments were able to fully reconstruct the signal, the limits
from $K$ meson decays would be better than the laboratory
constraints by at least an order of magnitude in the corresponding
mass region. In fact, the constraints on ${|V_{e 4}|}^2$ from the
kaon decay mode $K^+ \rightarrow e^+ e^+ \pi^-$ would be more
stringent than even the constraints from $0\nu\beta\beta$ shown in
Fig.~\ref{fig:Uepeak}. Usually $0\nu\beta\beta$ experiments have
the best sensitivity as they have an advantage of a large
``effective luminosity" resulting from the large number of nuclei
available for decay. The meson (and tau) experiments on the other
hand have a small luminosity coming from a limited number of
mesons (taus) produced in accelerators compared to the number of
nuclei in $0\nu\beta\beta$ experiments.  It is interesting to note
that the resonant enhancement in the case of the $K$ meson decay
is able to match or improve over the large ``effective luminosity"
of $0\nu\beta\beta$ experiments.  In conclusion, the constraints
on mixing from $\lv$ meson decays are competitive with the
precision EW constraints and all the laboratory constraints,
potentially even $0\nu\beta\beta$, in some mass regions. But
again, we emphasize that the aim of our analysis is to study $\lv$
processes and hence Majorana neutrinos.

We have also taken into account the fact that, for small mixing,
only part of the heavy sterile neutrinos produced will decay in
the detector. We have considered $L_{\mathrm{exp}}=10$~m, $|V_{e
4}|=|V_{\mu 4}|=|V_{\tau 4}|$ and the gamma factor of $N_4$,
$\gamma =1$, for simplicity. In this case, as discussed for the
$\Delta L=2$ tau decays, the bounds get sensibly weakened. An
estimate of these bounds is reported in Figs.~\ref{ve4fig},
\ref{ve4mu4fig} and \ref{vmu4fig} by thin black lines. We see that
the bounds get significantly weakened by few orders of magnitude
for $K \rightarrow e e \pi$, $K \rightarrow e \mu \pi $ and $K
\rightarrow \mu \mu \pi$ and a careful analysis of these searches
should be performed to find the detailed bounds on the mixing
angles.

\begin{figure}[tb]
\center
\includegraphics[width=0.6\textwidth]{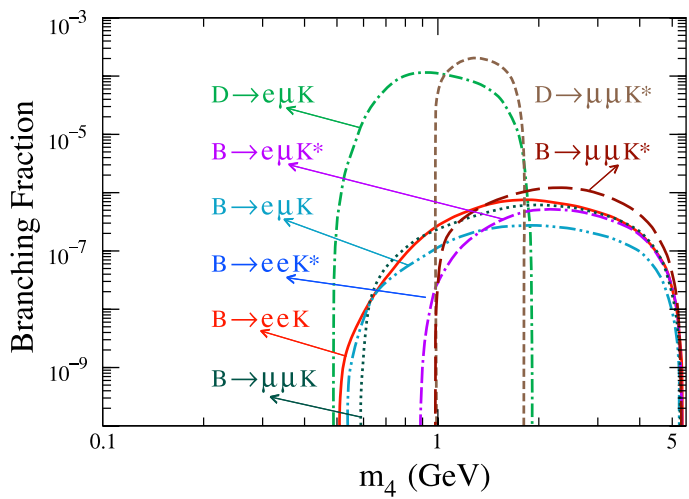}
\caption{Branching fraction versus heavy neutrino mass $m_4$ for
decay modes $M_1^+ \rightarrow \ell_1^+ \ell_2^+ M_2^-$ not yet
constrained by direct experimental searches. The regions below the
curve are theoretically allowed.} \label{nocons}
\end{figure}

The sensitivity of current direct experimental searches are not
adequate to constrain mixings for some decay modes. The
theoretically allowed branching fraction versus mass $m_4$ for
such modes is given in Fig.~\ref{nocons}. As we can deduce from
Table \ref{mixingrmd} and Fig.~\ref{nocons} all the modes are very
close to start being probed by direct experimental searches. The
experimental bounds on branching fractions can improve in future
and similar to tau decay modes, an order of magnitude improvement
in the experimental branching fraction will give approximately an
order of magnitude improvement in the constraints for the mixing
parameters ${|V_{\ell_1 4} V_{\ell_2 4}|}$. Currently we do not
have any constraints on the mixing parameter ${|V_{\tau 4}|}^2$
from $\lv$ rare meson decay modes. Only very weak constraints for
$\mathrm{BR}(B \to X \tau^+ \tau^-) < \cal{O}(\mbox{5}\%)$ exist
in a theoretical analysis \cite{Grossman}. The similar signature
$B^+ \rightarrow \tau^+ \tau^+ M^-$ is a  possible decay mode that
would bound ${|V_{\tau 4}|}^2$ and should be pursued.


\section{Collider Signatures}
\label{colsig}

\begin{figure}[tb]
\hspace{1.5cm}\includegraphics[width=6truecm,clip=true]{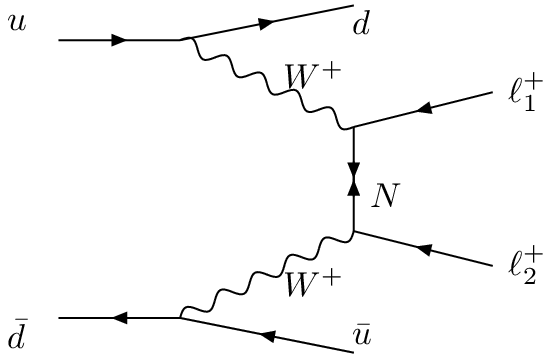}
\includegraphics[width=6truecm,clip=true]{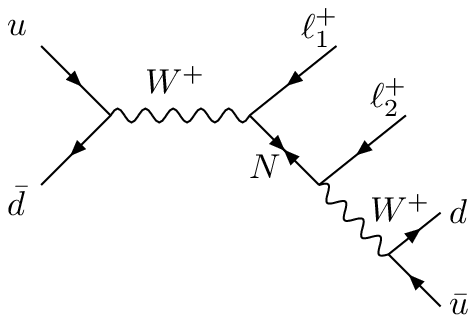}
 \caption{(a) Left: Feynman diagram for like-sign dilepton signature via $WW$ fusion in hadronic collisions; (b) right: the exchanged coherent diagram which is same as heavy neutrino production and decay.}
\label{feynmandiagram2}
\end{figure}

In this section we study heavy Majorana neutrinos at hadron
colliders. The most distinctive channels of the signal  involve
like-sign di-leptons. It was first proposed in Ref.~\cite{goran}
in the context of the left-right symmetric model, and subsequently
studied in  Ref.~\cite{NCollider,HZ,more,delAguila:2007em} We
discuss the signatures for a heavy Majorana neutrino and the
sensitivity to probe the parameters $m_4$ and $V_{\ell 4}$ at the
Tevatron and the LHC.

As for the production of a heavy Majorana neutrino at hadron
colliders, the representative diagrams at the parton level are
depicted in Fig.~\ref{feynmandiagram2}, with the exchange of final
state leptons implied. The first diagram is via $WW$ fusion with a
$t$-channel heavy neutrino $N_4$ exchange, directly analogous to
the process of $0\nu\beta\beta$. The second diagram is via
$s$-channel $N_4$ production and subsequent decay. Although in our
full calculations, we have coherently counted for all the
contributing diagrams  of like-sign dilepton production including
possible identical particle crossing, it is informative to
separately discuss these two classes of diagrams due to their
characteristically different kinematics.

\begin{figure}[tb]
\center
\includegraphics[width=11truecm,clip=true]{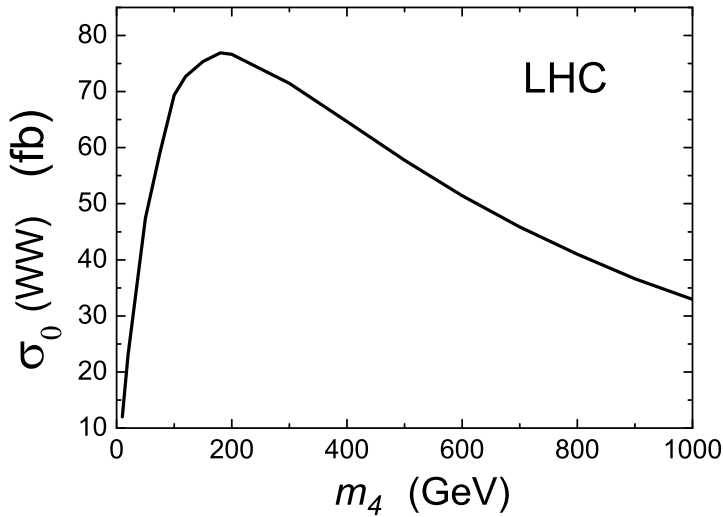}
 \caption{The bare cross section $\sigma_0(WW)$ versus mass of the heavy neutrino $m_4$.}
\label{Fig:csforwwfusion}
\end{figure}

The scattering amplitude for the process in
Fig.~\ref{feynmandiagram2}(a) is proportional to $V_{\ell_1 4}
V_{\ell_2 4}$ and the cross section can be expressed as
\beq \sigma \left( pp\rightarrow W^\pm W^\pm \to \ell_1 ^\pm
\ell_2 ^\pm X\right) =\left( 2- \delta _{\ell_1 \ell_2 }\right)
\left| V_{\ell_1 4}V_{\ell_2 4}\right| ^{2}\sigma_0(WW),
\label{cs} \eeq
where $\sigma_0(WW)$ is the ``bare cross section", independent of
the mixing parameters. We show the bare cross section at the LHC
energy of 14 TeV versus the heavy neutrino mass in
Fig.~\ref{Fig:csforwwfusion}. This cross section  can be at the
order of tens of femtobarns. However, due to the large suppression
of the small flavor mixing to the fourth power, the cross section
is rather small. This process was calculated in
Ref.~\cite{AliBorisov} under the effective vector boson
approximation. The authors of Ref.~\cite{AliBorisov} obtained
significantly more optimistic results than ours. Further scrutiny
indicated that they missed a factor of ${G^2_F m^4_W}/{8}$ and
their result should be scaled down by this factor. The
corresponding curve for the Tevatron is not shown in
Fig.~\ref{Fig:csforwwfusion} as the bare cross section is smaller
by nearly two orders of magnitude. Including the small mixing
element (to the fourth power) further reduces the cross section
drastically with no hope of detection at the Tevatron via this
mode.

\begin{figure}[tb]
\center
\includegraphics[width=12truecm,clip=true]{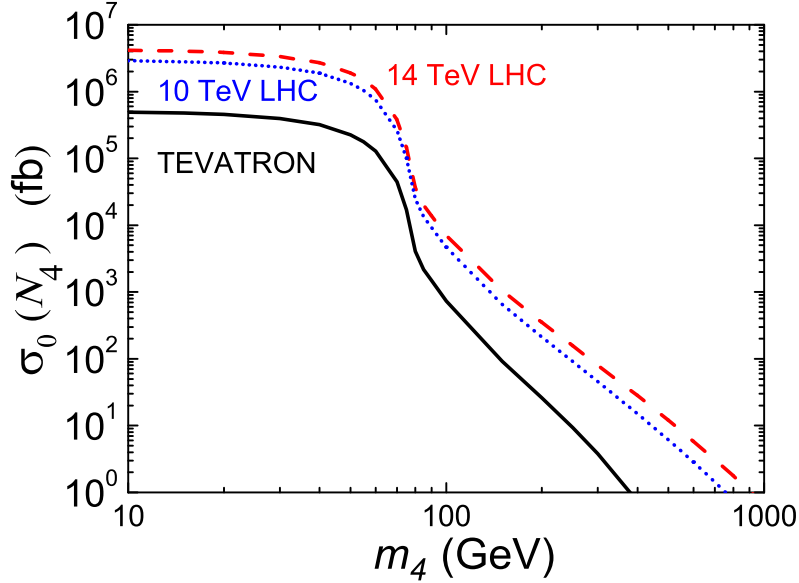}
\caption{The bare cross section $\sigma_0(N_4)$ versus mass of
heavy Majorana neutrino $m_4$ for the Tevatron ($p\bar p$ at 1.96
TeV, solid curve) and the LHC ($pp$ at 10 and 14 TeV, dotted and
dashed curves, respectively). } \label{Fig:KvsMass}
\end{figure}

By far, the dominant production process of heavy Majorana neutrino
in hadronic collisions is the diagram shown in
Fig.~\ref{feynmandiagram2}(b). We calculate the exact process, but
it turns out to be an excellent approximation to parameterize the
cross section as
\beq \sigma(pp\rightarrow \ell_1^\pm\ \ell_2^\pm\ W^\mp)\approx
\left( 2 - \delta _{\ell_1 \ell_2 }\right) \sigma(pp\rightarrow
\ell_1^\pm N_4)Br(N_4 \rightarrow \ell_2^\pm W^\mp)\propto
\frac{|V_{\ell_1 4} V_{\ell_2 4}|^2}{\sum_{\ell =e}^{\tau}
\left|V_{\ell 4}\right|^{2}}. \label{nw} \eeq
This observation allows us to study the process in a
model-independent way. We can rewrite the cross section in a
factorized form
\beq \sigma(pp\rightarrow \ell_1^\pm\ \ell_2^\pm\ W^\mp \rightarrow \ell_1^\pm\ \ell_2^\pm\ j\ j')= \left( 2
- \delta _{\ell_1 \ell_2 }\right)\ S_{\ell_1 \ell_2}\
\sigma_0(N_4), \label{eq:bare} \eeq
where $\sigma_0(N_4)$, called the ``bare cross section",  is only
dependent on the mass of heavy neutrino and is independent of all
the mixing parameters when the heavy neutrino decay width is
narrow. As seen in Fig.~\ref{Fig:Ndecaywidth}, this is indeed the
case for $m_4 \lsim 1$ TeV once we fold in the constraints
$|V_{\ell4}|^2< {\cal O}(10^{-3})$ from precision EW measurements.
We calculate  the exact cross section for the dilepton production
and use the definition Eq.~(\ref{eq:bare}) to find the bare cross
sections $\sigma_0(N_4)$, which are shown in
Fig.~\ref{Fig:KvsMass} at the Tevatron and the LHC energies versus
the mass of the heavy Majorana neutrino. Due to the fact that the
LHC will start its operation at 10 TeV, we have calculated the
cross sections at both 10 and 14 TeV c.m. energy. The production rate
is increased at the higher energy by a factor of 1.5, 2.0, 2.5 for
$m_4=100,\ 550$ and 1000 GeV, respectively. We will mainly present
our results at 14 TeV for the rest of the paper. An obvious
feature of the cross sections is the transition near the $W$ mass.
For $m_4 < M_W$ the cross section is nearly a constant due to an
on-shell $W$ production via the Drell-Yan mechanism with its
subsequent leptonic decay to $\ell^\pm N_4$. For  $m_4 > M_W$ the
cross section falls off sharply versus $m_4$ and the on-shell
decay goes like $N_4\to \ell^\pm W^\mp \to \ell^\pm\ j_1 j_2$.

The flavor information of the final state leptons is parameterized
by \beq
 S_{\ell_1\ell_2}=\frac{ \left|V_{\ell_1 4} V_{\ell_2 4}
 \right| ^{2}}{\sum_{\ell=e}^{\tau} \left|V_{\ell 4}\right| ^{2}},
\eeq In general the two final state charged leptons can be of any
flavor combination, namely,
\beq e^\pm e^\pm,\ \  e^\pm \mu^\pm,\ \  e^\pm \tau^\pm,\ \
\mu^\pm \mu^\pm ,\ \  \mu^\pm \tau^\pm \quad  {\rm and}\quad
\tau^\pm \tau^\pm. \eeq
The constraint from $0 \nu \beta \beta$ as given in
Eq.~(\ref{0nubb}) is very strong and makes it difficult to observe
like-sign di-electrons $e^\pm e^\pm$. The events with $\tau$
leptons will be challenging to reconstruct experimentally. We will
thus concentrate on clean dilepton channels of $\mu^\pm\mu^\pm$
and  $\mu^\pm e^\pm$, although we will comment on our proposal to
include the $\tau$ modes. The corresponding mixing parameters in
our notation will be
\beq
 S_{\mu\mu}
 =\frac{\left|V_{\mu 4}\right| ^{4}}{\sum_{\ell=e}^{\tau} \left|V_{\ell 4}\right| ^{2}},\quad
 S_{e \mu}=\frac{\left|V_{e 4}  V_{\mu 4} \right| ^{2}}{\sum_{\ell=e}^{\tau} \left|V_{\ell 4}\right| ^{2}},
\eeq
respectively. Given the smallness of $|V_{e 4}|^2$, we can further
simplify our study by exploring only two cases: an optimistic case
$|V_{\mu 4}|^2\gg |V_{\tau 4}|^2,\ |V_{e 4}|^2$ and a generic case
$|V_{\mu 4}|^2\approx |V_{\tau 4}|^2 \gg |V_{e 4}|^2$, which lead
to
\beq S_{\mu\mu} =\left\{
\begin{array}{c}
|V_{\mu 4}|^2\quad {\rm (optimistic)} \\
{1\over 2} |V_{\mu 4}|^2 \quad {\rm (generic)}
\end{array}
\right. , \qquad S_{\mu e} =\left\{
\begin{array}{c}
|V_{e 4}|^2 \quad {\rm (optimistic)} \\
{1\over 2}|V_{e 4}|^2  \quad  {\rm (generic)}
\end{array}
\right. . \label{cases} \eeq

\subsection{Search for Like-sign Dilepton Signals at the Tevatron}
\label{tev}

We now consider the search for $N_4$ at the Fermilab Tevatron,
which is currently running at a c.m.~energy of 1.96 TeV in $p\bar
p$ collisions. We concentrate on the clean like-sign $\mu^\pm
\mu^\pm$ mode
\beq p\bar p \to  \mu^\pm \mu^\pm\ j_1 j_2\ X , \eeq
where $X$ is some inclusive hadronic activities common in hadronic
collisions. To quantify the signal observability, we first impose
the basic acceptance cuts  on leptons and jets to simulate the
CDF/D0 detector coverage
\bea p_T^\mu >5 {\,\rm GeV},\quad| \eta^\mu| <2.0,\quad
 p_T^j > 10 {\,\rm GeV},\quad  |\eta^j|<3.0.
 \label{eq:basic}
\eea
We also smear the lepton momentum by a tracking resolution and the
jet energy by hadronic calorimeter resolution as
\bea {\Delta p_T^\mu \over p_T^\mu} = 1.5\times 10^{-3}\
p_T^\mu,\quad {\Delta E_j\over E_j} = {75\% \over \sqrt{E_j} }
\oplus 3\% , \label{eq:smear} \eea
where $p_T^\mu$ and $E_j$ are in units of GeV. 
\begin{figure}[tb]
\includegraphics[width=7.8truecm,clip=true]{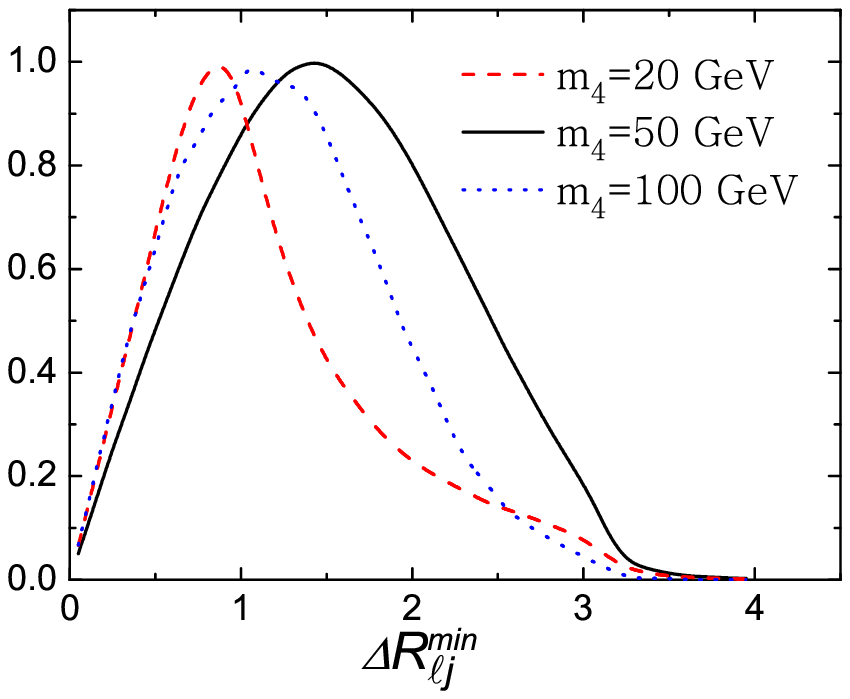}
\includegraphics[width=7.8truecm,clip=true]{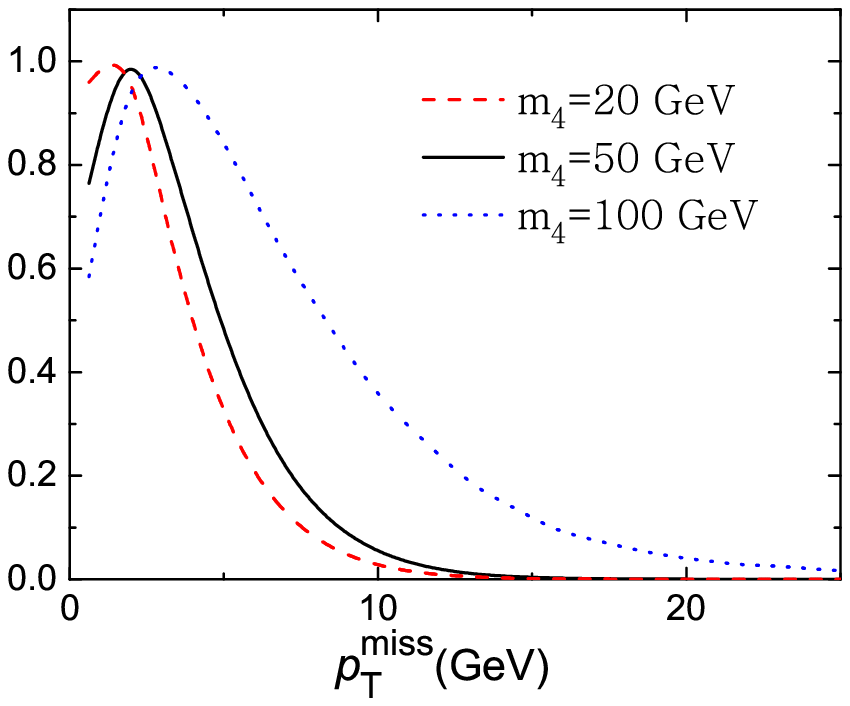}
\includegraphics[width=7.8truecm,clip=true]{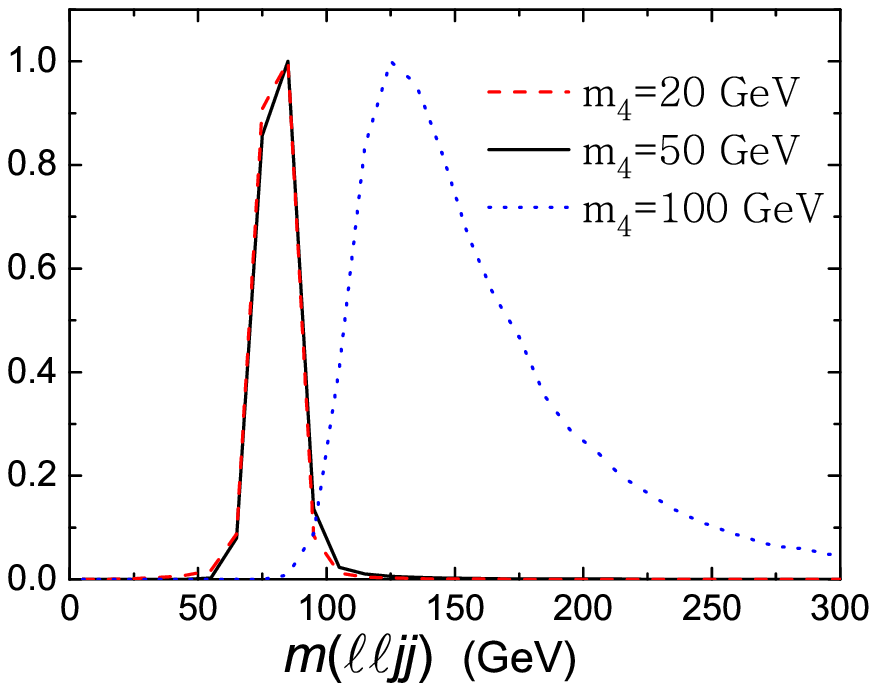}
\includegraphics[width=8truecm,clip=true]{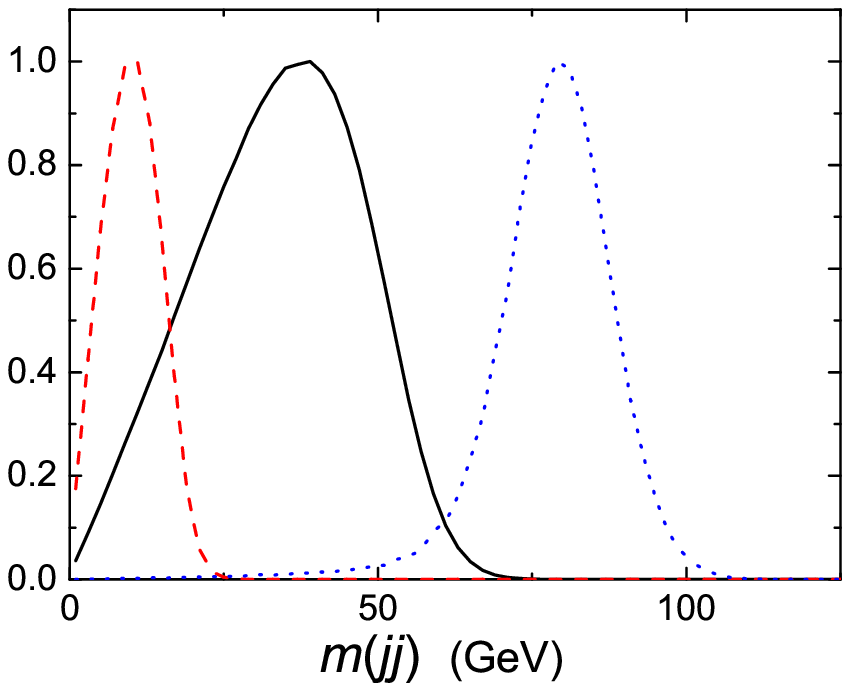}
\caption{Normalized distributions $\sigma^{-1} d\sigma/dX$ for
$m_4=20,\ 50$ and 100 GeV at the Tevatron for (a) upper left: the
minimal isolation $\Delta R_{\ell j}^{min}$; (b) upper right: the
missing transverse momentum $\ptmiss$; (c) bottom left: the $2\ell
2j$ system invariant mass $m(\ell\ell jj)$; (d) bottom right: the
di-jet invariant mass $m(jj)$. } \label{Fig:DR}
\end{figure}

The signal events we are searching for have very unique
kinematical features. For the purpose of illustration, we choose
$m_4=20,\ 50\  \gev$  (below $m_W$ threshold) and 100 GeV (above
$m_W$). First of all, there are  two well-isolated like-sign
charged leptons. This is shown  in Fig.~\ref{Fig:DR}(a) by a
normalized distribution of the minimal isolation $\Delta R_{\ell
j} =\sqrt{\Delta \eta^2 + \Delta \phi^2}$. Second, there is
essentially no missing transverse energy. However, realistically,
the detectors have finite resolutions as simulated  by the
Gaussian smearing given in Eq.~(\ref{eq:smear}). Consequently,
there is always some misbalance in the energy-momentum
measurements, which is attributed to the missing transverse
energies and is plotted in Fig.~\ref{Fig:DR}(b). Thirdly, due to
the existence of an on-shell $W^\pm$ in the signal process, one
would expect to reconstruct it by an invariant mass either from
the $2\ell 2j$ system $m(\ell\ell jj)$ (in the case of DY
production) or from the  di-jets $m_(jj)$ (in the case of $N_4$
decay). This is demonstrated in  Figs.~\ref{Fig:DR}(c) and (d),
respectively. The above kinematical features motivate us to impose
the following event selection cuts
\bea && \Delta R^{min}_{\ell j}> 0.5,
\label{eq:R} \\
&&  60\ {\gev} <  \ {\rm either}\ m(\ell\ell jj)\ \  {\rm or}\
m(jj) < 100\ \gev ,
\label{eq:m}\\
&&  p\!\!\!\slash_T < 20\ \gev . \label{eq:pt} \eea
These cuts are highly  efficient in selecting the signal events.
We illustrate this in Table \ref{Tab:eff}, in which we calculate
the signal rates with the consecutive cuts for $m_4=60$ GeV and
$\left|V_{\mu 4}\right|^{2}=\left|V_{\tau 4}\right| ^{2}=5\times
10^{-3} \gg \left|V_{e 4}\right|^{2}$. Note that the choice of
mixing elements used in the illustration is motivated by
constraints from precision EW measurements. However this is for
illustration purposes only and in our full analysis we have  kept
$S_{\mu\mu}$ as a free parameter.

At the Tevatron energies, the SM contribution to the like-sign
dilepton events is rather small. The leading background of this
type comes  from the top-quark production and its cascade decay
via the chain
 \bea
&&  t \to W^+ b \to \ell^+ \nu_\ell\  b, \\
&& \bar t \rightarrow W^- \bar b \to W^-\ \bar c\ \nu_\ell\
\ell^+.
 \eea
The background rates and survival probabilities with the
consecutive cuts are also given in Table \ref{Tab:eff}. We see
that the $t\bar t$ background is essentially eliminated by the
selective cuts. We have also considered  other SM backgrounds
coming from the production of $W^\pm W^\pm jj,\ W^\pm Z jj$. After
the selective cuts, all these backgrounds are negligibly small.

\bet \caption{ The representative signal and background cross
sections at  the Tevatron, for $\mu^\pm \mu^\pm j j $ and the
efficiencies with the consecutive cuts. For illustration, we have
used $m_4=60$ GeV, $\left|V_{\mu 4}\right|^{2}=\left|V_{\tau
4}\right| ^{2}=5\times 10^{-3} \gg \left|V_{e 4}\right|^{2}$. }
\vskip 0.2cm
\begin{tabular}{c|c|c|c|c|c}
 \hline
& No cut & Basic cut (\ref{eq:basic}) &$+\Delta R$ (\ref{eq:R}) &
$+m(jj), m(\ell\ell jj)$ (\ref{eq:m}) & +$p\!\!\!\slash_T$ (\ref{eq:pt}) \\
 \hline
Signal  & & & & & \\
 $\sigma$ (fb) &319&108&99&96&96\\
eff.&-&33\%&92\%&97\%&100\% \\
 \hline
 $t\bar t$ Bkg & & & & & \\
 $\sigma$ (fb) &78.4&58.2&1.85&0.04&0.005\\
 eff.&-&74\%&3.2\%&2.2\%&12.5\%\\
 \hline
\end{tabular}
\label{Tab:eff} \eet

\begin{figure}[tb]
\center
\includegraphics[width=10truecm,clip=true]{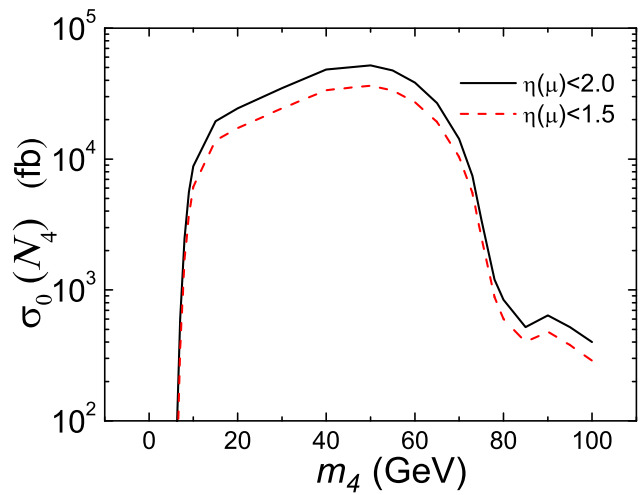}
 \caption{ $\sigma_0(N_4)$ with varying heavy neutrino mass $m_4$ after all the cuts. The two cases correspond to muon rapidity acceptance at D0 and CDF. }
\label{sigma0}
\end{figure}

In Fig.~\ref{sigma0}, we plot the bare cross section $\sigma_0(N_4)$
with the basic cuts of Eq.~(\ref{eq:basic}) as well as the selection
cuts Eqs.~(\ref{eq:R})$-$(\ref{eq:pt}).  The reduction in rate is
mainly due to the basic acceptance cuts. For comparison, we have
also included two choices of pseudo-rapidity cut $|\eta(\mu)| < 2$
and $|\eta(\mu)|<1.5$. We now consider the statistical significance
of the signal observation. In the absence of background events, we
use Poisson statistics to determine the search sensitivity. We take
a signal with ${95\%}$ Confidence Level (as this is very close to
$2\sigma$ we call it a 2$\sigma$ effect henceforth) to be 3 events.
We can thus translate this to the sensitivity to the mixing
parameter
\beq
 (2-\delta_{\ell_1\ell_2})  \sigma _0 (N_4)\ S_{\ell_1 \ell_2}\ {\it L} \geq 3 ,
 \label{lumscl}
\eeq
where $\it L$ is the integrated luminosity.

\begin{figure}[tb]
\includegraphics[width=7.8truecm,clip=true]{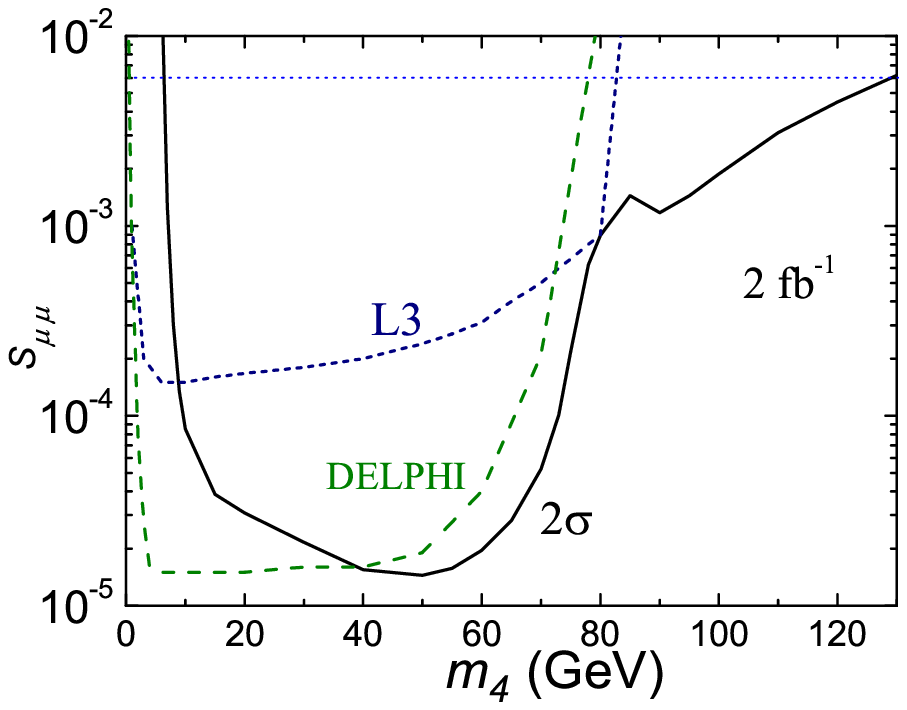}
\includegraphics[width=7.8truecm,clip=true]{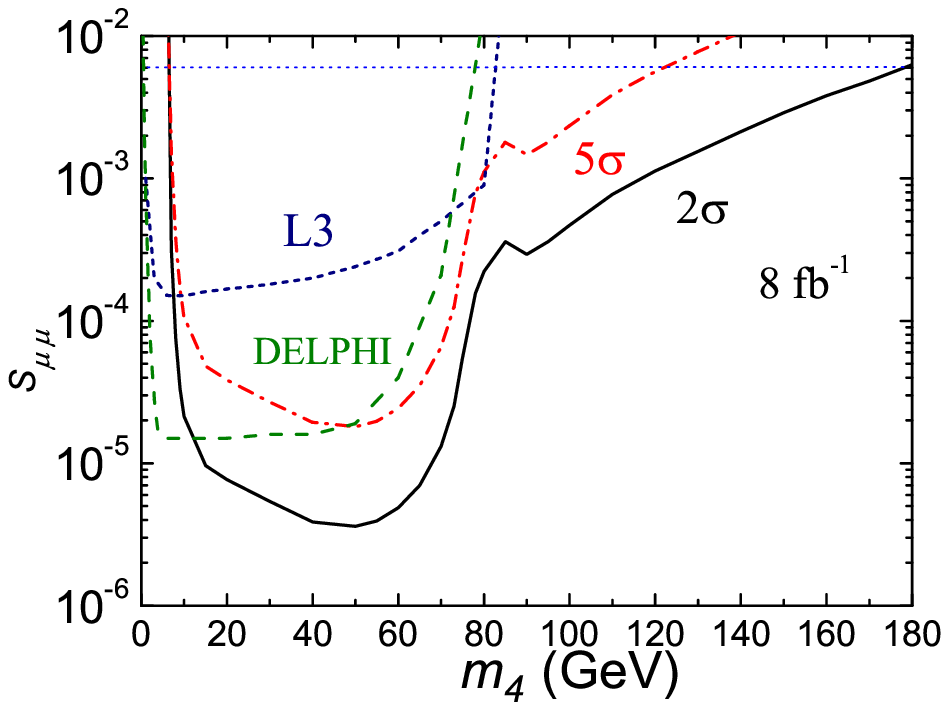}\\
\includegraphics[width=7.8truecm,clip=true]{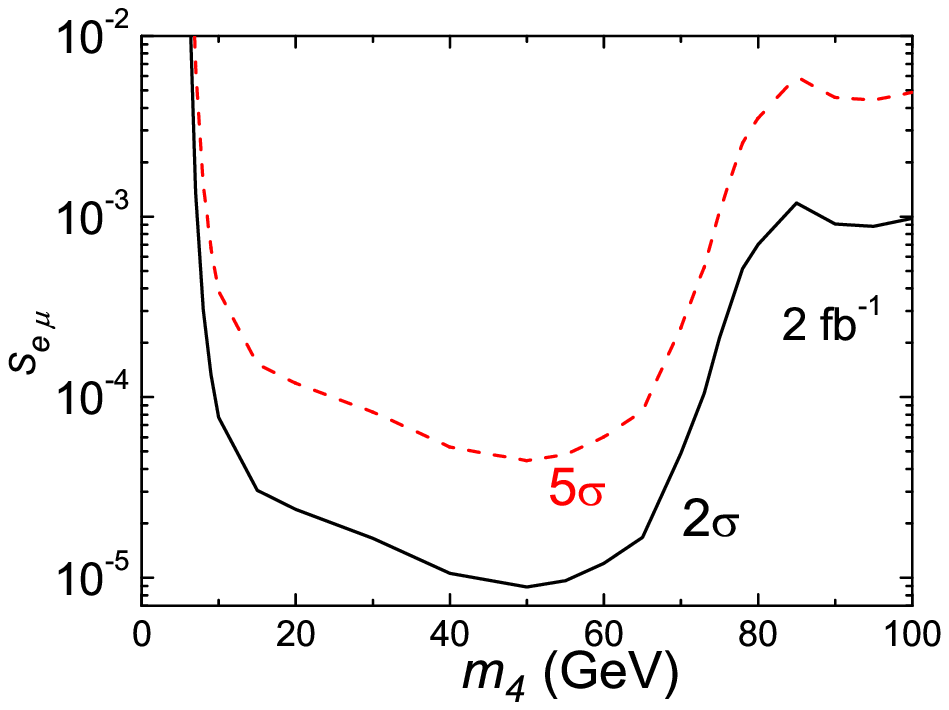}
\includegraphics[width=7.8truecm,clip=true]{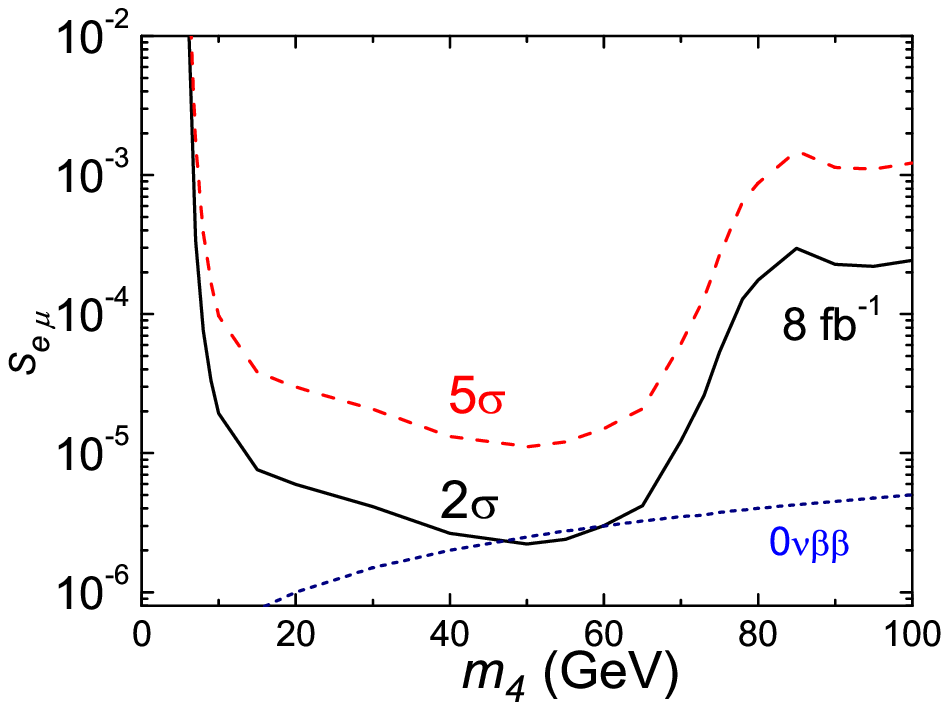}
 \caption{The Tevatron sensitivity to the mixing parameters  versus $m_4$
(a) upper-left: $2\sigma$ and  $5\sigma$ sensitivity of
$S_{\mu\mu}$
 with 2 fb$^{-1}$ integrated luminosity;
 (b) upper-right: same as (a) but  with 8 fb$^{-1}$ integrated luminosity;
(c) lower-left: $2\sigma$ and  $5\sigma$ sensitivity of
$S_{e\mu}$
 with 2 fb$^{-1}$ integrated luminosity;
 (d) lower-right: same as (c) but with 8 fb$^{-1}$ integrated luminosity.
 The horizontal dotted lines in (a) and (b) are the constraint on
 $S_{\mu\mu}\simeq  |V_{\mu4}|^2 < 6\times10^{-3}$
 from an analysis of precision EW measurements \cite{Nardi:1994iv}. The  DELPHI~\cite{Abreu:1996pa} and L3~\cite{Adriani:1992pq} bounds are
 also given here for comparison. The lower dotted curve in (d) is the bound on
 $S_{e \mu}\simeq  |V_{e 4}|^2$  from $0\nu\beta\beta$.}
\label{smumu}
\end{figure}

The CDF collaboration at the Tevatron has successfully
studied the events with like-sign dileptons  in a different
context  \cite{Abulencia:2007rd}. Given our event selection, in Fig.~\ref{smumu} we estimate the sensitivity reach for the
mixing parameters  versus $m_4$ at the $2\sigma$ (solid curves)
and $5\sigma$ (dashed curves) level at the Tevatron.
In Figs.~\ref{smumu}(a$-$b) (upper-left and upper-right), the
sensitivity is shown for $S_{\mu\mu}$  with 2 and 8 fb$^{-1}$
integrated luminosity. The horizontal dotted lines are the
constraint on
 $S_{\mu\mu}\simeq  |V_{\mu4}|^2 < 6\times10^{-3}$
 from an analysis of precision EW measurements \cite{Nardi:1994iv}. The  DELPHI~\cite{Abreu:1996pa} and L3~\cite{Adriani:1992pq} bounds are
 also given for comparison.  We find that the Tevatron has the potential to reach
 the following sensitivity for the mass of the heavy neutrino
\beq m_4 \sim \left\{
\begin{array}{c}
40 - 130\ {\gev}\  \quad {\rm for}\ 2\sigma\ \ {\rm with\ 2\  fb}^{-1}; \\
10 - 180\ {\gev}\ (50-120\ {\gev}) \quad {\rm for}\ 2\sigma\ (5\sigma)\ {\rm with\ 8\  fb}^{-1}.\\
\end{array}
\right.  \label{TeVmass} \eeq Alternatively, the sensitivity for
the mixing parameter can be
\beq S_{\mu\mu} \sim \left\{
\begin{array}{c}
2\times 10^{-5}\  \qquad {\rm for}\ 2\sigma\ \ {\rm with\ 2\  fb}^{-1}; \\
5\times 10^{-6}\  (2\times 10^{-5}) \quad {\rm for}\ 2\sigma\ (5\sigma)\ {\rm with\ 8\  fb}^{-1}.\\
\end{array}
\right. \label{TeVmix} \eeq

Similar to Figs.~\ref{smumu}(a$-$b),  Figs.~\ref{smumu}(c$-$d)
(lower-left and lower-right) show the results for $S_{e\mu}$
instead. The lower dotted curve in Fig.~\ref{smumu}(d) is the bound on
 $S_{e \mu}\simeq  |V_{e 4}|^2$  from $0\nu\beta\beta$.
We have assumed the same detection efficiencies for $\mu$ and $e$.
With this assumption, the slightly better reach for $S_{e\mu}$
compared to $S_{\mu\mu}$ is due the factor of two difference in
total rate with identical and nonidentical particles as evident
from Eq.~(\ref{eq:bare}).
With 2 fb$^{-1}$ luminosity, the sensitivity to $|V_{e 4}|^2$ is
not close to the stringent bound from the $0\nu \beta\beta$ decay
as seen in Fig.~\ref{fig:Uepeak}. We see from Fig.~\ref{smumu}(d)
that with 8 fb$^{-1}$ luminosity, the Tevatron sensitivity for
$S_{e \mu}$ may reach the level of the current bound from
$0\nu\beta\beta$.  From Eq.~(\ref{lumscl}),
it is straightforward to obtain future sensitivity to mixing
parameters ($S_{\mu\mu}, S_{e\mu}$) by a simple scaling of the
luminosity.

\subsection{Search for Like-sign Dilepton Signals  at the LHC}
\label{lhc}

At the LHC with a c.m.~energy of 14 TeV in $pp$ collisions, we
adopt the basic acceptance cuts  on leptons and jets as
\bea p_T^\ell >10 {\,\rm GeV},\quad |\eta^\ell | <2.5,\quad
 p_T^j  > 15 {\,\rm GeV},\quad  |\eta^j | < 2.5.
 \label{eq:basic-lhc}
\eea
The efficiency of these cuts  increases with heavy neutrino mass
and is $50\%$ for $m_4=200\ \gev$ and $80\%$ for $m_4=800\ \gev$.
The smearing parameters to simulate the ATLAS/CMS detectors are
\cite{ATLAS}
\bea {\Delta p_T^\mu \over p_T^\mu} = 36\times 10^{-5}\
p_T^\mu,\quad {\Delta E_j\over E_j} = {1 \over \sqrt{E_j} } \oplus
5\% , \label{eq:smear-lhc} \eea
where $p_T^\mu$ and $E_j$ are in units of GeV.

\begin{figure}[tb]
\includegraphics[width=7.8truecm,clip=true]{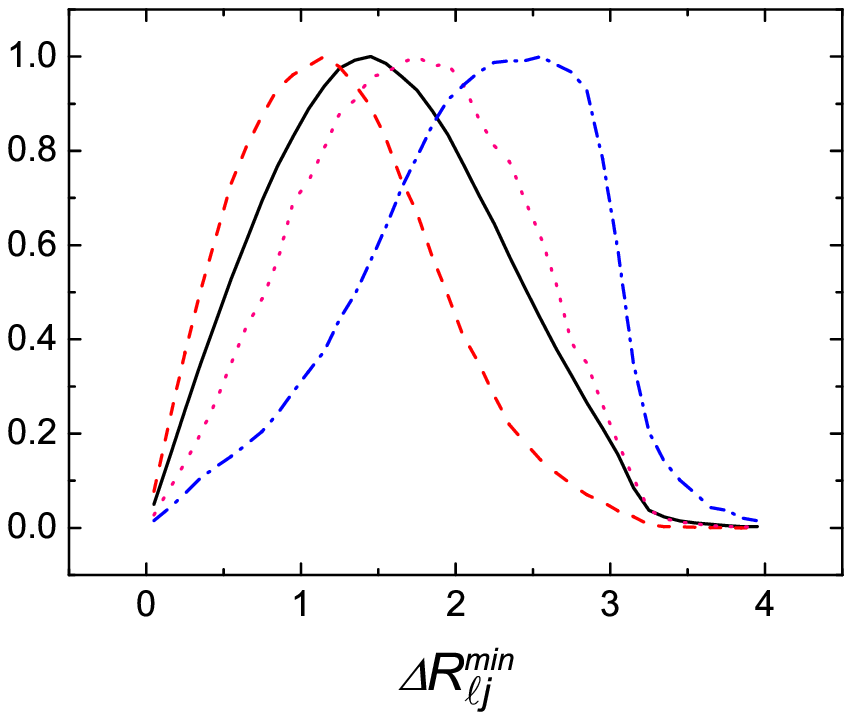}
\includegraphics[width=7.8truecm,clip=true]{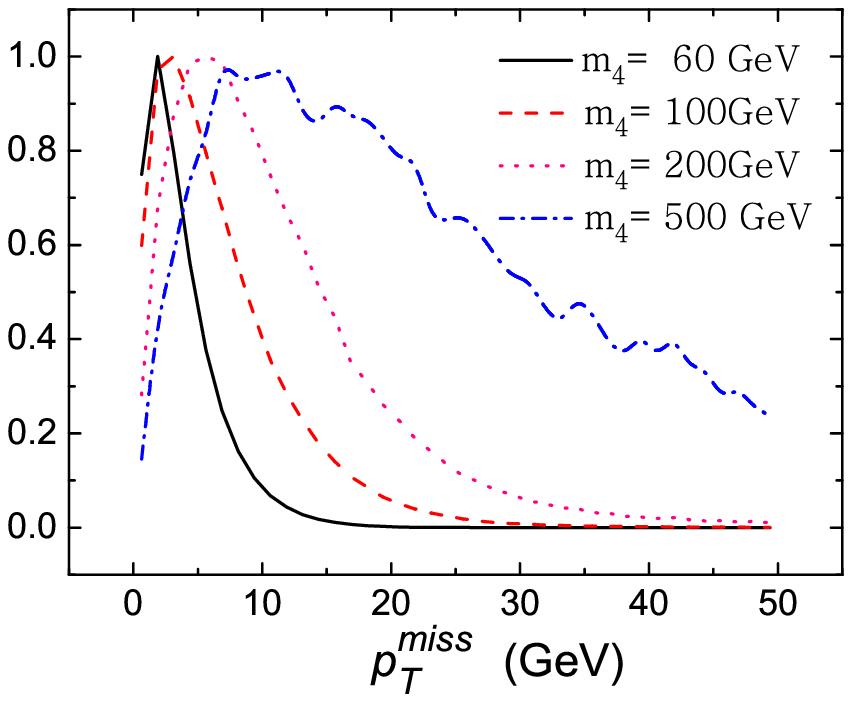}
\includegraphics[width=7.8truecm,clip=true]{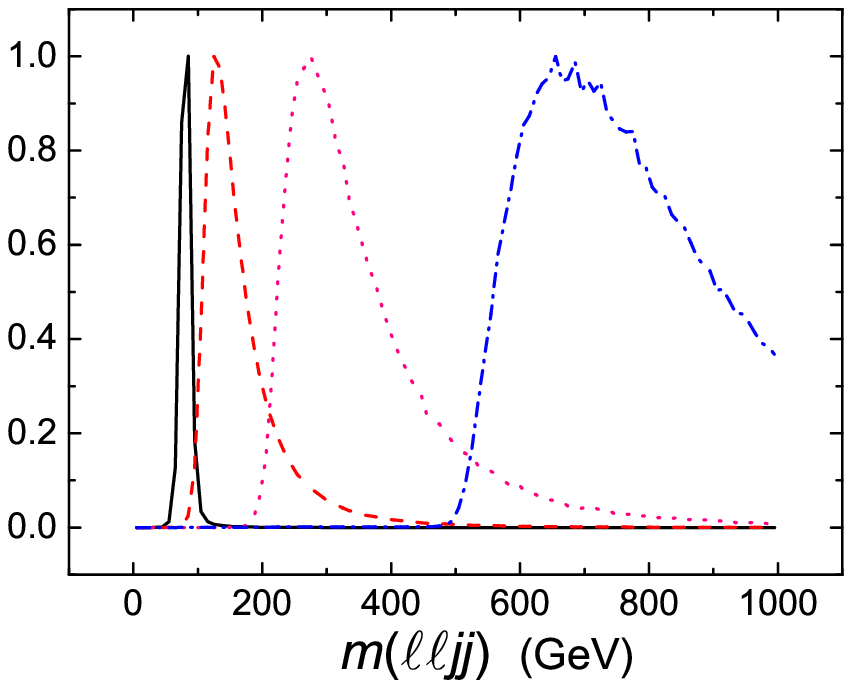}
\includegraphics[width=8truecm,clip=true]{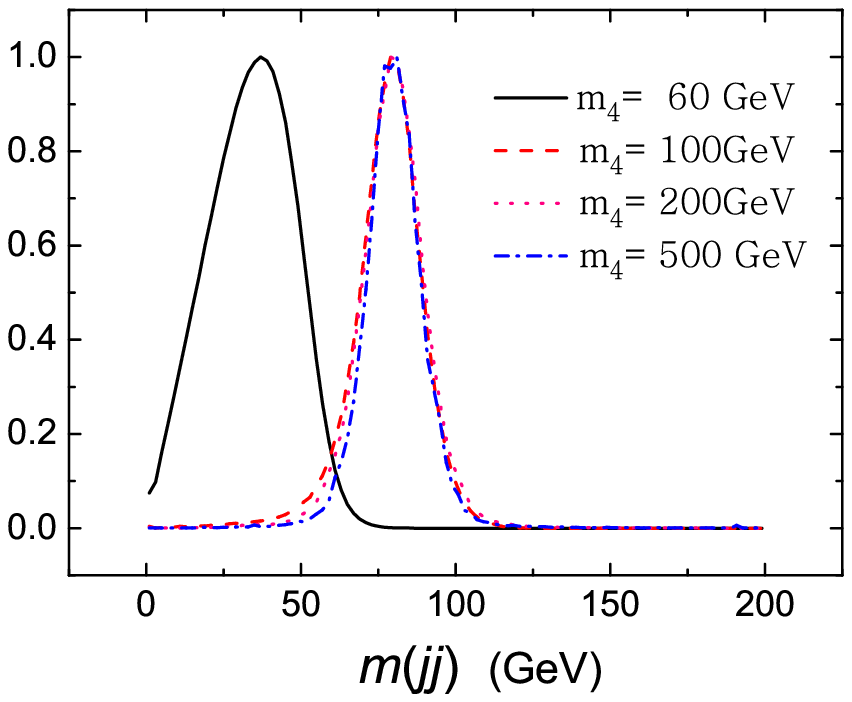}
\caption{Normalized distributions $\sigma^{-1} d\sigma/dX$ for
$m_4=60,\ 100,\ 200$ and 500 GeV at the LHC for (a) upper left:
the minimal isolation $\Delta R_{\ell j}^{min}$; (b) upper right:
the missing transverse momentum $\ptmiss$; (c) bottom left: the
$2\ell 2j$ system invariant mass $m(\ell\ell jj)$; (d) bottom
right: the  di-jet invariant mass $m(jj)$. } \label{Fig:DR-lhc}
\end{figure}

We again present the characteristic kinematical distributions for
the signal. Fig.~\ref{Fig:DR-lhc}(a) shows the normalized
distribution of the minimal isolation $\Delta R_{\ell j}$. The
simulated missing transverse momentum after the energy-momentum
smearing is plotted in Fig.~\ref{Fig:DR-lhc}(b). The invariant
masses of  the $2\ell 2j$ system $m(\ell\ell jj)$ and the  di-jets
$m_(jj)$ are demonstrated in Figs.~\ref{Fig:DR-lhc}(c) and (d),
respectively. We thus design the selection cuts at the LHC as
\bea && \Delta R^{min}_{\ell j}> 0.5,
\label{eq:R-lhc} \\
&&  60\ {\gev} <  \ {\rm either}\ m(\ell\ell jj)\ \  {\rm or}\
m(jj) < 100\ \gev ,
\label{eq:m-lhc}\\
&&  p\!\!\!\slash_T < 25\ \gev . \label{eq:pt-lhc} \eea
These cuts are highly  efficient in selecting the signal events.
We illustrate this in Table \ref{Tab:lhceff}, in which we
calculate the signal rates with the consecutive cuts for $m_4=200$
GeV and $\left|V_{\mu 4}\right|^{2}=\left|V_{\tau 4}\right|
^{2}=5\times 10^{-3} \gg \left|V_{e 4}\right|^{2}$. Again the
choice of mixing elements is motivated by constraints from
precision EW measurements. However as discussed earlier this is
for illustration purposes only and in our full analysis we have
kept $S_{\mu\mu}$ and $S_{\mu e}$ as free parameters.

\bet \caption{The representative signal and background cross
sections at  the LHC,   for $\mu^\pm \mu^\pm j j $ and the
efficiencies with the consecutive cuts. For illustration, we have
used $m_4=200$ GeV, $\left|V_{\mu 4}\right|^{2}=\left|V_{\tau
4}\right| ^{2}=5\times 10^{-3} \gg \left|V_{e 4}\right|^{2}$, and
$m_H=120,\ 300$ GeV.
} \vskip 0.2cm
\begin{tabular}{c|c|c|c|c|c}
 \hline
&No cut & Basic cut & $+p\!\!\!\slash_T$ cut & $+\Delta R$ cut & $+m(jj), m(\ell\ell jj)$ cut  \\
& & (\ref{eq:basic-lhc}) & (\ref{eq:pt-lhc}) & (\ref{eq:R-lhc}) & (\ref{eq:m-lhc})  \\
 \hline
Signal  & & & & &\\
 $\sigma$~(fb) &0.86&0.42&0.37&0.35&0.33\\
eff.&-&48\%&88\%&96\%&94\% \\
 \hline
 $t\bar t$ Bkg & & & & & \\
 $\sigma$ (fb) &29.6&16.9&2.7&0.075&0.002\\
eff.&-&57\%&16\%&2.8\%&2.7\%\\
 \hline
$W^{\pm}W^{\pm}W^{\mp} $&$m_H=$120 GeV~ & & & & \\
$\sigma$&1.01&0.42&0.057&0.052&0.050\\
eff.&-&42\%&14\%&91\%&96\%\\
& $m_H=$300 GeV & & & & \\
 $\sigma$ (fb) &1.28&0.58&0.066&0.061&0.058\\
eff.&-&45\%&11\%&92\%&95\%\\
 \hline
$W^{\pm}W^{\pm} j j $& $m_H=$120 GeV & & & & \\
  $\sigma$ (fb)&4.2&1.3&0.29&0.17&0.019\\
eff.&-&31\%&22\%&59\%&11\%\\
 &$m_H=$300 GeV & & & & \\
  $\sigma$ (fb) &4.4&1.4&0.34&0.19&0.025\\
eff.&-&32\%&24\%&56\%&13\%\\
 \hline
\end{tabular}
\label{Tab:lhceff} \eet

\begin{figure}[tb]
\center
\includegraphics[width=10truecm,clip=true]{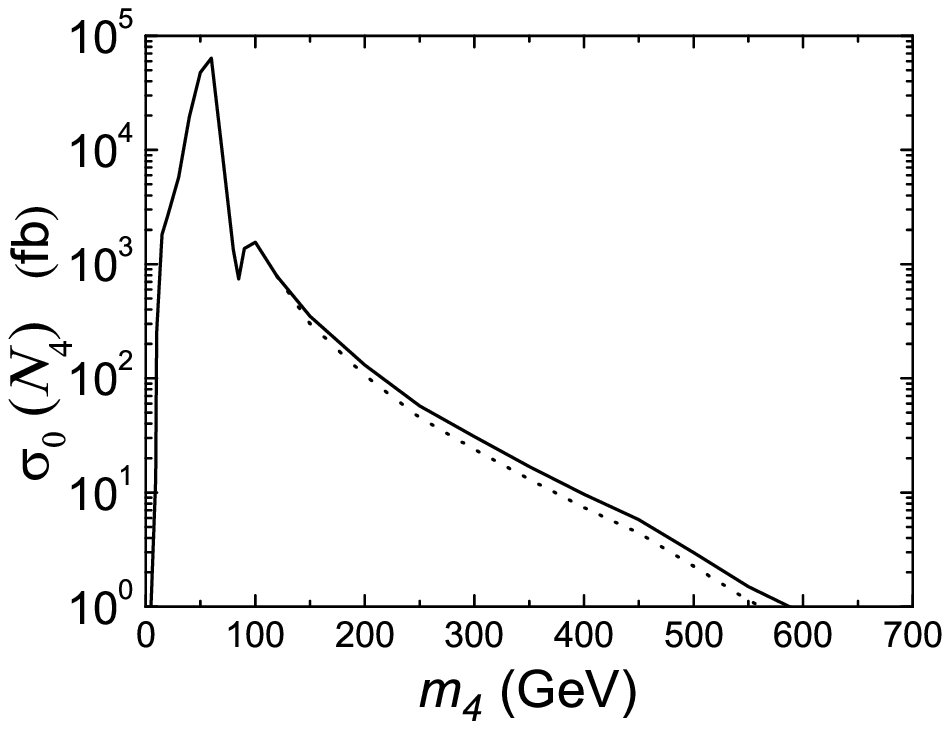}
 \caption{ The bare cross section $\sigma_0(N_4)$  versus heavy neutrino mass $m_4$ after all the cuts at the LHC (14 TeV). The solid (dotted) line correspond to the exclusion (inclusion) of the Higgs decay channel for $m_H = 120$ GeV.}
\label{sigma0-lhc}
\end{figure}

In Fig.~\ref{sigma0-lhc}, we plot the bare cross section
$\sigma_0(N_4)$ with the basic cuts of Eq.~(\ref{eq:basic-lhc}) as
well as the selection cuts
Eqs.~(\ref{eq:R-lhc})$-$(\ref{eq:pt-lhc}) at 14 TeV. The solid
(dotted) curves correspond to the bare cross section without
(with) the Higgs decay channel  for $m_H = 120$ GeV. The reduction
in rate is mainly due to the basic acceptance cuts. We note that
the cross section with the cuts at 14 TeV is higher than that at
10 TeV by a factor of 1.4$-$1.6 for $m_4=100 - 500$ GeV. The
sensitivity reach for the mixing parameters to be presented later
will be scaled down roughly according to this factor for LHC with c.m. energy of  10 TeV.

As discussed in the previous section, a large SM  background comes
mainly from top quark production and decay  via the chain decay
$t\rightarrow b\rightarrow c\ \ell^+ \ \nu_{\ell}$. Fortunately,
after all the selective cuts in
Eqs.~(\ref{eq:basic-lhc})$-$(\ref{eq:pt-lhc}), the top-quark decay
background is essentially eliminated and has no remaining events
for the expected luminosity of 100 fb$^{-1}$ at LHC.

There are several other SM backgrounds coming from like-sign $W$
boson production at the LHC energies. First of all, the triple
gauge-boson production process
\beq pp \rightarrow W^{\pm}W^{\pm}W^{\mp}\rightarrow
\ell^{\pm}\ell^{\pm}\nu\nu\ jj , \eeq
leads to the irreducible background with two like-sign leptons
plus jets. Next, the same final state can be produced via the
process
\beq pp \rightarrow W^{\pm}W^{\pm}\ jj \rightarrow
\ell^{\pm}\ell^{\pm}\nu\nu\ jj , \eeq
where the two jets may come from either QCD scattering or from the
gauge-boson fusion process. However these backgrounds have two
missing neutrinos and can be suppressed by a combination of
cuts on the missing transverse energy and invariant mass. We also
analysed the backgrounds coming from $Z$ boson production
\beq pp \rightarrow jjZZ,\quad pp \rightarrow jjZW. \eeq
in which some charged leptons are missing in the detection so that
they lead to like-sign dilepton events. The backgrounds are very
small after the  cuts.

We list the number of background events and efficiency of cuts in
Table~\ref{Tab:lhceff} for a luminosity of 100 fb$^{-1}$ at LHC.
The total background is about $7-8$ events for 100 fb$^{-1}$ at
the LHC. The main background is from the $W^{\pm}W^{\pm}W^{\mp} $
channel and can be further suppressed if a tighter  missing energy
cut could be exploited. For instance,  the background events may
be reduced by half, leaving about $3-4$ events with $\ptmiss < 15$
GeV.

\begin{figure}[tb]
\center
\includegraphics[width=8.5truecm,clip=true]{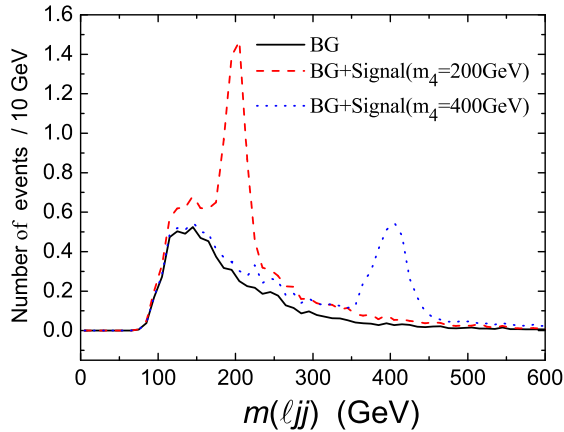}
 \caption{Invariant mass distributions of $m(\ell jj)$ for the signal with $m_4=200,\ 400$ GeV and background processes.
 }
\label{fig:ljj}
\end{figure}

\begin{figure}[tb]
\center
\includegraphics[width=8.5truecm,clip=true]{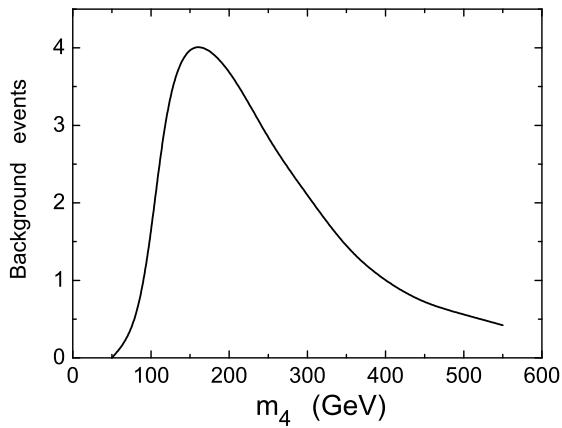}
 \caption{ Number of background events vs mass of the heavy neutrino, $m_4$.}
\label{bgvsmn}
\end{figure}

\begin{figure}[tb]
\includegraphics[width=7.8truecm,clip=true]{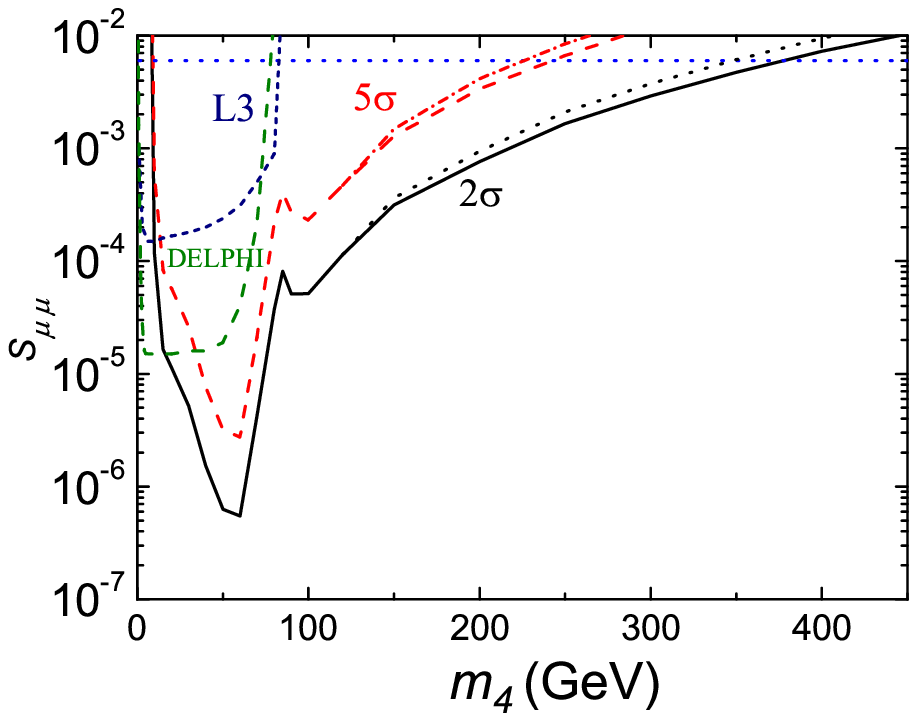}
\includegraphics[width=7.8truecm,clip=true]{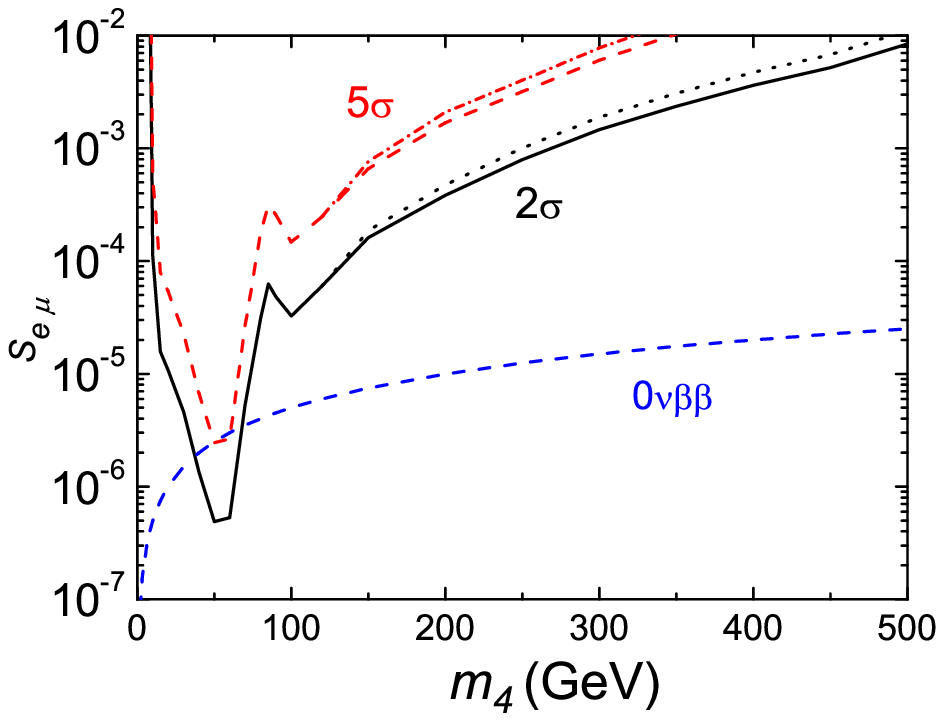}
 \caption{ (a) Left: $2\sigma$ and  $5\sigma$ sensitivity for
  $S_{\mu\mu}$
  versus $m_4$  at the LHC with 100 fb$^{-1}$ integrated luminosity; (b) right: same as (a) but for $S_{e\mu}$ . The solid and dashed (dotted and dash dotted) curves correspond to limits with the exclusion (inclusion) of the Higgs decay channel for $m_H = 120$ GeV. The horizontal dotted line corresponds to the constraint on
$S_{\mu\mu}\simeq  |V_{\mu4}|^2 < 6 \times 10^{-3}$ from precision
EW measurements
 \cite{Nardi:1994iv}. }
\label{Fig:DetectLimit}
\end{figure}

The last but not least important feature of the signal is the
direct reconstruction of the resonant  mass of $N_4$ in the final
state $\ell^\pm jj$. This is shown in Fig.~\ref{fig:ljj} for the
SM background and the signal with $m_4=200,\ 400$ GeV. We see the
effective reconstruction of the resonant mass. For a given mass
$m_4$ in the search, one can further make the event selection on
$m(\ell jj)$
\beq 0.8\ m_4<m(\ell jj)<1.2\ m_4 , \label{eq:mcut} \eeq
to estimate the significance of the signal observation. This loose
cut has little effect on  the signal, but reduces the total
background to $0-4$ events for 100 fb$^{-1}$ in the range of $m_4$
as shown in Fig.~\ref{bgvsmn}. We once again adopt  Poisson
statistics to determine the search sensitivity. The number of
signal events needed for $2\sigma$ significance would be $3-11$;
and $15-44$ for $5\sigma$ significance. In
Fig.~\ref{Fig:DetectLimit}(a) and Fig.~\ref{Fig:DetectLimit}(b),
we summarize the sensitivity  for $S_{\mu\mu}$ and $S_{e \mu}$
versus $m_4$, respectively. The solid (dashed) curves correspond
to $2 \sigma$ ($5 \sigma$) limits on $S_{\ell \ell'}$ with the
exclusion of the Higgs decay channel. The dotted (dash dotted)
curves are similar but with the inclusion of the Higgs decay
channel for $m_H = 120$ GeV. The horizontal dotted line
corresponds to constraints on $|V_{\mu4}|^2 < 6 \times 10^{-3}$
from precision EW measurements \cite{Nardi:1994iv}. In
Fig.~\ref{Fig:DetectLimit}(b) the dashed line at the bottom
corresponds to the limit from $0\nu\beta\beta$.

%
%
%
In the optimistic case, we assume that $\left|V_{\tau 4}\right|
^{2}\ll \left|V_{\mu 4}\right| ^{2}$ and $ S_{\mu\mu}\simeq
\left|V_{\mu 4}\right| ^{2} \leq 6 \times 10^{-3} $. The
detection sensitivity on heavy neutrino mass can be
\bea m_4 \sim \left\{
\begin{array}{c}
 375\ \rm{GeV }\,\,\,\,\,\, \rm{for}\,\,\,2\sigma ;\\
 250\ \rm{GeV }\,\,\,\,\,\, \rm{for}\,\,\,5\sigma.
\end{array}
\right. \eea
Or alternatively, the mixing parameter can be probed to \bea
S_{\mu\mu}\sim \left\{
\begin{array}{c}
7\times 10^{-7} \,\,\,\,\,\, \rm{for}\,\,\,2\sigma ;\\
3\times 10^{-6} \,\,\,\,\,\, \rm{for}\,\,\,5\sigma .
\end{array}
\right. \eea

In particular, even with the very stringent bound on $|V_{e4}|^2$
from $0\nu\beta\beta$ as indicated by the dashed curve in
Fig.~\ref{Fig:DetectLimit}(b), one may still have $2\sigma$
sensitivity if $m_4\approx m_W$.

\begin{figure}[tb]
\includegraphics[width=7.8truecm,clip=true]{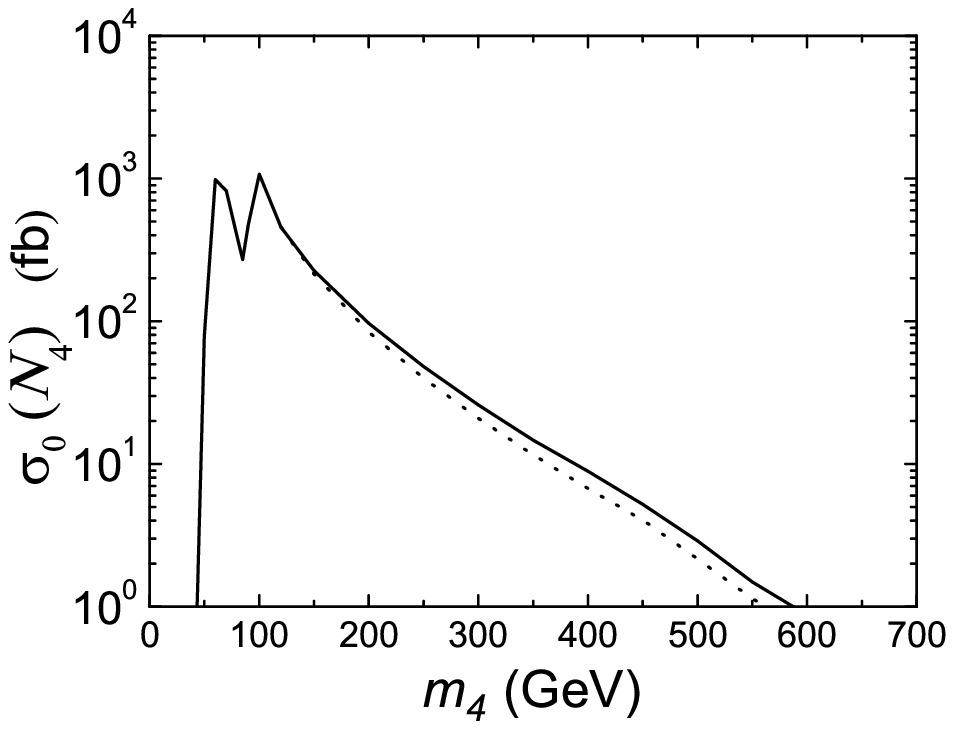}
\includegraphics[width=7.8truecm,clip=true]{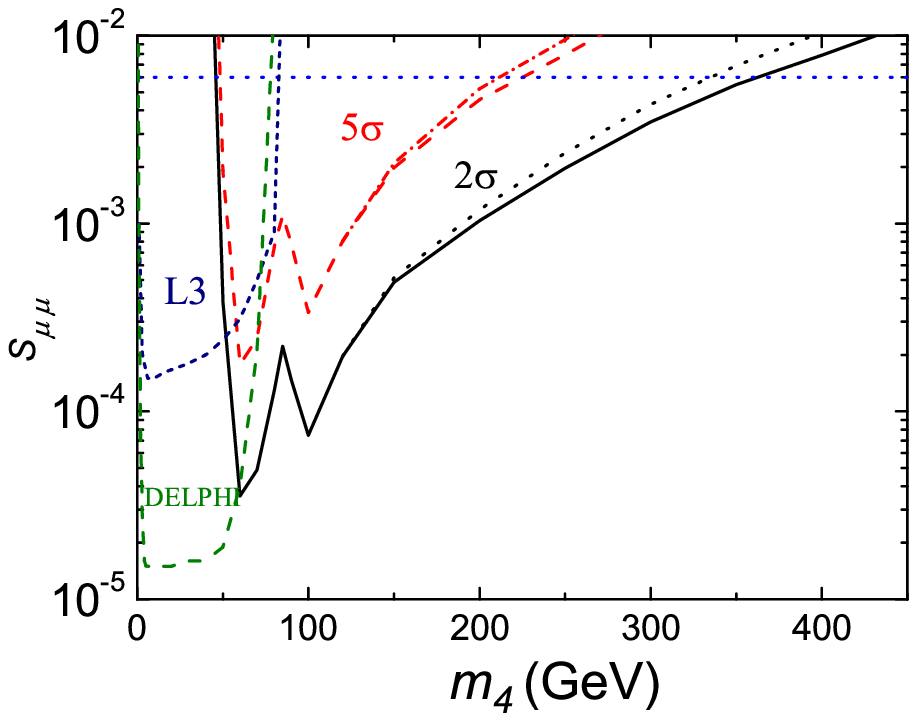}
 \caption{ (a) Left: same as Fig.~\protect\ref{sigma0-lhc} but  with the tighter cuts of Eqs.~(\protect\ref{eq:drcut}) and (\protect\ref{eq:ptcut}); (b)  right: same as
 Fig.~\protect\ref{Fig:DetectLimit}(a) but  with the tighter cuts of Eqs.~(\protect\ref{eq:drcut}) and (\protect\ref{eq:ptcut}). }
\label{Fig:Newcuts}
\end{figure}

Our calculations for hadron colliders have been based on
parton-level simulations. A recent study \cite{delAguila:2007em}
pointed out that there may be other backgrounds to be considered
when detector effects are included. One of them is the faked
like-sign dileptons from the $b\bar b$ cascade decay. The other is
due to the QCD multi-jet radiation to degrade the reconstruction
of $W\to jj$. Those backgrounds can not be easily simulated in
particular at the parton-level. A preliminary analysis including
full CMS detector simulations cannot support their claim
\cite{NRCMS}.
Nevertheless, we may consider to design more stringent acceptance
cuts to further discriminate against the backgrounds. First,
common wisdom suggests to tighten up the charged lepton isolation
requirement
\beq
 \Delta R^{min}_{\ell j} >0.8 ,
 \label{eq:drcut}
\eeq
which would  remove the backgrounds from $b, c$ decays
substantially, but a full assessment can be made only when real
data become available and after the detectors are fully
understood. Next, we may increase the jet threshold to suppress
the initial state QCD jet radiation to purify the $W\to jj$
sample. Our estimate based on a PYTHIA simulation shows that the
kinematics of a DY-type electroweak process can be largely
preserved with the appropriate jet threshold. We thus examine the
cut
\beq p_T^j > 25\ {\rm GeV}, \label{eq:ptcut} \eeq
which results in only about $17\%$ of the events with potential
jet contamination. The results with the tightened cuts are given
in Fig.~\ref{Fig:Newcuts}. We see that the stringent cuts severely
hurt the low mass region, but the effect on the high mass region
is modest.

\subsection{Like-sign Dilepton Signals with $\tau$ in the Final States}

So far we have only presented the results with electron and muon
final states and ignored the taus. This is due to the experimental
challenge of $\tau$ reconstruction. Given the importance to cover
all the lepton flavors, one must strive to include taus on the
search list. Besides the known experimental practice for $\tau$
identification at the Tevatron \cite{Abulencia:2005kq}, there are
proposals to identify $\tau$ events in connection with the
neutrino sector \cite{taurec}. The central issue is to reconstruct
the missing momenta from $\tau$ decays. We can generalize our
requirement for the charged leptons to the isolated charged tracks
presumably from the $\tau$ decays ($e,\mu$, or one-prong and
three-prong charged hadrons)
\beq p_T({\rm track}) >10 {\,\rm GeV},\quad |\eta({\rm track})|
<2.5. \eeq
This assures that the parent taus are very energetic. For
events with one $\tau$ and no other sources of missing particles,
the missing momentum will be along the direction of the charged
track. We thus have
\beq {\vec{p}}\ ({\rm invisible}) =  \kappa\vec{p}\ ({\rm track}),
\eeq
where the proportionality constant $\kappa$ is determined from the
$\etmiss$ measurement by assigning $\etmiss = \kappa p_T({\rm
track})$. For events with two taus, we generalize it to
\beq {\vec{p}}\ ({\rm invisible}) =  \kappa_1 \vec{p}\ ({\rm
track}_1) + \kappa_2 \vec{p}\ ({\rm track}_2). \eeq
As long as the two $\tau$ tracks are not linearly dependent,
$\kappa_1$ and $\kappa_2$ can be determined again from the
$\etmiss$ measurement. The missing momenta, as well as the $\tau$
kinematics, are  thus fully reconstructed.

Although we believe that the $N_4$ signals in the modes of $e^\pm
\tau^\pm,\ \mu^\pm \tau^\pm$ and $\tau^\pm \tau^\pm$ would be very
promising for observation, the background analyses will be
considerably more involved due to the complication of $\tau$
reconstruction. Since our simulations are performed at the parton
level, we are unable to adequately address the background
suppression and to quantify the signal observability. We thus
leave this for future studies.

\section{Summary and Conclusions}
\label{concld}

The observation of a $\lv$ process would show that neutrino is a
Majorana particle unambiguously. Apart from light neutrinos, $\lv$
processes involving SM particles can receive a contribution from
heavy Majorana neutrinos due to mixing. In fact, this contribution
can be resonantly enhanced for appropriate masses of the heavy
neutrino. In the absence of observation of $\lv$ interactions, the
rates for these processes can constrain the mixing elements
${|V_{\ell_1 4} V_{\ell_2 4}|}$ as a function of the mass $m_4$ of
the heavy Majorana neutrino. We  considered  two classes of $\lv$
violating processes: (a) low energy $\dl=2$ tau decays and rare
meson decays and (b) collider signals for like-sign dilepton
production with no missing energy implying the existence of
Majorana neutrinos. We emphasize the necessity of involving two
charged leptons and no neutrinos in the initial and final states,
to be conclusive about lepton-number violation.

For the low energy interactions we evaluated the transition rates
and branching fractions as a function of the mass and mixing of
the heavy neutrinos. We then translated  the current experimental
bounds from direct searches into limits on ${|V_{\ell_1 4}
V_{\ell_2 4}|}$ as a function of the mass $m_4$ of the heavy
neutrino. Amongst the rare meson decays, the $K^+ \rightarrow
\ell_1^+ \ell_2^+ \pi^-$ decay mode currently  gives the most
sensitive experimental limits on ${|V_{\ell_1 4} V_{\ell_2 4}|}$.
Potentially, these constraints are six orders of magnitude more
stringent than the constraints from precision electroweak data
which limit $|V_{\ell 4}|^2$ to few times $10^{-3}$. As the
intermediate heavy sterile neutrino is a real particle which might
exit the detector if the decay length is longer than the detector
size, for very small mixing angles the bounds get weakened but are
still much more stringent than the electroweak precision
constraints. This effect should be taken into account and a
detailed analysis of past experimental data is required in order
to find the precise limits on the mixing angles. Next in
sensitivity are the $D$ and $D_s$ meson decay modes with
constraints of the order of $10^{-3}$. Again, these are
competitive with if not better than constraints from EW precision
data. The other processes (in other mass ranges) have very weak
experimental limits, weaker than EW precision data and essentially
do not impose any meaningful bounds on ${|V_{\ell_1 4} V_{\ell_2
4}|}$. This implies that more accurate experimental studies on
those rare decays should be strongly encouraged. In particular,
many interesting processes of $D,\ B$ decays have not even been
experimentally probed as well as those with a $\tau$ lepton in the
final state. Among the $\tau$-decay modes the best limits come
from $\tau^- \rightarrow \ell^+ \pi^- \pi^-$. The other
$\tau$-decay modes have sensitivity of order $10^{-3}$ to
$10^{-5}$. Again, the constraints from $\tau$ decay modes are
competitive with or better than constraints from precision EW data
by 2 to 3 orders of magnitude. The experimental bound on $\lv$
processes is expected to improve in the future. The future
sensitivity of the square of the mixing parameter will increase
approximately by an order of magnitude for every order of
magnitude improvement in experimental bounds on branching
fractions. We have shown that the low energy $\dl=2$ $\tau$ decays
and rare meson decays can be very strong probes to discover or
constrain the mass and mixing of heavy Majorana neutrinos. Even
those decay modes which do not impose strong constraints should
not be neglected. It only implies that a large range of the
parameter space is available for exploration.

In addition to analyzing the $\lv$ tau and meson decay modes and
precision EW measurements we also compiled the constraints on the
mixing elements ($|V_{e4}|^2, |V_{\mu4}|^2$ and $|V_{\tau4}|^2$)
from peak searches, accelerator experiments, reactor experiments
and others - collectively called laboratory constraints. In the
absence of detection of $\lv$, the laboratory constraints and the
ones from $\dl = 2$ processes can be compared. The constraints on
mixing from $\lv$ tau decays are always competitive with or better
than laboratory constraints in the corresponding mass region while
the constraints from $\lv$ meson decays are competitive with
laboratory constraints only in some mass regions. We note that we
explore more combinations of mixing elements and also provide
better constraints on mixing in some mass regions. More
importantly, a detection in one of the experiments analyzed to
obtain laboratory constraints implies the existence of a sterile
neutrino while a detection in one of the $\dl = 2$ tau or meson
decay modes studied in our analysis would imply $\lv$ and hence
the existence of a Majorana neutrino.
We pointed out the fact, often ignored in the literature when
analyzing low-energy processes, that a heavy neutrino might decay
outside the detector if it becomes long-lived for low mass (less
than a GeV) and/or very small mixing.

For the collider signals of heavy Majorana neutrinos we looked for
the definitive lepton-number violating like-sign dilepton
production and no missing energy. Such signals have low
backgrounds and have the potential for discovery of heavy Majorana
neutrinos. At the Tevatron, with the current and future integrated
luminosities, we find that  the mass of the heavy Majorana neutrinos can be probed up to
\beq m_4 \sim \left\{
\begin{array}{c}
\nonumber
10 - 130\ {\gev}\ (10-75\ {\gev}) \quad {\rm for}\ 2\sigma\ (5\sigma)\ {\rm with\ 2\  fb}^{-1}; \\
\nonumber
10 - 180\ {\gev}\ (10-120\ {\gev}) \quad {\rm for}\ 2\sigma\ (5\sigma)\ {\rm with\ 8\  fb}^{-1}.\\
\end{array}
\right. \eeq Alternatively, the sensitivity for the mixing
parameter can be
\beq S_{\mu\mu} \sim |V_{\mu 4}|^2 \sim \left\{
\begin{array}{c}
\nonumber
2\times 10^{-5}\  (10^{-4}) \qquad {\rm for}\ 2\sigma\ (5\sigma)\ {\rm with\ 2\  fb}^{-1}; \\
\nonumber
5\times 10^{-6}\  (2\times 10^{-5}) \quad {\rm for}\ 2\sigma\ (5\sigma)\ {\rm with\ 8\  fb}^{-1}.\\
\end{array}
\right. \eeq
%
%
This will surpass the  DELPHI~\cite{Abreu:1996pa} and L3
\cite{Adriani:1992pq, l3} $95 \%$ C.L. bounds.

The sensitivity for heavy Majorana neutrinos can be extended
significantly at the LHC. With 100 fb$^{-1}$ of integrated
luminosity, \bea m_4 \sim
 375\ (250)\ \rm{GeV }\,\,\,\,\,\, \rm{for}\,\,\,2\sigma \,\,(5\sigma) ;
\eea
or alternatively, the mixing parameter can be probed to \bea
S_{\mu\mu}\sim |V_{\mu 4}|^2 \sim 7\times 10^{-7} \ \  (3\times
10^{-6}) \,\,\,\,\, \rm{for}\,\,\,2\sigma \ (5\sigma). \eea
The sensitivity at LHC will go well beyond the  DELPHI and L3 $95
\%$ C.L. bounds in both mass reach and mixing, and beyond the
current bound on $|V_{e 4}|^2$ from $0\nu \beta\beta$. In summary,
there is a rich avenue of possibilities for discovering or
constraining the elusive Majorana neutrinos.


\begin{acknowledgments}
We would like to thank Pavel Fileviez Perez for comments on the
draft. This research was supported in part by the U.S.~DOE under
Grants No.DE-FG02-95ER40896, W-31-109-Eng-38, and in part by the
Wisconsin Alumni Research Foundation. Fermilab is operated by
Fermi Research Alliance, LLC under Contract No. DE-AC02-07CH11359
with the United States Department of Energy. The work at KITP  was
supported in part by the National Science Foundation under Grant
No. PHY05-51164. The work of B.~Z. is supported by the National
Science Foundation of China under Grant No. 10705017. SP would
like to thank the Theoretical Physics Department at Fermilab and
the PH-TH Unit at CERN for hospitality.
\end{acknowledgments}

\vspace{1cm}
\appendix

\section{Lepton mixing formalism}
\label{appmix}

In this appendix, we illustrate the parameterization for the
lepton sector, although we have not followed the relations
literally, assuming that new physics beyond this minimal formalism
exists. The leptonic content in the theory includes three
generations of left-handed SM $SU(2)_L$ doublets and $n$
right-handed SM singlets:
\beq L_{aL}=\left(
\begin{array}{c} \nu_a\\
l_a
\end{array}
\right)_L ,\quad   N_{b R}, \eeq
where $a=1,2,3$ and $b=1,2,3, \cdots, n$  ($n\ge 2$ for at least
two massive neutrinos).

The leptonically universal gauge interactions involving neutrinos
are of the form
\bea -{\cal L} = \left( \frac{g}{\sqrt{2}}W^+_\mu
\sum_{a=1}^{3}\overline{{\nu_a}_L}\ \gamma^\mu l_{aL} + \mathrm{h.c.}
\right) + \frac{g}{2\cos_W}Z_\mu \sum_{a=1}^{3}
\overline{{\nu_a}_L}\ \gamma^\mu {\nu_a}_L . \label{A-Lweak} \eea
The gauge-invariant Yukawa interactions are
\bea -{\cal L}_Y = \left( \sum_{a,b=1}^{3}  f^l_{ab}\
\overline{L_{aL}}\ H {l_{bR}} + \sum_{a=1}^{3} \sum_{b=1}^{n} \
f^\nu_{a b}\ \overline{L_{aL}}\ \hat H N_{b R} \right) + \mathrm{h.c.}
\label{A-LYukawa} \eea
where $H$ is the SM Higgs doublet and $\hat H = i\tau_2 H^*$.
After the Higgs field develops a vev $\langle H\rangle \to v/
\sqrt 2$, the Yukawa interactions lead to Dirac masses for the
leptons
\beq -{\cal L}_m^D =
 \left( \sum_{a,b=1}^{3} \overline{l_{aL}}\   m^l_{ab}\  {l_{bR}} +
\sum_{a=1}^{3} \sum_{b=1}^{n} \ \overline{\nu_{aL}}\ m^\nu_{a b} \
N_{b R}
 \right) + \mathrm{h.c.}
\label{A-Dmass} \eeq
where the mass matrices are given by the vev times the
corresponding Yukawa couplings $m^{l, \nu}_{ab}=f^{l,
\nu}_{ab}v/\sqrt 2$.

The  $3\times 3$ mass matrix $m^l$ can be diagonalized by two
unitary rotations among the gauge interaction eigenstates $l_L,\
l_R$
\bea
 O_L^\dag\ m^l\ O_R ={\rm diag}(m_e, m_\mu, m_\tau), \quad
 l_a  = O_{a \ell}\ \ell,
\label{A-lmix} \eea
where $\ell=e,\mu,\tau$ are the mass eigenstates, which define the
charged lepton flavors. The Dirac masses as well as interactions
with the Higgs boson for the charged leptons now have the standard
form
\beq -{\cal L}_Y^\ell =  \sum_{\ell=e}^{\tau} m_\ell\  (1+{H\over
v})\ \overline{\ell}\  {\ell}. \label{A-LH} \eeq
If the Yukawa interactions of Eq.~(\ref{A-LYukawa}) are the whole
source for neutrino mass, then we would have min($n,3$) massive
Dirac neutrinos.

To complete the neutrino mass sector,  there is also a possible
heavy Majorana mass term
\beq -{\cal L}_m^M  = \frac{1}{2}\sum_{b,b'=1}^{n}
\overline{N^c_{b L}}\ B_{b b'}\ N_{b' R} +\mathrm{h.c.} \eeq
where a charge conjugate state is defined as
$\psi^c=C\bar{\psi}^T$ ($\overline{\psi^c}={\psi}^T C$), and a
chiral state satisfies $(\psi^c)_\tau = (\psi_{-\tau})^c,$ with
$\tau=-,+$ for $L,R$. The full neutrino mass terms thus read
\bea -{\cal L}_{m}^\nu &=&\frac{1}{2} \left(\ \sum_{a=1}^{3}
\sum_{b=1}^{n} \ ( \overline{\nu_{aL}}\ m^\nu_{a b} \ N_{b R} +
\overline{N^c_{bL}}\ m^{\nu}_{ ba} \ \nu^c_{a R} ) +
\sum_{b,b'=1}^{n} \  \overline{N^c_{b L}}\ B_{b b'}\ N_{b' R}
\right)+ \mathrm{h.c.} \nonumber \\
&=& \frac{1}{2}\left( \overline{\nu_L}\ \ \overline{N^c_L} \right)
 \left(
 \begin{array}{cc} 0_{3\times 3} & m^\nu_{3\times n} \\ m^{\nu T}_{n\times 3}
 & B_{n\times n} \end{array}
 \right)
 \left( \begin{array}{c} \nu^c_R\\
N_R
\end{array} \right)
+ \mathrm{h.c.} \label{A-numass} \eea
where we have used the identity $\overline{\nu_{aL}}\ m_{ab}\ N_{b
R}=\overline{N^c_{b L}}\ m_{ba}\ \nu^c_{aR}$,

The mass matrix can be diagonalized by one unitary transformation
\beq { \mathbb L}^\dag \left(\begin{array}{cc} 0 & m^\nu \\ m^{\nu
T} & B
\end{array}\right) {\mathbb L}^* = \left(\begin{array}{cc} m^\nu_{diag} & 0 \\ 0 & M^N_{diag}
\end{array}\right)
\eeq
where the mass eigenvalues are of the order
\bea m^\nu_{diag} \approx {m_\nu^2\over B}, \quad
M^N_{diag}\approx B. \eea
${\mathbb L}$ is a $(3+n)\times(3+n)$ unitary matrix and can be
parameterized  as
\beq
{\mathbb L}=\left(\begin{array}{cc} U_{3\times 3} & V_{3\times n} \\
X_{n\times 3} & Y_{n\times n}
\end{array}\right).
\eeq
The relation between the gauge interaction eigenstates and the
mass eigenstates are given by
\beq
\left( \begin{array}{c}  \nu_L \\
N^c_L  \end{array} \right) = {\mathbb L}
\left( \begin{array}{c}  \nu_L \\
N^c_L \end{array} \right)_m, \eeq
 with\ the mass eigenstates $\nu_m\ (m=1,2,3), \ N_{m'}\ (m'=4,\cdots, 3+n).$
The diagonalized (Majorana) mass terms of Eq.~(\ref{A-numass})
thus read
\bea -{\cal L}_{m}^\nu = {1\over 2} \left(\sum_{m=1}^3 m^\nu_m\
\overline{\nu_{m L}}\  \nu^c_{m R} + \sum_{m'=4}^{3+n} M^N_{m'}\
\overline{N^c_{m'L}}\ N_{m'R} \right) + \mathrm{h.c.}\ , \eea
with the mixing relations  between the gauge and mass eigenstates
\bea \nu_{aL} &=& \sum_{m=1}^3 U_{a m}\nu_{m L}+\sum_{m'=4}^{3+n}
V_{am'} N^c_{m'L}, \ \
N^c_{bL} = \sum_{m=1}^3 X_{b m}\nu_{m L} + \sum_{m'=4}^{3+n} Y_{b m'} N^c_{m' L}, \\
\nu^c_{a R} &=& \sum_{m=1}^3 U^*_{a m}\nu^c_{m
R}+\sum_{m'=4}^{3+n} V^*_{am'} N_{m'R}, \ \ N_{bR} = \sum_{m=1}^3
X^*_{b m}\nu_{m R}^c + \sum_{m'=4}^{3+n} Y^*_{b m'} N_{m' R}.
\label{A-numix} \eea
Note that the unitarity condition for $\mathbb L$ leads to the
relations
\bea
U U^\dag  + V V^\dag = U^\dag U +  X^\dag X = I_{3\times 3},\\
X X^\dag  + Y Y^\dag =  V^\dag V + Y^\dag Y = I_{n\times n}. \eea
Parametrically, $U U^\dag$ and $\ Y^\dag Y \sim {\cal{O}}(1),$
 $V V^\dag$ and $X^\dag X \sim {\cal{O}}(m_\nu/M_N).$

In terms of the mass eigenstates, the gauge interaction lagrangian
Eq.~(\ref{A-Lweak}) can be written as
\bea
\nonumber -{\cal L} &=& \frac{g}{\sqrt{2}} W^+_\mu \left(
\sum_{\ell=e}^\tau \sum_{m=1}^3 (U^\dag O_L)_{m \ell}\
\overline{\nu_m} \gamma^\mu P_L \ell + \sum_{\ell=e}^\tau
\sum_{m'=4}^{3+n}
(V^\dag O_L)_{m' \ell}\ \overline{N^c_{m'}} \gamma^\mu P_L \ell  \right)+ \mathrm{h.c.} \\
\nonumber &+& \frac{g}{2\cos_W}Z_\mu  \left( \sum_{m_1,m_2=1}^3
(U^\dag U)_{m_1 m_2}\ \overline{\nu_{m_1}}\gamma^\mu P_L \nu_{m_2}
+ \sum_{m_1',m_2'=4}^{3+n} (V^\dag
V)_{m_1'm_2'}\overline{N^c_{m_1'}} \gamma^\mu
P_L N^c_{m_2'}\right)\\
&+& \frac{g}{2\cos_W}Z_\mu  \left( \sum_{m_1=1}^3
\sum_{m_2'=4}^{3+n} (U^\dag V)_{m_1,m_2'}
\overline{\nu_{m_1}}\gamma^\mu P_L N^c_{m_2'} +  \mathrm{h.c.} \right)
\label{A-NCcoupling} . \eea
To make the couplings more intuitive, we define the combination
matrices by
\bea U^{l\nu} = O_L^\dag U,\quad  V^{lN} = O_L^\dag V,\quad U^{\nu
N} = U^\dag V,\quad U^{\nu\nu} = U^\dag U,\quad  V^{NN} = V^\dag V
. \label{mixings} \eea
We thus rewrite the gauge interaction lagrangian by one mixing
matrix for each term
\bea
\nonumber -{\cal L} &=& \frac{g}{\sqrt{2}} W^+_\mu \left(
\sum_{\ell=e}^\tau \sum_{m=1}^3 U^{l\nu *}_{\ell m}\
\overline{\nu_m} \gamma^\mu P_L \ell + \sum_{\ell=e}^\tau
\sum_{m'=4}^{3+n}
V^{lN *}_{ \ell m'}\ \overline{N^c_{m'}} \gamma^\mu P_L \ell  \right)+ \mathrm{h.c.} \\
\nonumber &+& \frac{g}{2\cos_W}Z_\mu  \left( \sum_{m_1,m_2=1}^3
U^{\nu\nu}_{ m_1 m_2}\ \overline{\nu_{m_1}}\gamma^\mu P_L
\nu_{m_2} + \sum_{m_1',m_2'=4}^{3+n} V^{NN}_{m_1'm_2'}\
\overline{N_{m_1'}} \gamma^\mu
P_L N_{m_2'}\right) \\
&+& \frac{g}{2\cos_W}Z_\mu  \left( \sum_{m_1=1}^3
\sum_{m_2'=4}^{3+n} U^{\nu N}_{m_1 m_2'}\
\overline{\nu_{m_1}}\gamma^\mu P_L N^c_{m_2'} +  \mathrm{h.c.} \right)
\label{NCcoupling1}. \eea
These couplings along with the mixing matrices Eq.~(\ref{mixings})
give the most general leptonic interactions of the charged and
neutral currents in terms of the mass eigenstates.
Alternatively, the neutral current interactions can be aligned
along with that of the charged currents when rotating left-handed
neutrinos in the same way as the charged leptons,
\bea
 \nu_{aL}  = (O_L)_{a \ell}\ \nu_{\ell L},\ {\rm or}\
 \nu_{\ell L} &=& \sum_{m=1}^3 (O_L^\dag U)_{\ell m}\nu_{m L}
 +\sum_{m'=4}^{3+n} (O_L^\dag V)_{\ell m'} N^c_{m'L}.
\eea
It may be convenient in certain practical calculations to  rewrite
the neutral current interactions in terms of  their flavor
eigenstates
\bea
\nonumber
-{\cal L} &=& \frac{g}{\sqrt{2}} W^+_\mu \left(
\sum_{\ell=e}^\tau \sum_{m=1}^3 U^*_{\ell m}\  \overline{\nu_m}
\gamma^\mu P_L \ell + \sum_{\ell=e}^\tau \sum_{m'=4}^{3+n}
V^{*}_{ \ell m'}\ \overline{N^c_{m'}} \gamma^\mu P_L \ell  \right)+ \mathrm{h.c.} \\
&+& \frac{g}{2\cos \theta_W}Z_\mu \left( \sum_{\ell=e}^\tau
\sum_{m=1}^3 U^{*}_{\ell m}\  \overline{\nu_m} \gamma^\mu P_L\
\nu_\ell + \sum_{\ell=e}^\tau \sum_{m'=4}^{3+n} V^{*}_{ \ell m'}\
\overline{N^c_{m'}} \gamma^\mu P_L\ \nu_\ell  \right) + \mathrm{h.c.} + ...
\qquad \label{NCcoupling2} \eea
where we have dropped the superscripts for  $U,\ V$ defined in
Eq.~(\ref{mixings}), for simplicity as adopted throughout the text.

\begin{figure}[tb]
\center
\includegraphics[width=4.8truecm,clip=true]{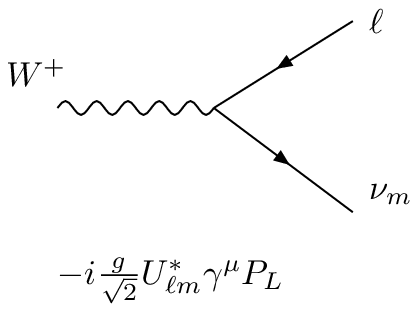}
\includegraphics[width=4.8truecm,clip=true]{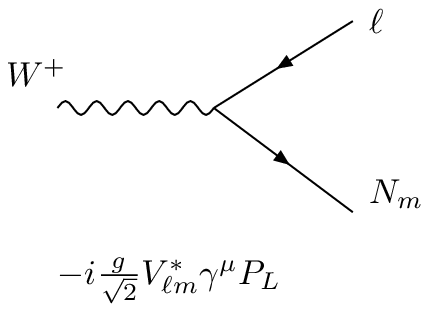}
\includegraphics[width=4.8truecm,clip=true]{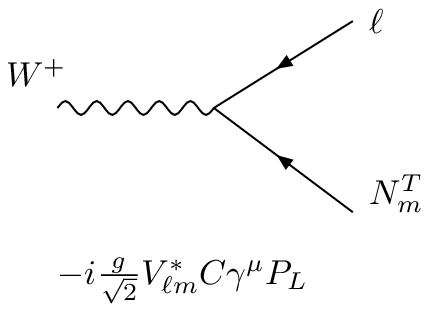}
 \caption{Feynman rules for the charged current vertices in terms of the neutrino mass eigenstates, as given in Eq.~(A.22). }
\label{Fig:wlnu}
\end{figure}

\begin{figure}[tb]
\center
\includegraphics[width=5.5truecm,clip=true]{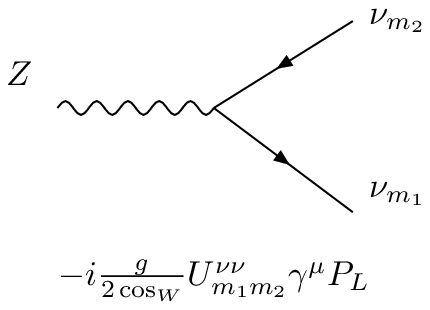}
\includegraphics[width=5.5truecm,clip=true]{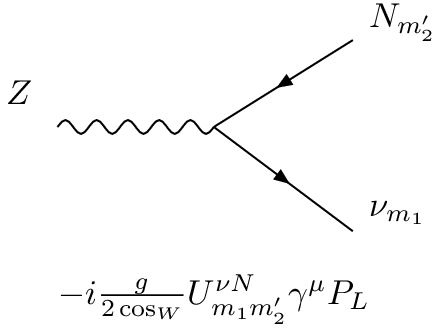}
\includegraphics[width=5.5truecm,clip=true]{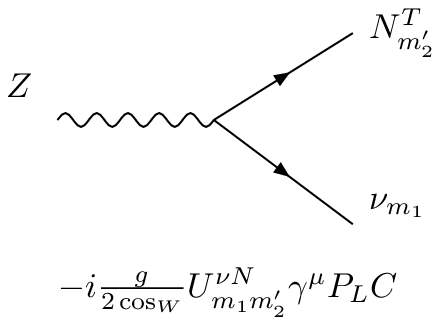}
\includegraphics[width=5.5truecm,clip=true]{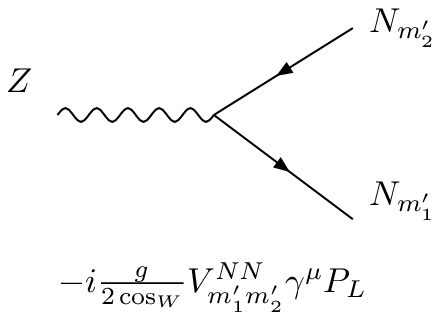}
 \caption{Feynman rules for the neutral current vertices in terms of the neutrino mass eigenstates, as given in Eqs.~(A.20). }
 \label{Fig:znunu}
\end{figure}

For the reader's convenience, we give most of the corresponding
Feynman rules for the interaction vertices, listed in
Fig.~\ref{Fig:wlnu} for the charged currents,  and in
Fig.~\ref{Fig:znunu} for the neutral currents. The Feynman rules
for the other diagrams can be easily deduced from the ones that
are explicitly written down in Fig.~\ref{Fig:wlnu} and
Fig.~\ref{Fig:znunu}.

\begin{figure}[tb]
\center
\includegraphics[width=5.5truecm,clip=true]{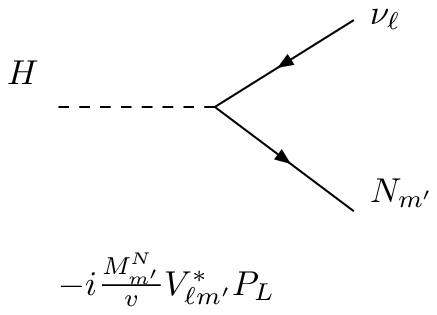}
\includegraphics[width=5.5truecm,clip=true]{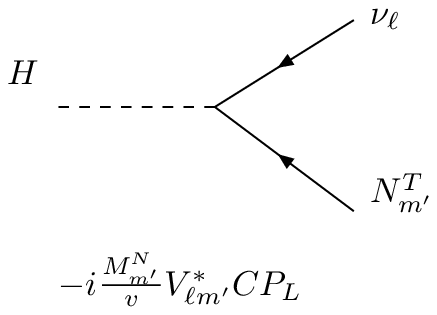}
 \caption{Feynman rule for the Higgs vertex in terms of the heavy neutrino mass eigenstates, as given in Eq.~(A.23).}
\label{Fig:H}
\end{figure}

Finally, the heavy neutrino interactions with the Higgs boson read
\bea -{\cal L_H}  =
 \frac{H}{v}  \sum_{\ell=e}^\tau \sum_{m'=4}^{3+n}
V^{*}_{ \ell m'}\  M^N_{m'}\ \overline{N^c_{m'}} P_L\ \nu_\ell  +
 \mathrm{h.c.} + ... \label{HCcoupling} \eea
The corresponding Feynman rule for the interaction vertex is given
in Fig.~\ref{Fig:H}.

\section{ General amplitude of $\dl=2$ processes}
\label{genamp}

The charged current interaction lagrangian in terms of neutrino
mass eigenstates is
\beq \label{3} {\cal L}_{cc} = -\frac{g}{\sqrt{2}} W^+_\mu \Bigl (
\sum_{\ell = e}^{\tau} \sum_{m = 1}^{3} U^{l \nu *}_{\ell m}\
\overline{\nu_m} \gamma^\mu P_L \ell + \sum_{\ell = e}^{\tau}
\sum_{m' = 4}^{3+n} V^{l N *}_{\ell m'}\ \overline{N^c_{m'}}
\gamma^\mu P_L \ell \Bigr ) +  \mathrm{h.c.} \eeq
where $P_L = \frac{1}{2} (1 - \gamma_5)$. The leptonic $\dl=2$
subprocess $W^-W^- \rightarrow \ell_1^- \ell_2^-$ is induced by
the product of  two charged  currents
\beq \label{4} {{\cal M}_{lep}^{\mu\nu}} \propto \sum_{m = 1}^{3}
U^{l \nu}_{\ell_1 m} U^{l \nu}_{\ell_2 m}\ (\overline{\ell_1}
\gamma^\mu P_L \nu_m) (\overline{\ell_2} \gamma^\nu P_L \nu_m) +
\sum_{m' = 4}^{3+n} V^{l N}_{\ell_1 m'} V^{l N}_{\ell_2 m'}\
(\overline{\ell_1} \gamma^\mu P_L N_{m'})(\overline{\ell_2}
\gamma^\nu P_L N_{m'}), \eeq
which can be rewritten using charge conjugation as
\beq \label{5} {{\cal M}_{lep}^{\mu\nu}} \propto {\sum_{m = 1}^3}
U^{l \nu}_{\ell_1 m} U^{l \nu}_{\ell_2 m}\ (\overline{\ell_1}
\gamma^\mu P_L \nu_m) (\overline{\nu_m} \gamma^\nu P_R \ell^c_2) +
\sum_{m' = 4}^{3+n} V^{l N}_{\ell_1 m'} V^{l N}_{\ell_2 m'}\
(\overline{\ell_1} \gamma^\mu P_L N_{m'})(\overline{N_{m'}}
\gamma^\nu P_R \ell^c_2). \eeq
The Majorana neutrino fields can be contracted to form a neutrino
propagator, and the transition matrix element is thus given by
\bea \label{6} \nonumber {{\cal M}_{lep}^{\mu\nu}} &=&
\frac{g^2}{2}{\sum_{m = 1}^3} U^{l \nu}_{\ell_1 m}U^{l
\nu}_{\ell_2 m}\ ({\overline{\ell_1}}
 \gamma^\mu P_L )\frac{\qslash + m_{\nu_m}}{q^2 - m_{\nu_m}^2+i\Gamma_{\nu_m}m_{\nu_m}}(\gamma^\nu P_R { \ell^c_2}   )\\
&+& \frac{g^2}{2}{\sum_{m'=4}^{3+n}} V^{l N}_{\ell_1 m'}V^{l
N}_{\ell_2 m'}\ ({\overline{\ell_1}}
 \gamma^\mu P_L )\frac{\qslash + m_{N_{m'}}}{q^2 - m_{N_{m'}}^2+i\Gamma_{N_{m'}}m_{N_{m'}}}(\gamma^\nu P_R { \ell^c_2}   ),
\eea
where $q$ is the momentum exchange carried by the neutrino. The
$\qslash$ term vanishes due to the chirality flip.  Including the
crossed diagram ($\ell_1 \leftrightarrow \ell_2$) the leptonic
amplitude then becomes
\bea \label{9} \nonumber
{{\cal M}_{lep}^{\mu\nu}} &=& \frac{g^2}{2}{\sum_{m=1}^3}U^{l \nu}_{\ell_1 m}U^{l \nu}_{\ell_2 m}\ {m_{\nu_m}}{\overline{u_1}} \Biggl (\frac{\gamma^\mu \gamma^\nu}{q^2 - m^2_{\nu_m}+i\Gamma_{\nu_m}m_{\nu_m}} + \frac{\gamma^\nu \gamma^\mu}{q'^2 - m^2_{\nu_m}+i\Gamma_{\nu_m}m_{\nu_m}} \Biggr ) P_R v_2\\
\nonumber
&+& \frac{g^2}{2}{\sum_{m'=4}^{3+n}} V^{l N}_{\ell_1 m'} V^{l N}_{\ell_2 m'}\ {m_{N_{m'}}} \times\\
& &{\overline{u_1}} \Biggl ( \frac{\gamma^\mu \gamma^\nu}{q^2 -
m^2_{N_{m'}}+i\Gamma_{N_{m'}}m_{N_{m'}}}+\frac{\gamma^\nu
\gamma^\mu}{q'^2 - m^2_{N_{m'}}+i\Gamma_{N_{m'}}m_{N_{m'}}} \Biggr
) P_R v_2  . \eea
For the light Majorana neutrinos, namely, $m = 1,2,3$ the masses
$m_{\nu_m} \sim \cal{O}(\mbox{eV})$ \cite{Seljak} and for the
heavy Majorana neutrinos , the masses $m_{N_{m'}} \sim
\cal{O}(\mev -\gev)$ for the low energy processes we consider. The
heavy Majorana neutrino contribution has a resonant enhancement
when $q^2,q'^2 \approx m^2_{N_{m'}}$ and is the dominant one. The light Majorana neutrino contribution however encounters a severe suppression due to the small neutrino mass like ${m^2_{\nu_m}}/{M^2_W}$. Hence we
can neglect the contributions of the light Majorana neutrinos and
the $\sum_{m = 1}^3$ part of the amplitude drops out.

In principle all the heavy Majorana neutrinos will contribute to
the amplitude but in our analysis we only consider the
contribution of one of the heavy neutrinos, in particular the
lightest one for simplicity. 
So the amplitude can now be written as
\beq \label{10} {{\cal M}_{lep}^{\mu\nu}} = \frac{g^2}{2}V_{\ell_1
4}V_{\ell_2 4}\ {m_4}{ \overline{u_1}} \Biggl (\frac{\gamma^\mu
\gamma^\nu}{q^2 - m_4^2+i\Gamma_{N_4}m_4} + \frac{\gamma^\nu
\gamma^\mu}{q'^2 - m_4^2+i\Gamma_{N_4}m_4}\Biggr ) P_R v_2  . \eeq
We can rewrite the above amplitude as
\bea \nonumber {{\cal M}_{lep}^{\mu\nu}}&=& \frac
{g^2}{2}V_{\ell_1 4}V_{\ell_2 4}\ {m_4}  \frac {\overline{u_1}
\gamma^\mu \gamma^\nu P_R v_2}{q^2 - m_4^2+ i\Gamma_{N_4}m_4}
+ \frac {g^2}{2}V_{\ell_1 4}V_{\ell_2 4}\ {m_4}  \frac {\overline{u_1} \gamma^\nu \gamma^\mu P_R v_2}{q'^2 - m_4^2+i\Gamma_{N_4}m_4}  \\
&=&{\cal M}_1 + {\cal M}_2. \eea
When $q^2 \approx m^2_4$, ${\cal M}_1$ has a resonant contribution
and when $q'^2 \approx m^2_4$, ${\cal M}_2$ has a resonant
contribution. In general, $q \ne q'$, and it is convenient to
split up the individual resonant contributions by the
Single-Diagram-Enhanced multi-channel integration method
\cite{maltoni}. To do this, define the functions
\beq \label{fdefn} f_i = \frac {|{\cal M}_i|^2}{\sum_{i}|{\cal
M}_i|^2}\Bigg {|}\sum_{i}{\cal M}_i\Bigg{|}^2 \eeq
Then the amplitude squared is given by
\beq \Bigg{|}\sum_{i}{\cal M}_i\Bigg{|}^2 = \sum_i f_i \eeq
The amplitude squared splits up into the functions $f_i$ defined
above and the phase space integration can be done for each $f_i$
separately. This helps to make convenient simplifications for the
phase space integration and the computation can be carried out in
parallel. The contributions from each $f_i$ can be added up after
phase space integration.

\section{Decay modes of heavy Majorana neutrino}
\label{decmod}

In this section we will discuss in detail the decay modes of the
heavy Majorana neutrino $N_4$, with mass $m_4$ much smaller than
the mass of the W boson, $m_W$. From EW precision measurements the mixing elements $|V_{\ell 4}|^2 \lsim {\cal O} (10^{-3})$ and the higher order terms in mixing would be very small and can be ignored. Hence the widths are presented only up to leading terms in mixing. The charged current and neutral
current vertices of $N_4$ with the mixing elements are given in
Fig.~\ref{Fig:wlnu} and Fig.~\ref{Fig:znunu}. With increasing mass
of the heavy neutrino new decay channels open up and can be
classified into two body and three body decays. The decay width
scales as the third and the fifth power of the mass($m_4$) for two
and three body decays respectively.

1) $N_4 \rightarrow \ell^- P^+$ where $\ell = e, \mu, \tau$ and
$P^+$ is a charged pseudoscalar meson. This decay mode  has
charged current interactions only as shown in Fig.~\ref{Fig:wlnu}
and the decay width is given by
\bea \label{app2bodyCC} \nonumber
\Gamma^{\ell P} &\equiv& \Gamma(N_4 \rightarrow \ell^- P^+)= \frac {G^2_F}{16 \pi}f^2_P\ |V_{q \bar q'}|^2\ |V_{\ell 4}|^2\ m^3_4 \mbox{ } I_1(\mu_\ell,\mu_P), \\
\nonumber
I_1(x,y)&=& [(1+x-y)(1+x)-4x]\lambda^{\frac {1}{2}}(1, x, y),\\
\lambda(a,b,c)&=& a^2 + b^2 +c^2 -2ab -2 bc -2 ca, \eea
where $f_P$ is the meson decay constant and $V_{q \bar q'}$ are
the CKM matrix elements. $\mu_\ell$ and $\mu_P$ are the masses
scaled by the mass of the heavy neutrino and are given by $\mu_i =
m^2_i/m^2_4$.

2) $N_4 \rightarrow \nu_\ell P^0$ where $\nu_\ell = \nu_e, \nu_\mu, \nu_\tau$  and $P^0$ is a neutral pseudoscalar
meson. This decay mode  has neutral current interactions only as
shown in Fig.~\ref{Fig:znunu} and the
decay width is given by
\beq \label{app2bodyNC} \Gamma^{\nu_\ell P} \equiv
\Gamma(N_4 \rightarrow \nu_\ell P^0)= \frac {G^2_F}{64 \pi}f^2_P\ |V_{\ell 4}|^2\ m^3_4\ (1-\mu_P)^2,
\eeq
where $f_P$ is the meson decay constant, $\mu_P$ is the mass of
the neutral meson scaled by the mass of the heavy neutrino and is
given by $\mu_P = m^2_P/m^2_4$. The mass of the light neutrino
$\sim {\cal O}(\ev)$ \cite{Atre} is much smaller than the
mass of $N_4 \sim {\cal O}(\mev - \gev)$ and can be neglected to a
very good approximation. We have set the mass of the light
neutrino to zero in the expression for
the width above and henceforth.

3) $N_4 \rightarrow \ell^- V^+$ where $\ell = e, \mu, \tau$ and
$V^+$ is a charged vector meson. This decay mode  has charged
current interactions only as shown in Fig.~\ref{Fig:wlnu} and the
decay width is given by
\bea \label{app2bodyCCV} \nonumber
\Gamma^{\ell V} &\equiv& \Gamma(N_4 \rightarrow \ell^- V^+)= \frac {G^2_F}{16 \pi}f^2_V\ |V_{q \bar q'}|^2\  |V_{\ell 4}|^2\ m^3_4 \mbox{ } I_2(\mu_\ell,\mu_V), \\
I_2(x,y)&=& [(1+x-y)(1+x + 2y)-4x]\lambda^{\frac {1}{2}}(1, x, y),
\eea
where $f_V$ is the vector meson decay constant and $V_{q \bar q'}$
are the CKM matrix elements. $\mu_\ell$ and $\mu_V$ are the masses
of the lepton and the vector meson scaled by the mass of the heavy
neutrino and are given by $\mu_i = m^2_i/m^2_4$.

4) $N_4 \rightarrow \nu_\ell V^0$ where $\nu_\ell = \nu_e, \nu_\mu, \nu_\tau$ and $V^0$ is a neutral vector
meson. This decay mode  has neutral current interactions only as
shown in Fig.~\ref{Fig:znunu} and the
decay width is given by
\bea \label{app2bodyNCV} \nonumber
\Gamma^{\nu_\ell V} &\equiv& \Gamma(N_4 \rightarrow \nu_\ell V^0) = {\frac {G^2_F}{2 \pi}} {\kappa^2_V}\ {f^2_V}\ {{|V_{\ell 4}|}^2}\ {m^3_4}\ I_3(\mu_{\nu_\ell},\mu_V),\\
I_3(x,y)&=& (1+2y)(1-y)\lambda^{\frac {1}{2}}(1, x, y), \eea
where $f_V$ is the meson decay constant, $\mu_V$ is the mass of
the neutral meson scaled by the mass of the heavy neutrino and is
given by $\mu_V = m^2_V/m^2_4$. $\kappa_V$ is the vector coupling
associated with the meson and is expressed in terms of $x_w =
\sin^2\theta_w$, where $\theta_w$ is the Weinberg angle. The
values of $\kappa$ for the various vector mesons are: $\kappa =
\frac{1}{3}x_w$ for $\rho^0$ and $\omega$; $\kappa =
(-\frac{1}{4}+ \frac{1}{3}x_w)$ for $K^{*0},\overline{K}^{*0}$ and
$\phi$; and $\kappa = (\frac{1}{4}- \frac{2}{3}x_w)$ for
$D^{*0},\overline{D}^{*0}$ and $J/\psi$.

5) $N_4 \rightarrow \ell^-_1 \ell^+_2 \nu_{\ell_2}$ where $\ell_1, \ell_2
= e, \mu, \tau$ with $\ell_1 \ne \ell_2$. This decay mode has
charged current interactions only as shown in Fig.~\ref{Fig:wlnu} and the decay width is given by
\bea \label{app3bodycc} \nonumber
\Gamma^{\ell_1 \ell_2 \nu_{\ell_2}} &\equiv& \Gamma(N_4 \rightarrow \ell^-_1 \ell^+_2 \nu_{\ell_2}) = \frac{G^2_F}{192 \pi^3} m^5_4\ {|V_{\ell_1 4}|}^2\ I_1 (x_{\ell_1},x_{\nu_{\ell_2}},x_{\ell_2}),\\
I_1(x,y,z) &=& 12 \int\limits_{(x+y)^2}^{(1-z)^2}
\frac{ds}{s}(s-x^2-y^2)(1+z^2-s)\lambda^{\frac{1}{2}}(s,x^2,y^2)
\lambda^{\frac{1}{2}}(1,s,z^2), \eea
where $I_1(0,0,0)=1$, $x_i$ are the masses scaled by the mass of
the heavy neutrino and are given by $x_i = m_i/m_4$. The mass of
the light neutrino $\sim {\cal O}(\ev)$ is much smaller than
the  mass of $N_4 \sim {\cal O}(\mev - \gev)$ and hence can be
neglected compared to the mass of $N_4$. We have set the mass of the light neutrino to zero with very good approximation in the expression for the width above and henceforth.

6) $N_4 \rightarrow \nu_{\ell_1} \ell^-_2 \ell^+_2 $ where $\ell_1, \ell_2 = e, \mu, \tau$. Both charged current and
neutral current interactions as shown in Fig.~\ref{Fig:wlnu} and
Fig.~\ref{Fig:znunu} are relevant for this mode and the decay width is given by
\bea
\label{app3bodyNC1}
\nonumber
\Gamma^{\nu_{\ell_1} \ell_2 \ell_2}&\equiv& \Gamma(N_4 \rightarrow \nu_{\ell_1} \ell^-_2 \ell^+_2 )= \frac {G^2_F}{96 \pi^3}{|V_{\ell_1 4}|}^2\ m^5_4 \times \biggl [ \Bigl ( g^\ell_L g^\ell_R +  \delta_{\ell_1 \ell_2} g^\ell_R \Bigr ) I_2(x_{\nu_{\ell_1}}, x_{\ell_2},  x_{\ell_2}) \\
&+& \Bigl ( {(g^{\ell}_L)}^2 + {(g^{\ell}_R)}^2  + \delta_{\ell_1 \ell_2} (1 + 2g^\ell_L ) \Bigr )I_1(x_{\nu_{\ell_1}}, x_{\ell_2},  x_{\ell_2}) \biggr ] ,\\
I_2(x,y,z) &=& 24yz \int\limits_{(y+z)^2}^{(1-x)^2}
\frac{ds}{s}(1+x^2-s)\lambda^{\frac{1}{2}}(s,y^2,z^2)\lambda^{\frac{1}{2}}(1,s,x^2),
\eea
where $I_2(0,0,0) = 1$, $I_1(x,y,z)$ has been defined in
Eq.~(\ref{app3bodycc}), $x_i$ are the masses scaled by the mass of
the heavy neutrino and are given by $x_i = m_i/m_4$,
$g^\ell_L = -\frac{1}{2} + x_w$, $g^\ell_R = x_w$ and $x_w = \sin^2\theta_w = 0.231$, where $\theta_w$
is the Weinberg angle.

7) $N_4 \rightarrow \nu_{\ell_1} \nu \overline{\nu} $
where $\nu_{\ell_1} = \nu_e, \nu_\mu, \nu_\tau$. This decay mode  has neutral current
interactions only as shown in Fig.~\ref{Fig:znunu}. Using the
massless approximation for the neutrinos as
described above
the decay width has a simple form given by
\beq \label{app3bodyNC2} \Gamma^{\nu_{\ell_1} \nu \nu} \equiv
\sum_{\ell_2=e}^\tau \Gamma(N_4 \rightarrow \nu_{\ell_1} \nu_{\ell_2}
\overline{\nu_{\ell_2}} )= \frac {G^2_F}{96\pi^3}|V_{\ell_1 4}|^2\ m^5_4. \eeq

All the decay modes listed above contribute to the total decay
width of the heavy Majorana neutrino which is given by:
\bea \label{apptotwid} \nonumber
\Gamma_{N_4}& =& \sum_{\ell, P}{\Gamma^{\nu_\ell P}} + \sum_{\ell, V} {\Gamma^{\nu_\ell V}} + \sum_{\ell,P} {2 \Gamma^{\ell P}} +  \sum_{\ell,V} {2 \Gamma^{\ell V}} \\
&+&  \sum_{\ell_1,\ell_2(\ell_1 \ne \ell_2)}{2\Gamma^{\ell_1
\ell_2 \nu_{\ell_2}}} + \sum_{\ell_1, \ell_2} {\Gamma^{\nu_{\ell_1} \ell_2 \ell_2}} +  \sum_{\nu_{\ell_1}}
{\Gamma^{\nu_{\ell_1} \nu \nu}},
 \eea
where $\ell, \ell_1, \ell_2 = e, \mu, \tau$. For a
Majorana neutrino, the $\Delta L = 0$ process $N_4 \rightarrow
\ell^- P^+$ as well as its charge conjugate $|\Delta L| = 2$
process $N_4 \rightarrow \ell^+ P^-$ are possible and have the
same width, $\Gamma^{\ell P}$. Hence the factor of 2 associated
with the decay width of this mode in Eq.~(\ref{apptotwid}).
Similarly, the $\Delta L = 0$ and its charge conjugate $|\Delta L|
= 2$ process are possible for the decay modes $N_4 \rightarrow
\ell^- V^+$ and $N_4 \rightarrow \ell^-_1 \ell^+_2 \nu_{\ell_2}$ and
hence have a factor of 2 associated with their width in
Eq.~(\ref{apptotwid}).

As mentioned earlier, new channels open with increasing mass of
the heavy neutrino. For the low energy $\lv$ tau decays and rare
meson decays we consider, the mass of the heavy neutrino is in the
range
 $140\  \mev \lsim m_4 \lsim 5278\  \mev$. For this mass range we list all the possible decay channels for $N_4$ in Table \ref{newchannel}. The mass and decay constants of pseudoscalar and vector mesons used in the calculation of partial widths given in Eqs.~(\ref{app2bodyCC} -\ref{app3bodyNC2}) are listed in Table \ref{constants} in Appendix \ref{apprmd}.

\bet[tb] \caption{Decay modes of heavy Majorana neutrino based on
its mass $m_4$.} \label{newchannel}
\begin{center}
\begin{tabular}{|l|l|   |l|l|}
\hline \hline
Mass of heavy  & Decay mode of & Mass of heavy  & Decay mode of \\
neutrino ($\mev$)  & heavy neutrino & neutrino ($\mev$) & heavy neutrino \\
\hline \hline
$\gsim \sum_{m}\nu_m = 10^{-6}$ & $N_4 \rightarrow \nu_{\ell_1} \nu_{\ell_2} \overline{\nu_{\ell_2}} $  &$ > m_\mu+m_\tau = 1880  $ & $N_4 \rightarrow  \mu^- \tau^+ \nu_\tau + c.c $     \\
 &   & & $N_4 \rightarrow  \tau^- \mu^+ \nu_\mu + c.c $     \\
\hline
$ > 2m_e = 1.02  $ & $N_4 \rightarrow \nu_\ell e^- e^+ $  & $>m_\tau + m_{\pi} = 1920  $  & $N_4 \rightarrow \tau^- \pi^+ + c.c$   \\
\hline
$ > m_e +m_\mu = 106  $ & $N_4 \rightarrow e^- \mu^+ \nu_m + c.c$  & $> m_e + m_{D_s}= 1970  $ & $N_4 \rightarrow e^- D^+_s + c.c$  \\
 & $N_4 \rightarrow \mu^- e^+ \nu_e + c.c$  &  &   \\
\hline
$ > m_{\pi^0} = 135  $ & $N_4 \rightarrow \nu_\ell \pi^0$  &$> m_\mu + m_{D}= 1980  $  & $N_4 \rightarrow \mu^- D^+ + c.c$   \\
\hline
$ > m_e + m_{\pi} = 140  $ & $N_4 \rightarrow e^- \pi^+ +c.c $  &$ > m_{D^{*0}} = 2010  $  &  $N_4 \rightarrow \nu_\ell D^{*0} $     \\
\hline
$ > 2m_\mu = 211  $ & $N_4 \rightarrow \nu_\ell \mu^- \mu^+ $  & $ > m_{\overline{D}^{*0}} = 2010  $  &  $N_4 \rightarrow \nu_\ell \overline{D}^{*0} $    \\
\hline
$ > m_\mu + m_{\pi} = 245  $ & $N_4 \rightarrow \mu^- \pi^+ +c.c$  & $> m_e + m_{D^*}= 2010  $ & $N_4 \rightarrow e^- D^{*^+} + c.c$       \\
\hline
$ > m_e + m_{K} = 494  $ & $N_4 \rightarrow e^- K^+ +c.c$  &$> m_\mu + m_{D_s}= 2070  $  & $N_4 \rightarrow \mu^- D^+_s + c.c$     \\
\hline
$ > m_{\eta} = 548  $ & $N_4 \rightarrow \nu_\ell \eta $  &$> m_e + m_{D^*_s}= 2110  $ & $N_4 \rightarrow e^- D^{*+}_s + c.c$      \\
\hline
$ > m_\mu + m_K = 599  $ & $N_4 \rightarrow \mu^- K^+ + c.c$  & $> m_\mu + m_{D^*}= 2120  $  & $N_4 \rightarrow \mu^- D^{*+} + c.c$           \\
\hline
$ > m_{\rho^0} = 776  $ & $N_4 \rightarrow \nu_\ell \rho^0 $  & $> m_\mu + m_{D^*_s}= 2220  $  & $N_4 \rightarrow \mu^- D^{*+}_s + c.c$   \\
\hline
$ > m_e + m_{\rho} = 776  $ & $N_4 \rightarrow e^- \rho^+ +c.c$  &$> m_\tau + m_K= 2270  $  & $N_4 \rightarrow \tau^- K^+ + c.c$    \\
\hline
$ > m_{\omega} = 783  $ & $N_4 \rightarrow \nu_\ell \omega$  & $> m_\tau + m_{\rho}= 2550  $ & $N_4 \rightarrow \tau^- \rho^+ + c.c$     \\
\hline
$ > m_\mu + m_{\rho} = 882  $ & $N_4 \rightarrow \mu^- \rho^+ +c.c$  & $> m_\tau + m_K^*= 2670  $  & $N_4 \rightarrow \tau^- K^{*+} + c.c$     \\
\hline
$ > m_e + m_{K^*} = 892  $ & $N_4 \rightarrow e^- K^{*+} +c.c$  & $> m_{\eta_c}= 2980  $ & $N_4 \rightarrow \nu_\ell \eta_c $   \\
\hline
$ > m_{K^{*0}} = 896  $ & $N_4 \rightarrow \nu_\ell K^{*0} $  &$> m_{J/\psi} = 3100  $ & $N_4 \rightarrow \nu_\ell J/\psi $     \\
\hline
$ > m_{\overline{K}^{*0}} = 896  $ & $N_4 \rightarrow \nu_\ell \overline{K}^{*0} $  & $> 2 m_\tau = 3550  $  & $N_4 \rightarrow \nu_\ell \tau^- \tau^+ $     \\
\hline
$ > m_{\eta'} = 958  $ & $N_4 \rightarrow \nu_\ell \eta' $  & $> m_\tau + m_D= 3650  $  & $N_4 \rightarrow \tau^- D^+ + c.c$    \\
\hline
$ > m_\mu + m_{K^*} = 997  $ & $N_4 \rightarrow \mu^- K^{*+} +c.c$  &  $> m_\tau + m_{D_s} = 3750  $  & $N_4 \rightarrow \tau^- D^+_s + c.c$    \\
\hline
$ > m_{\phi} = 1019  $ & $N_4 \rightarrow \nu_\ell \phi$  & $> m_\tau + m_{D^*} = 3790  $  & $N_4 \rightarrow \tau^- D^{*+} + c.c$    \\
\hline
$ > m_e+m_\tau = 1780  $ & $N_4 \rightarrow  e^- \tau^+ \nu_\tau+ c.c $  & $> m_\tau + m_{D^*_s} = 3890  $  & $N_4 \rightarrow \tau^- D^{*+}_s + c.c$     \\
 & $N_4 \rightarrow \tau^- e^+ \nu_e+c.c $  &   &    \\
\hline
$ > m_e + m_{D} = 1870  $ & $N_4 \rightarrow e^- D^+ + c.c$  & &   \\
\hline \hline
\end{tabular}
\end{center}
\eet

\section{ Lepton-number violating tau decay}
\label{apptd}

The decay amplitude for lepton number violating tau decays can be
separated into leptonic and hadronic parts,
\beq \label{13} {i\cal M} = {({\cal M}_{lep})_{\mu\nu}}{({\cal
M}_{had})^{\mu\nu}}. \eeq
For the tree level amplitude, the hadronic part can be expressed
in terms of the decay constants of the mesons in a model
independent way. The box diagram includes hadronic matrix elements
which cannot be simplified in terms of decay constants and needs
to be evaluated in a model dependent way. We expect the tree level
amplitude to dominate and do not include the box diagram. It has
been argued that in certain cases for rare meson decays
sub-leading contributions may be appreciable \cite{Ivanov,
AliBorisov}. Even in such a scenario the difference will not be
important at the current level of sensitivities and we include the
more conservative limit from tree level diagrams only. The tau
decays and the rare meson decays are crossed versions of each
other and the above arguments are true for both.

The leptonic part of the subprocess $\tau^- \rightarrow \ell^+
W^{-*} W^{-*}$ is obtained by crossing the amplitude in (\ref{10})
\beq \label{20} {{\cal M}_{lep}^{\mu\nu}} = \frac{g^2}{2}V^*_{\tau
4}V^*_{\ell 4}\ { \overline{v_\tau}}\frac{m_4}{q^2 - m_4^2
+i\Gamma_{N_4}m_4} \gamma^\mu \gamma^\nu P_R v_\ell   . \eeq
Combining the hadronic and leptonic parts, the decay amplitude for
\beq \tau^-(p_1) \rightarrow \ell^+(p_2)\  M_1^-(q_1)\  M_2^-(q_2)
\eeq
is given by
\bea {i\cal M}& = &{({\cal M}_{lep})_{\mu\nu}}{{\cal
M}_{M_1}^{\mu}}{{\cal M}_{M_2}^{\nu}}
+ (M_1 \leftrightarrow M_2)\\
\nonumber
 & = & 2G_F^2 V^{CKM}_{M_1} V^{CKM}_{M_2} f_{M_1} f_{M_2} {V_{\tau 4}^*}{V^*_{\ell 4}}\   m_4
 \Biggl [\frac {\overline{v_\tau} \qslash_1 \qslash_2 P_R v_\ell}{(p_1 - q_1)^2 - m_4^2 +i\Gamma_{N_4}m_4}\Biggr ]\\
\label{23}
&+& 2G_F^2 V^{CKM}_{M_1} V^{CKM}_{M_2} f_{M_1} f_{M_2} {V_{\tau 4}^*}{V^*_{\ell 4}}\   m_4 \Biggl [\frac {\overline{v_\tau} \not q_2 \not q_1 P_R v_\ell}{(p_1 - q_2)^2 - m_4^2 +i\Gamma_{N_4}m_4}\Biggr ],\\
&=& {\cal M}_1 + {\cal M}_2, \eea
where $V^{CKM}_{M_i}$ are the quark flavor-mixing matrix elements
for the mesons and $f_{M_i}$ are meson decay constants. Then the
functions, $f_1$ and $f_2$ defined in Eq.~(\ref{fdefn}) are given
by
\bea \label{f1_tau}
f_1 &=& \Biggl (\frac {F_\tau A}{a^2_1 + b^2}\Biggr )\Biggl [ \frac {(a^2_2 + b^2)A + (a_1 a_2 +b^2)C}{(a^2_2 + b^2)A + (a^2_1 + b^2)B} + (q_1\leftrightarrow q_2)\Biggr ],\\
\label{f2_tau}
f_2 &=& f_1(q_1 \leftrightarrow q_2),\\
\nonumber A(p_i,q_j) &=& 8(p_1\cdot q_1) (p_2\cdot q_2) (q_1\cdot
q_2) -
      4m^2_{M_1} (p_1\cdot q_2) (p_2\cdot q_2)\\
\label{eq:A}
      & - & 4m^2_{M_2} (p_1\cdot q_1) (p_2\cdot q_1) + 2m^2_{M_1} m^2_{M_2} (p_1\cdot p_2),\\
\label{eq:B}
  B(p_i,q_j) &=& A(q_1 \leftrightarrow q_2),\\
            C(p_i,q_j) &=& 4(p_1 \cdot p_2) (q_1\cdot q_2)^2 -  A(p_i,q_j) ,\\
\label{eq:C}
D(p_i,q_j) &=& C(q_1 \leftrightarrow q_2) ,\\
\label{eq:D}
F_\tau &=& 4 G^4_F f_{M_1}^2 f_{M_2}^2 |V^{CKM}_{M_1} V^{CKM}_{M_2}|^2 {|V_{\tau 4}V_{\ell 4}|}^2 m^2_4,\\
 a_{1,2}(p_i,q_j)& =& (p_1-q_{1,2})^2 - m^2_4;\ \ \ b = \Gamma_{N_4} m_4.
\eea
The decay width for the $\lv$ tau decay is then given by
\bea \label{LVPtauRate} \Gamma^{\tau}_{\lv} &=&  (1 - {1\over 2}
\delta_{M_1M_2}) \frac{1}{128\pi^5 m_\tau}
\Biggl [\int f_1 dPS_{31}+ \int f_2 dPS_{32}\Biggr ]  ,\\
dPS_{31}&=&\frac{\pi^2}{4 m^2_\tau}\lambda^{\frac{1}{2}}(m^2_\tau,m^2_{M_1},m^2_{c1})\lambda^{\frac{1}{2}}(m^2_{c1},m^2_\ell,m^2_{M_2})\frac {dm^2_{c1}}{m^2_{c1}} dy_1 dy_2 dy_3 dy_4,\\
dPS_{32} &=& dPS_{31}(q_1 \leftrightarrow q_2), \eea
where $dPS_{31}$ and $dPS_{32}$ are the  phase space factors
obtained by conveniently clustering two different sets of
particles to enable applying the narrow-width approximation
easily. $y_1$ to $y_4$ are rescaled angular variables with
integration limits $0 \le y_i \le1$. As seen in Sec.~\ref{smlm4},
the width of the heavy neutrino is very small compared to the mass
and hence we can apply the narrow-width approximation.
\beq \label{narwid_tau} \int \frac {dm^2_{c_i}} {(m^2_{c_i} -
m^2_4)^2 + \Gamma^2_{N_4}m^2_4} \Bigg{|}_
{{\Gamma_{N_4}\rightarrow\ 0}} = \int\delta (m^2_{c_i} -
m^2_4)dm^2_{c_i}\frac {\pi}{\Gamma_{N_4} m_4} \eeq
Applying the narrow-width approximation as described above and
integrating over the $\delta-$function we get
\bea \label{fidPS3i_tau} \nonumber
\int f_1 dPS_{31}&=&\int \Biggl (\frac{F_\tau A \pi^3}{4 m^2_\tau m^3_4 \Gamma_{N_4}}\Biggl[\frac {(a^2_2 + b^2)A + b^2(B+C+D)}{(a^2_2 + b^2)A + b^2 B}\Biggr ]\\
& &\lambda^{\frac{1}{2}}(m^2_\tau,m^2_{M_1},m^2_4)\lambda^{\frac{1}{2}}(m^2_4,m^2_\ell,m^2_{M_2})\Biggr ) dy_1 dy_2 dy_3 dy_4,\\
\int f_2 dPS_{32}&=&\int f_1 dPS_{31}(q_1 \leftrightarrow q_2)
\eea
Now, we can find the decay rate from Eq.~(\ref{LVPtauRate}).
Normalized to the $\tau$ decay width $\Gamma_\tau = G_F^2
m_\tau^5/192\pi^3$, the corresponding branching fraction is
$\mathrm{Br} = \Gamma^{\tau}_{\!\!\! \mbox{}_{\lv}} /
\Gamma_\tau$. The masses and decay constants of mesons are listed
in Table \ref{constants}. The CKM matrix elements and $\tau$ mass
are taken from the Particle Data Group (PDG) \cite{PDG}:
$$m_\tau = 1777\ {\mev},\ |V_{ud}| = 0.9738,\ |V_{us}| = 0.2200.$$

\section{ Rare meson decay}
\label{apprmd}

The rare meson decays
$$ M_1^+(q_1) \rightarrow \ell^+(p_1)\  \ell^+(p_2)\  M_2^-(q_2)$$
have the same Feynman diagrams as tau decay. The meson $M_2$ can
be a pseudoscalar or vector meson. The decay amplitude when $M_2$
is a pseudoscalar meson is given by
\bea \nonumber
i{\cal M}^P & = &  2G_F^2 V^{CKM}_{M_1} V^{CKM}_{M_2} f_{M_1} f_{M_2}{V_{\ell_1 4}}{V_{\ell_2 4}}\ m_4 \Biggl [ \frac {\overline{u_{\ell_1}} \qslash_1 \qslash_2 P_R v_{\ell_2}}{(q_1-p_1)^2-{m_4}^2 + i\Gamma_{N_4} m_4}\Biggr ] \\
\nonumber
 &+& 2G_F^2 V^{CKM}_{M_1} V^{CKM}_{M_2} f_{M_1} f_{M_2}{V_{\ell_1 4}}{V_{\ell_2 4}}\ m_4 \Biggl [\frac {\overline{u_{\ell_1}} \not q_2 \not q_1 P_R v_{\ell_2}}{(q_1-p_2)^2-{m_4}^2 + i\Gamma_{N_4}m_4} \Biggr ]\\
&=& {\cal M}^P_1 + {\cal M}^P_2. \eea
Next we consider the case where $M_2$ is a vector meson. The decay
amplitude is given by
\bea \nonumber
i{\cal M}^V & = &  2G_F^2 V^{CKM}_{M_1} V^{CKM}_{M_2} f_{M_1} f_{M_2}{V_{\ell_1 4}}{V_{\ell_2 4}}\ m_4\ m_{M_2} \Biggl [ \frac {\overline{u_{\ell_1}} \qslash_1 \not \epsilon^\lambda (q_2) P_R v_{\ell_2}}{(q_1-p_1)^2-{m_4}^2 + i\Gamma_{N_4} m_4}\Biggr ] \\
\nonumber
 &+& 2G_F^2 V^{CKM}_{M_1} V^{CKM}_{M_2} f_{M_1} f_{M_2}{V_{\ell_1 4}}{V_{\ell_2 4}}\ m_4\ m_{M_2}\Biggl [\frac {\overline{u_{\ell_1}} \not \epsilon^\lambda (q_2) \not q_1 P_R v_{\ell_2}}{(q_1-p_2)^2-{m_4}^2 + i\Gamma_{N_4}m_4} \Biggr ]\\
&=& {\cal M}^V_1 + {\cal M}^V_2. \eea

Similar to tau decay, we define the functions $f^P_1, f^P_2$ and
$f^V_1, f^V_2 $ for the pseudoscalar and vector mesons
respectively as given in Eq.~(\ref{fdefn}). $f^P_i$ and $f^V_i$
turn out to have the same form and are given below.
\bea \label{f1_rmd}
f^{P(V)}_1 &=& \Biggl (\frac {F_M A^{P(V)}}{a^2_1 + b^2}\Biggr )\Biggl [ \frac {(a^2_2 + b^2)A^{P(V)} + (a_1 a_2 +b^2)C^{P(V)}}{(a^2_2 + b^2)A^{P(V)} + (a^2_1 + b^2)B^{P(V)}} + (p_1\leftrightarrow p_2)\Biggr ],\\
\label{f2_rmd}
f^{P(V)}_2 &=& f^{P(V)}_1(p_1 \leftrightarrow p_2),\\
\nonumber A^P(p_i,q_j) &=& 8(p_1\cdot q_1) (p_2\cdot q_2)
(q_1\cdot q_2) -
      4m^2_{M_1} (p_1\cdot q_2) (p_2\cdot q_2)\\
\label{eq:AP}
      & - & 4m^2_{M_2} (p_1\cdot q_1) (p_2\cdot q_1) + 2m^2_{M_1} m^2_{M_2} (p_1\cdot p_2),\\
\nonumber A^V(p_i,q_j) &=& 8(p_1\cdot q_1) (p_2\cdot q_2)
(q_1\cdot q_2) -
      4m^2_{M_1} (p_1\cdot q_2) (p_2\cdot q_2)\\
\label{eq:AV}
      &+ & 4m^2_{M_2} (p_1\cdot q_1) (p_2\cdot q_1) - 2m^2_{M_1} m^2_{M_2} (p_1\cdot p_2),\\
 \label{eq:BV}
  B^{P(V)}(p_i,q_j) &=& A^{P(V)}(p_1 \leftrightarrow p_2),\\
      C^P(p_i,q_j) &=& 4(p_1 \cdot p_2) (q_1\cdot q_2)^2 -  A^P(p_i,q_j) ,\\
\label{eq:CP}
            C^V(p_i,q_j) &=& 4(p_1 \cdot p_2) (q_1\cdot q_2)^2 - 4 m^2_{M_1} m^2_{M_2}(p_1 \cdot p_2)-  A^V(p_i,q_j) ,\\
\label{eq:CV}
D^{P(V)}(p_i,q_j) &=& C^{P(V)}(p_1 \leftrightarrow p_2) ,\\
\label{eq:DV}
F_M &=& 4 G^4_F f_{M_1}^2 f_{M_2}^2 |V^{CKM}_{M_1} V^{CKM}_{M_2}|^2 {|V_{\ell_1 4}V_{\ell_2 4}|}^2 m^2_4,\\
 a_{1,2}(p_i,q_j)& =& (q_1-p_{1,2})^2 - m^2_4;\ \ \  b = \Gamma_{N_4} m_4.
\eea

The decay rate for $\lv$ rare meson decay is then given by
\bea \label{rmdRate} \Gamma^{M_1}_{\lv} &=&  (1 - {1\over 2}
\delta_{\ell_1 \ell_2}) \frac{1}{64\pi^5 m_{M_1}}
\Biggl [\int f^{P(V)}_1 dPS_{31}+ \int f^{P(V)}_2 dPS_{32}\Biggr ]  ,\\
dPS_{31}&=&\frac{\pi^2}{4 m^2_{M_1}}\lambda^{\frac{1}{2}}(m^2_{M_1},m^2_{\ell_1},m^2_{c1})\lambda^{\frac{1}{2}}(m^2_{c1},m^2_{\ell_2},m^2_{M_2})\frac {dm^2_{c1}}{m^2_{c1}} dy_1 dy_2 dy_3 dy_4,\\
dPS_{32} &=& dPS_{31}(p_1 \leftrightarrow p_2), \eea
where $dPS_{31}$ and $dPS_{32}$ are the  phase space factors
obtained by conveniently clustering two different sets of
particles to enable applying the narrow-width approximation
easily. $y_1$ to $y_4$ are rescaled angular variables with
integration limits $0 \le y_i \le1$. The width of the heavy
neutrino is very small compared to the mass and hence we can apply
the narrow-width approximation.
\beq \label{narwid_rmd} \int \frac {dm^2_{c_i}} {(m^2_{c_i} -
m^2_4)^2 + \Gamma^2_{N_4}m^2_4}\Bigg{|}_
{{\Gamma_{N_4}\rightarrow\ 0}} = \int\delta (m^2_{c_i} -
m^2_4)dm^2_{c_i}\frac {\pi}{\Gamma_{N_4} m_4} \eeq
Applying the narrow-width approximation as described above and
integrating over the $\delta-$function we get
\bea \label{fidPS3i_rmd}
\nonumber
\int f^{P(V)}_1 dPS_{31}&=&\int \Biggl (\frac{F_M A^{P(V)} \pi^3}{4 m^2_{M1} m^3_4 \Gamma_{N_4}}\Biggl[\frac {(a^2_2 + b^2)A^{P(V)} + b^2(B^{P(V)}+C^{P(V)}+D^{P(V)})}{(a^2_2 + b^2)A^{P(V)} + b^2 B^{P(V)}}\Biggr ]\\
& &\lambda^{\frac{1}{2}}(m^2_{M_1},m^2_{\ell_1},m^2_4)\lambda^{\frac{1}{2}}(m^2_4,m^2_{\ell_2},m^2_{M_2})\Biggr ) dy_1 dy_2 dy_3 dy_4,\\
\int f^{P(V)}_2 dPS_{32}&=& \int f^{P(V)}_1 dPS_{31}(p_1
\leftrightarrow p_2),
\eea
Now, we can find the decay rate from Eq.~(\ref{rmdRate}). The
branching fraction is then given by $\mathrm{Br} = \tau^{}_{M_1}
\Gamma^{M_1}_{\!\!\! \mbox{}_{\lv}}$.

The CKM matrix elements and the lifetimes of mesons used in our
calculations are taken from PDG \cite{PDG} and are listed below.
\bea \nonumber
&& |V_{ub}| = 0.00367,\ |V_{cd}| = 0.224,\  |V_{cs}| = 0.996;\\
\nonumber && \tau_K = 1.2384\times 10^{-8}\ {\s},\ \tau_D =
1.040\times 10^{-12}\ {\s},\ \tau_{D_s} = 4.9\times 10^{-13}\
{\s},\  \tau_B = 1.671\times 10^{-12}\ {\s}. \eea
The mass and decay constants of pseudoscalar and vector mesons
used in our calculations are listed in Table \ref{constants}.
\bet[tb] \caption{Mass and decay constants of pseudoscalar and
vector mesons used. } \label{constants}
\begin{center}
\begin{tabular}{|c|c|c|c|c|c|}
\hline
Pseudoscalar & Mass  & Decay Constant & Vector  & Mass  & Decay Constant    \\
Meson&$(\mev)$\cite{PDG} & $(\mev)$ \cite{PDG}&Meson&$(\mev)$\cite{PDG}&$(\mev)$\cite{ebert}\\
\hline
$\pi^\pm$ & 139.6 & 130.7 & $\rho^{\pm}$ & 775.8 & 220 \\
\hline
$K^\pm$ & 493.7 & 159.8 & $K^{*\pm}$ & 891.66 & 217 \\
\hline
$D^\pm$ & 1869.4 & 222.6 \cite{artuso} & $D^*$ & 2010 & 310 \\
\hline
$D^\pm_s$ & 1968.3 & 266 & $D^{*\pm}_s$ & 2112.1 & 315 \\
\hline
$B^\pm$ & 5279  & 190 \cite{MILC} & $\omega$ & 782.59 & 195\\
\hline
$\pi^0$ & 135 & 130 & $K^{*0}, \overline{K}^{*0}$ & 896.10 & 217\\
\hline
$\eta$ & 547.8 & 164.7 \cite{feldmann} & $\phi$ & 1019.456 & 229\\
\hline
$\eta'$ & 957.8 & 152.9 \cite{feldmann} & $D^{*0}, \overline{D}^{*0}$  & 2006.7 & 310 \\
\hline
$\eta_c$ & 2979.6 & 335.0 \cite{edwards} & $J/\psi$ & 3096.916 & 459 \cite{wang}\\
\hline
\end{tabular}
\end{center}
\eet


\vspace{1cm}

\def\jnl#1#2#3#4{{#1}{\bf #2} (#4) #3}

\def\Zphys{{\em Z.\ Phys.} }
\def\jssc{{\em J.\ Solid State Chem.\ }}
\def\jpsJ{{\em J.\ Phys.\ Soc.\ Japan }}
\def\ptps{{\em Prog.\ Theoret.\ Phys.\ Suppl.\ }}
\def\PTP{{\em Prog.\ JMKZset.\ Phys.\  }}

\def\JMP{{\em J. Math.\ Phys.} }
\def\NPB{{\em Nucl.\ Phys.} B}
\def\NP{{\em Nucl.\ Phys.} }
\def\PLB{{\em Phys.\ Lett.} B}
\def\PL{{\em Phys.\ Lett.} }
\def\PRL{\em Phys.\ Rev.\ Lett. }
\def\PRB{{\em Phys.\ Rev.} B}
\def\PRD{{\em Phys.\ Rev.} D}
\def\PRe{{\em Phys.\ Rep.} }
\def\AP{{\em Ann.\ Phys.\ (N.Y.)} }
\def\RMP{{\em Rev.\ Mod.\ Phys.} }
\def\ZPC{{\em Z.\ Phys.} C}
\def\SCI{\em Science}
\def\CMP{\em Comm.\ Math.\ Phys. }
\def\MPLA{{\em Mod.\ Phys.\ Lett.} A}
\def\IJMPA{{\em Int.\ J.\ Mod.\ Phys.} A}
\def\IJMPB{{\em Int.\ J.\ Mod.\ Phys.} B}
\def\EPJC{{\em Eur.\ Phys.\ J.} C}
\def\PR{{\em Phys.\ Rev.} }
\def\JHEP{{\em JHEP} }
\def\cmp{{\em Com.\ Math.\ Phys.}}
\def\JPA{{\em J.\  Phys.} A}
\def\CQG{\em Class.\ Quant.\ Grav. }
\def\ATMP{{\em Adv.\ Theoret.\ Math.\ Phys.} }
\def\ibid{{\em ibid.} }

\def\journal#1#2#3#4{{ #1} {\bf #2}, #3 (#4)}
\def\epjc{Eur. Phys. J. C.}
\def\prl{Phys. Rev. Lett.}
\def\plb{Phys. Lett. B}
\def\npb{Nucl. Phys. B}
\def\npps{Nucl. Phys. B (Proc. Suppl.)}
\def\ptp{Prog. Theor. Phys.}
\def\ptps{Prog. Theor. Phys. Suppl.}
\def\mpl{Mod. Phys. Lett.}
\def\zp{Z. Phys.}
\def\prd{Phys. Rev. D}
\def\prp{Phys. Rep.}
\def\nc{Nuovo Cim.}
\def\jhep{JHEP}
\def\yf{Yad. Fiz.}
\def\tmf{Teo. Mat. Fiz.}
\def\jetp{JETP Lett.}
\def\ijmp{Int. Jour. Mod. Phys.}
\def\jp{J. Phys.}
\def\aj{Astrohys. J}


\newpage

\begin{thebibliography}{99}
\small \baselineskip=14pt



\bibitem{BargerReview}
For earlier comprehensive discussions of neutrino physics see {\it
e.g.}, {\it Massive Neutrinos in Physics and Astrophysics} by R.
N. Mohapatra and P. B. Pal (World Scientific 2004); {\it Physics
of Neutrinos and Applications to Astrophysics} by M. Fukugita and
T. Yanagida (Springer-Verlag 2003). For recent reviews, see {\it
e.g.}, V.~Barger, D.~Marfatia, and K.~Whisnant,
Int.~J.~Mod.~Phys.~{\bf E12}, 569 (2003); B.~Kayser, p.~145 in PDG
in Phys.~Lett.~{\bf B592}, 1 (2004); M.~C.~Gonzalez-Garcia and
M.~Maltoni, {\tt arXiv:0704.1800 [hep-ph]}; R.~N.~Mohapatra and
A.~Y.~Smirnov, Ann.\ Rev.\ Nucl.\ Part.\ Sci.\  {\bf 56} (2006)
569; A.~Strumia and F.~Vissani, {\tt arXiv:hep-ph/0606054}.

\bibitem{seesaw}
P. Minkowski, Phys. Lett. {\bf B67}, 421 (1977); T. Yanagida, in
{\it Proc. of the Workshop on Grand Unified Theory and Baryon
Number of the Universe}, KEK, Japan, 1979; M. Gell-Mann, P. Ramond
and R. Slansky in {\it Sanibel Symposium}, February 1979,
CALT-68-709 [{\tt retroprint arXiv:hep-ph/9809459}],
 and in {\it Supergravity}, eds. D. Freedman {\it et al}. (North Holland, Amsterdam, 1979);
 S.~L. Glashow in {\it Quarks and Leptons, Cargese,} eds. M. Levy {\it et al.} (Plenum, 1980, New York), p. 707;
 R. N. Mohapatra and G. Senjanovic,  Phys. Rev. Lett. {\bf 44}, 912 (1980).

\bibitem{LRModels}
J. C. Pati and A. Salam, Phys. Rev. {\bf D10}, 275 (1974); R. N.
Mohapatra and J. C. Pati, Phys. Rev. {\bf D11}, 566, 2558 (1975);
G. Senjanovic and R. N. Mohapatra, Phys. Rev. {\bf D12}, 1502
(1975).

\bibitem{SO10SUSYGUT}
J. A. Harvey, P. Ramond and D. B. Reiss, Nucl. Phys. {\bf B199},
223 (1982); S. Dimopoulos, L. J. Hall and S. Raby, Phys. Rev.
Lett. {\bf 68}, 1984 (1992); L. J. Hall and S. Raby, Phys. Rev.
{\bf D51}, 6524 (1995) [{\tt arXiv:hep-ph/9501298}].

\bibitem{MGUT}
I.~Dorsner and P.~Fileviez~P\'erez, Nucl.\ Phys.\  B {\bf 723}
(2005) 53 [{\tt arXiv:hep-ph/0504276}]; see also: I.~Dorsner,
P.~Fileviez~P\'erez and R.~Gonzalez Felipe, Nucl.\ Phys.\  B {\bf
747} (2006) 312 [{\tt arXiv:hep-ph/0512068}]; P.~Fileviez~P\'erez,
AIP Conf.\ Proc.\  {\bf 903} (2006) 385 [{\tt
arXiv:hep-ph/0606279}]; I.~Dorsner, P.~Fileviez~P\'erez and
G.~Rodrigo, Phys.\ Rev.\  D {\bf 75} (2007) 125007 [{\tt
arXiv:hep-ph/0607208}].

\bibitem{ZeeModel}
A. Zee, Phys. Lett. {\bf B93}, 389 (1980) [{\it Erratum - ibid.}
{\bf B95}, 461 (1980)]; Phys. Lett {\bf B161}, 141 (1985).

\bibitem{MaModels}
E. Ma and U. Sarkar, Phys. Rev. Lett. {\bf 80}, 5716 (1998) [{\tt
arXiv:hep-ph/9802445}]; E. Ma and G. Rajasekaran, Phys. Rev. {\bf
D64}, 113012 (2001) [{\tt arXiv:hep-ph/0106291}]; E. Ma, Mod.
Phys. Lett. {\bf A17}, 289 (2002) [{\tt arXiv:hep-ph/0201225}]; K.
S. Babu, E. Ma and J. W. Valle, Phys. Lett. {\bf B552}, 207 (2003)
[{\tt arXiv:hep-ph/0206292}]; E. Ma, Mod. Phys. Lett. {\bf A17},
2361 (2002) [{\tt arXiv:hep-ph/0211393}].

\bibitem{RparitySUSY}
C. S. Aulakh and R. N. Mohapatra,  Phys. Lett. {\bf B119}, 136
(1982); L. J. Hall and M. Suzuki, Nucl. Phys. {\bf B231}, 419
(1984); G. G. Ross and J. W. F. Valle, Phys. Lett. {\bf B151}, 375
(1985); J. Ellis, G. Gelmini, C. Jarlskog, G. G. Ross and J. W. F.
Valle, Phys. Lett. {\bf B150}, 142 (1985); S. Dawson, Nucl. Phys.
{\bf B261}, 297 (1985); M. Drees, S. Pakvasa, X. Tata and T. ter.
Veldhuis, Phys. Rev. {\bf D57}, R5335 (1998) [{\tt
arXiv:hep-ph/9712392}]; E. J. Chun, S. K. Kang, C. W. Kim and U.
W. Lee, Nucl. Phys. {\bf B544}, 89 (1999)  [{\tt
arXiv:hep-ph/9807327}]; V. Barger, T. Han, S. Hesselbach and D.
Marfatia, Phys. Lett. {\bf B538}, 346 (2002) [{\tt
arXiv:hep-ph/0108261}]; for a recent review see R. Barbieri {\it
et al.}, Phys.~Rept.~{\bf420}, 1 (2005) [{\tt
arXiv:hep-ph/0406039}];
  V.~Barger, P.~F.~Perez and S.~Spinner,
{\tt  arXiv:0812.3661 [hep-ph].}

\bibitem{ExtraDim}
N. Arkani-Hamed, S. Dimopoulos, G. Dvali and J. March-Russell,
Phys. Rev. {\bf D65}, 024032 (2002) [{\tt arXiv:hep-ph/9811448}];
  Y.~Grossman and M.~Neubert,
  Phys.\ Lett.\  B {\bf 474}, 361 (2000)
{\tt   [arXiv:hep-ph/9912408]; }
K. R. Dienes and I. Sarcevic, Phys. Lett. {\bf B500}, 133 (2001)
[{\tt arXiv:hep-ph/0008144}];
  S.~J.~Huber and Q.~Shafi,
  Phys.\ Lett.\  B {\bf 544}, 295 (2002)
{\tt  [arXiv:hep-ph/0205327];}
  M.~C.~Chen and H.~B.~Yu,
{\tt  arXiv:0804.2503 [hep-ph]:}
  G.~Perez and L.~Randall,
  {\tt arXiv:0805.4652 [hep-ph].}

\bibitem{TypeII}
W.~Konetschny and W.~Kummer, Phys.\ Lett.\  B {\bf 70} (1977) 433;
J.~Schechter and J.~W.~F.~Valle, Phys.\ Rev.\ D {\bf 22} (1980)
2227; T.~P.~Cheng and L.~F.~Li, Phys.\ Rev.\  D {\bf 22} (1980)
2860; G.~Lazarides, Q.~Shafi and C.~Wetterich, Nucl.\ Phys.\ B
{\bf 181} (1981) 287; R.~N.~Mohapatra and G.~Senjanovi\'c, Phys.\
Rev.\ D {\bf 23} (1981) 165.

\bibitem{TypeIII}
R.~Foot, H.~Lew, X.~G.~He and G.~C.~Joshi, Z.\ Phys.\ C {\bf 44}
(1989) 441; E.~Ma, Phys.\ Rev.\ Lett.\  {\bf 81} (1998) 1171 [{\tt
arXiv:hep-ph/9805219}]; B.~Bajc and G.~Senjanovi\'c, JHEP {\bf
0708} (2007) 014 [{\tt arXiv:hep-ph/0612029}];
P.~Fileviez~P\'erez, Phys.\ Lett.\  B {\bf 654} (2007) 189 [{\tt
arXiv:hep-ph/0702287}]; P.~Fileviez~P\'erez, Phys.\ Rev.\  D {\bf
76} (2007) 071701 [{\tt arXiv:0705.3589 [hep-ph]}];
  P.~Fileviez Perez,
{\tt  arXiv:0809.1202 [hep-ph].}

\bibitem{dim6}
S.~Weinberg, Phys.\ Rev.\ Lett.\  {\bf 43} (1979) 1566.

\bibitem{babuleung}
K.~S. Babu and C.~N. Leung, Nucl.~Phys.~{\bf B619}, 667 (2001)
[{\tt arXiv:hep-ph/0106054}].

\bibitem{nuless}
W. H. Furry, Phys. Rev. {\bf 56}, 1184 (1939); for early reviews
see, Primakoff and Rosen, Rep. Prog. Phys. {\bf 22}, 121 (1959);
Ann. Rev. Nucl. Part. Sci. {\bf 31}, 145 (1981).

\bibitem{DoiKotani}
M. Doi, T. Kotani and E. Takasugi, Prog. Theor. Phys. Suppl. {\bf
83}, 1 (1985).

\bibitem{ElliottEngel}
For recent review see {\it eg.} S.~R. Elliott and J. Engel, J.
Physics. {\bf G30} R183 (2004) [{\tt arXiv:hep-ph/0405078}].

\bibitem{TauDecay}
A. Ilakovac, B. A. Kniehl and A. Pilaftsis, Phys. Rev. {\bf D52},
3993 (1995) [{\tt arXiv:hep-ph/9503456}]; A. Ilakovac and A.
Pilaftsis, Nucl. Phys. {\bf B437}, 491 (1995) [{\tt
arXiv:hep-ph/9403398}]; A. Ilakovac, Phys. Rev. {\bf D54}, 5653
(1996)  [{\tt arXiv:hep-ph/9608218}]; V.~Gribanov, S.~Kovalenko
and I.~Schmidt, Nucl.\ Phys.\  B {\bf 607}, 355 (2001) [{\tt
arXiv:hep-ph/0102155}].

\bibitem{Atre}
A. Atre, V. Barger and T. Han, Phys. Rev. {\bf D71}, 113014 (2005)
[{\tt arXiv:hep-ph/0502163}].

\bibitem{Ng:1978ij}
J.~N.~Ng and A.~N.~Kamal, Phys.\ Rev.\  D {\bf 18}, 3412 (1978);
J.~Abad, J.~G.~Esteve and A.~F.~Pacheco, Phys.\ Rev.\  D {\bf 30},
1488 (1984); C.~Dib, V.~Gribanov, S.~Kovalenko and I.~Schmidt,
Phys.\ Lett.\  B {\bf 493}, 82 (2000) [{\tt
arXiv:hep-ph/0006277}].

\bibitem{AliBorisov}
A. Ali, A.~V. Borisov and N.~B. Zamorin, Eur. Phys. J. {\bf C21},
123 (2001) [{\tt arXiv:hep-ph/0104123}].

\bibitem{Barbero}
L.~S. Littenberg and R.~E. Shrock, Phys. Rev. {\bf D46}, R892
(1992); C. Barbero, G. Lopez Castro and A. Mariano, Phys. Lett.
{\bf B566}, 98 (2003) [{\tt arXiv:nucl-th/0212083}].

\bibitem{CSLim}
C.~S. Lim, E. Takasugi and M. Yoshimura, Prog.~Theor.~Phys. {\bf
113}, 1367 (2005), [{\tt arXiv:hep-ph/0411139}].

\bibitem{muep}
SINDRUM II Collaboration, J. Kaulard {\it et al.}, Phys. Lett.
{\bf B422}, 334 (1998); K. Zuber, {\tt arXiv:hep-ph/0008080}; P.
Domin, A. Faessler, S. Kovalenko and F. Simkovic, Phys. Rev. {\bf
C70}, 065501 (2004) [{\tt arXiv:nucl-th/0409033}].

\bibitem{mumup}
J.~H. Missimer, R.~N. Mohapatra and N. C. Mukhopadhyay, Phys. Rev.
{\bf D50}, 2067 (1994); F. Simkovic, A. Faessler,  S. Kovalenko
and P. Domin, Phys. Rev. {\bf D66}, 033005 (2002) [{\tt
arXiv:hep-ph/0112271}]; E. Takasugi, Nucl. Instrum. Meth. {\bf
A503}, 252 (2003); M. Aoki, Nucl. Instrum. Meth. {\bf A503}, 258
(2003).

\bibitem{TRizzo}
T. G. Rizzo, Phys. Lett. {\bf B116}, 23 (1982); C. A. Heusch and
P. Minkowski, Nucl. Phys. {\bf B416}, 3 (1994).

\bibitem{Dittmar:1989yg}
M.~Dittmar, A.~Santamaria, M.~C.~Gonzalez-Garcia and
J.~W.~F.~Valle, Nucl.\ Phys.\ B {\bf 332}, 1 (1990).

\bibitem{Rodejohann}
M. Flanz, W. Rodejohann and K. Zuber, Phys. Lett. {\bf B473}, 324
(2000), {\it Erratum - ibid.} {\bf B480}, 418 (2000)  [{\tt
arXiv:hep-ph/9911298}].

\bibitem{Flanz}
M. Flanz, W. Rodejohann and K. Zuber, Eur. Phys. J. {\bf C16}, 453
(2000) [{\tt arXiv:hep-ph/9907203}]; W. Rodejohann and K. Zuber,
Phys. Rev. {\bf D63}, 054031 (2001)  [{\tt arXiv:hep-ph/0011050}].

\bibitem{goran}
W.-Y.~Keung and G.~Senjanovic, Phys.~Rev.~Lett.~{\bf 50}, 1427
(1983); D.~Dicus, D.~Karatas, and P.~Roy, Phys.~Rev.~{\bf D44},
2033 (1991); A.~Datta, M.~Guchait, and A.~Pilaftsis,
Phys.~Rev.~{\bf D50}, 3195 (1994) [{\tt arXiv:hep-ph/9311257}].

\bibitem{NCollider}
F.~M.~L. Almeida, Y.~A. Coutinho, J.~A.~M. Simoes and M.~A.~B.
Vale, Phys.~Rev.~{\bf D62}, 075004 (2000) [{\tt
arXiv:hep-ph/0008201}]; O. Panella, M. Cannoni, C. Carimalo, and
Y.~N. Srivastava, Phys.~Rev.~{\bf D65}, 035005 (2002) [{\tt
arXiv:hep-ph/0107308}].

\bibitem{HZ}
T. Han and B. Zhang, Phys.~Rev.~Lett.~{\bf 97}, 171804 (2006)
[{\tt arXiv:hep-ph/0604064}].

\bibitem{more}
For a comparison for different colliders, see {\it e.g.}, F.~del
Aguila, J.~A.~Aguilar-Saavedra and R.~Pittau, J.\ Phys.\ Conf.\
Ser.\  {\bf 53}, 506 (2006) [{\tt arXiv:hep-ph/0606198}].

\bibitem{delAguila:2007em}
F.~del Aguila, J.~A.~Aguilar-Saavedra and R.~Pittau, JHEP {\bf
0710}, 047 (2007) [{\tt arXiv:hep-ph/0703261}].

\bibitem{NRCMS}
W.~Clarida, T.~Yetkin, R.~Vidal, W.~Wu, Tao Han, H.~Pi, and
E.~Yazgan, CMS Note 2008/054 (Dec. 2008) unpublished.

\bibitem{Jing}
S. Bar-Shalom, N. G. Deshpande, G. Eilam, J. Jiang and A. Soni,
Phys.~Lett.~{\bf B643}, 342 (2006) [{\tt arXiv:hep-ph/0608309}];
Z.-G.~Si and K.~Wang, {\tt arXiv:0810.5266}.

\bibitem{Soni}
D.~Atwood, S.~Bar-Shalom and A.~Soni, Phys.\ Rev.\  D {\bf 76},
033004 (2007) [{\tt arXiv:hep-ph/0701005}].

\bibitem{BarShalom:2008gt}
S.~Bar-Shalom, G.~Eilam, T.~Han and A.~Soni, Phys.~Rev.~D{\bf 77},
115019 (2008) [{\tt arXiv:0803.2835 [hep-ph]}].

\bibitem{Bajc:2007zf}
  I.~Dorsner and P.~Fileviez Perez,
  JHEP {\bf 0706}, 029 (2007)  {\tt [arXiv:hep-ph/0612216]};
  B.~Bajc, M.~Nemevsek and G.~Senjanovi\'c,
  Phys.\ Rev.\  D {\bf 76}, 055011 (2007) {\tt [arXiv:hep-ph/0703080].}

\bibitem{degouvea}
A.~de Gouvea, J.~Jenkins and N.~Vasudevan, Phys.\ Rev.\  D {\bf
75}, 013003 (2007) [{\tt arXiv:hep-ph/0608147}]; A.~de Gouvea,
{\tt arXiv:0706.1732 [hep-ph]}.

\bibitem{Pilaftsis}
A.~Pilaftsis and T.~E.~J.~Underwood, Nucl.\ Phys.\  B {\bf 692},
303 (2004) [{\tt arXiv:hep-ph/0309342}]; A.~Pilaftsis and
T.~E.~J.~Underwood, Phys.\ Rev.\  D {\bf 72}, 113001 (2005) [{\tt
arXiv:hep-ph/0506107}].

\bibitem{ETC}
T.~Appelquist and R.~Shrock, Phys.\ Rev.\ Lett.\  {\bf 90}, 201801
(2003) [{\tt arXiv:hep-ph/0301108}]; {\it ibidem}, Phys.\ Lett.\
B {\bf 548}, 204 (2002) [{\tt arXiv:hep-ph/0204141}]; {\it
ibidem}, in Neutrino Factories and Superbeams, NuFact03, A.I.P.
Conf. Proc, 721 (A.I.P., New York, 2004), p. 261; T.~Appelquist,
M.~Piai and R.~Shrock, Phys.\ Rev.\  D {\bf 69}, 015002 (2004)
[{\tt arXiv:hep-ph/0308061}]; T.~Appelquist, N.~D.~Christensen,
M.~Piai and R.~Shrock, Phys.\ Rev.\  D {\bf 70}, 093010 (2004)
[{\tt arXiv:hep-ph/0409035}].

\bibitem{largeangle}
K.~R.~S.~Balaji, A.~Perez-Lorenzana and A.~Y.~Smirnov, Phys.\
Lett.\  B {\bf 509}, 111 (2001) [{\tt arXiv:hep-ph/0101005}];
A.~Y.~Smirnov and R.~Zukanovich Funchal, Phys.\ Rev.\  D {\bf 74},
013001 (2006) [{\tt arXiv:hep-ph/0603009}].

\bibitem{dmstructure}
M.~Viel, J.~Lesgourgues, M.~G.~Haehnelt, S.~Matarrese and
A.~Riotto, Phys.\ Rev.\  D {\bf 71}, 063534 (2005) [{\tt
arXiv:astro-ph/0501562}]; M.~Viel, J.~Lesgourgues, M.~G.~Haehnelt,
S.~Matarrese and A.~Riotto, Phys.\ Rev.\ Lett.\  {\bf 97}, 071301
(2006) [{\tt arXiv:astro-ph/0605706}].

\bibitem{kusenkoSN}
G.~M.~Fuller, A.~Kusenko and K.~Petraki,
{\tt arXiv:0806.4273Ê[astro-ph]}.

 \bibitem{dodelson}
S. Dodelson and L.~M. Widrow, Phys. Rev. Lett. {\bf 72}, 17 (1994)
[{\tt arXiv:hep-ph/9303287}]; X.~D. Shi and G.~M. Fuller, Phys.
Rev. Lett. {\bf 82} [{\tt arXiv:astro-ph/9810076}], 2832 (1999);
A.~D. Dolgov and S.~H. Hansen, Astropart. Phys. {\bf 16}, 339
(2002).

\bibitem{fullerpatel}
K.~Abazajian, G.~M.~Fuller and M.~Patel, Phys.\ Rev.\  D {\bf 64},
023501 (2001) [{\tt arXiv:astro-ph/0101524}].

\bibitem{gelmini}
G. Gelmini, S. Palomares-Ruiz and S. Pascoli,
Phys.~Rev.~Lett.~{\bf 93}, 081302 (2004) [{\tt
arXiv:astro-ph/0403323}].

\bibitem{dolgov}
A.~D. Dolgov, Phys. Rept. {\bf 370}, 333 (2002) [{\tt
arXiv:hep-ph/0202122}].

\bibitem{kusenko}
A. Kusenko and G. Segre, Phys. Lett. {\bf B396}, 197 (1997) [{\tt
arXiv:hep-ph/9701311}]; G.~M. Fuller, A. Kusenko, I. Mocioiu and
S. Pascoli, Phys. Rev, {\bf D68}, 103002 (2003) [{\tt
arXiv:astro-ph/0307267}]; for a review, see A. Kusenko,
Int.~J.~Mod.~Phys.~{\bf D13}, 2065 (2004) [{\tt
arXiv:astro-ph/0409521}].

\bibitem{hansen}
S.~H. Hansen and Z. Haiman, Astrophys. J. {\bf 600}, 26 (2004)
[{\tt arXiv:astro-ph/0305126}]; P.~L.~Biermann and A.~Kusenko,
Phys.\ Rev.\ Lett.\  {\bf 96}, 091301 (2006) [{\tt
arXiv:astro-ph/0601004}].

\bibitem{fullerxray}
K.~Abazajian, G.~M.~Fuller and W.~H.~Tucker, Astrophys.\ J.\  {\bf
562}, 593 (2001) [{\tt arXiv:astro-ph/0106002}].

\bibitem{nuSMxray}
A.~Boyarsky, A.~Neronov, O.~Ruchayskiy and M.~Shaposhnikov, Phys.\
Rev.\  D {\bf 74}, 103506 (2006) [{\tt arXiv:astro-ph/0603368}];
A.~Boyarsky, A.~Neronov, O.~Ruchayskiy, M.~Shaposhnikov and
I.~Tkachev, Phys.\ Rev.\ Lett.\  {\bf 97}, 261302 (2006) [{\tt
arXiv:astro-ph/0603660}].

\bibitem{Boehm}
C.~Boehm, Y.~Farzan, T.~Hambye, S.~Palomares-Ruiz and S.~Pascoli,
{\tt arXiv:hep-ph/0612228}.

\bibitem{nuSM}
T.~Asaka, S.~Blanchet and M.~Shaposhnikov, Phys.\ Lett.\  B {\bf
631}, 151 (2005) [{\tt arXiv:hep-ph/0503065}]; T.~Asaka and
M.~Shaposhnikov, Phys.\ Lett.\  B {\bf 620}, 17 (2005) [{\tt
arXiv:hep-ph/0505013}]; T.~Asaka, M.~Laine and M.~Shaposhnikov,
JHEP {\bf 0701}, 091 (2007) [{\tt arXiv:hep-ph/0612182}].

\bibitem{Asaka:2006ek}
T.~Asaka, M.~Shaposhnikov and A.~Kusenko, Phys.\ Lett.\  B {\bf
638}, 401 (2006) [{\tt arXiv:hep-ph/0602150}].

\bibitem{nuSMconstraints}
F.~L.~Bezrukov and M.~Shaposhnikov, Phys.\ Rev.\  D {\bf 75},
053005 (2007) [{\tt arXiv:hep-ph/0611352}]; D.~Gorbunov and
M.~Shaposhnikov, JHEP {\bf 0710}, 015 (2007) [{\tt arXiv:0705.1729
[hep-ph]}].

\bibitem{Smirnov:2006bu}
A.~Y.~Smirnov and R.~Zukanovich Funchal, Phys.\ Rev.\  D {\bf 74},
013001 (2006) [{\tt arXiv:hep-ph/0603009}].

\bibitem{Kusenko:2007wv}
A.~Kusenko, {\tt arXiv:hep-ph/0703116}.

\bibitem{EGPP}
G.~Gelmini, E.~Osoba, S.~Palomares-Ruiz and S.~Pascoli,
JCAP {\bf 0810}, 029 (2008) [{\tt
arXiv:0803.2735 [astro-ph]}].

\bibitem{Chao:2008mq}
  W.~Chao, Z.~G.~Si, Z.~Z.~Xing and S.~Zhou,
  Phys.\ Lett.\  B {\bf 666}, 451 (2008)
{\tt  [arXiv:0804.1265 [hep-ph]].}

\bibitem{Wietfeldt:1995ja}
F.~E.~Wietfeldt and E.~B.~Norman, Phys.\ Rept.\  {\bf 273}, 149
(1996).

\bibitem{Armbruster:1995nr}
B.~Armbruster {\it et al.}  [KARMEN Collaboration], Phys.\ Lett.\
B {\bf 348}, 19 (1995).

\bibitem{Daum}
M.~Daum {\it et al.}, Phys.\ Rev.\ Lett.\  {\bf 85}, 1815 (2000)
[{\tt arXiv:hep-ex/0008014}].

\bibitem{Cirelli:2004cz}
M.~Cirelli, G.~Marandella, A.~Strumia and F.~Vissani, Nucl.\
Phys.\ B {\bf 708}, 215 (2005) [{\tt arXiv:hep-ph/0403158}].

\bibitem{Aguilar-Arevalo:2007it}
A.~A.~Aguilar-Arevalo {\it et al.}  [The MiniBooNE Collaboration],
{\tt arXiv:0704.1500 [hep-ex]}.

\bibitem{Maltoni:2007zf}
M.~Maltoni and T.~Schwetz, {\tt arXiv:0705.0107 [hep-ph]}.

\bibitem{Shrock:1980vy}
R.~E.~Shrock, Phys.\ Lett.\ B {\bf 96}, 159 (1980).

\bibitem{gNnuphi}
Z.~Chacko, L.~J.~Hall, S.~J.~Oliver and M.~Perelstein, Phys.\
Rev.\ Lett.\  {\bf 94}, 111801 (2005) [{\tt
arXiv:hep-ph/0405067}]; S.~Palomares-Ruiz, S.~Pascoli and
T.~Schwetz, JHEP {\bf 0509}, 048 (2005) [{\tt
arXiv:hep-ph/0505216}].

\bibitem{decayswrong}
D.~Errede {\it et al.}, Phys.\ Rev.\ D {\bf 37}, 577 (1988);
D.~Decamp {\it et al.}  [ALEPH Collaboration], Phys.\ Lett.\ B
{\bf 236}, 511 (1990); P.~Burchat {\it et al.}, Phys.\ Rev.\ D
{\bf 41}, 3542 (1990); B.~Adeva {\it et al.}  [L3 Collaboration],
Phys.\ Lett.\ B {\bf 251}, 321 (1990); P.~Achard {\it et al.}  [L3
Collaboration], Phys.\ Lett.\ B {\bf 517}, 67 (2001) [{\tt
arXiv:hep-ex/0107014}].

\bibitem{Nardi:1994iv}
E.~Nardi, E.~Roulet and D.~Tommasini, Phys.\ Lett.\  B {\bf 327},
319 (1994) [{\tt arXiv:hep-ph/9402224}]; E. Nardi, E. Roulet, and
D. Tommasini, Phys. Lett. B {\bf 344}, 225 (1995) [{\tt
arXiv:hep-ph/9409310}].

\bibitem{bergmann}
S.~Bergmann and A.~Kagan, Nucl.\ Phys.\  B {\bf 538}, 368 (1999)
[{\tt arXiv:hep-ph/9803305}].

\bibitem{delAguila:2008pw}
F.~del Aguila, J.~de Blas and M.~Perez-Victoria,
Phys.\ Rev.\  D {\bf 78} (2008) 013010
[{\tt arXiv:0803.4008 [hep-ph]}].

\bibitem{Ma:1980gm}
E.~Ma and A.~Pramudita, Phys.\ Rev.\  D {\bf 24}, 1410 (1981).

\bibitem{Langacker:1988up}
P.~Langacker and D.~London, Phys.\ Rev.\  D {\bf 38}, 907 (1988).

\bibitem{Tommasini:1995ii}
D.~Tommasini, G.~Barenboim, J.~Bernabeu and C.~Jarlskog, Nucl.\
Phys.\  B {\bf 444}, 451 (1995) [{\tt arXiv:hep-ph/9503228}].

\bibitem{Galeazzi:2001py}
M.~Galeazzi, F.~Fontanelli, F.~Gatti and S.~Vitale, Phys.\ Rev.\
Lett.\  {\bf 86}, 1978 (2001).

\bibitem{Hiddemann:1995ce}
K.~H.~Hiddemann, H.~Daniel and O.~Schwentker, J.\ Phys.\ G {\bf
21}, 639 (1995).

\bibitem{Holzschuh:1999vy}
E.~Holzschuh, W.~Kundig, L.~Palermo, H.~Stussi and P.~Wenk, Phys.\
Lett.\ B {\bf 451}, 247 (1999).

\bibitem{Holzschuh:2000nj}
E.~Holzschuh, L.~Palermo, H.~Stussi and P.~Wenk, Phys.\ Lett.\ B
{\bf 482}, 1 (2000).

\bibitem{Deutsch:1990ut}
J.~Deutsch, M.~Lebrun and R.~Prieels, Nucl.\ Phys.\ A {\bf 518},
149 (1990).

\bibitem{Back:2003ae}
H.~O.~Back {\it et al.}, JETP Lett.\  {\bf 78}, 261 (2003) [Pisma
Zh.\ Eksp.\ Teor.\ Fiz.\  {\bf 78}, 707 (2003)].

\bibitem{Hagner:1995bn}
C.~Hagner, M.~Altmann, F.~von Feilitzsch, L.~Oberauer, Y.~Declais
and E.~Kajfasz, Phys.\ Rev.\ D {\bf 52}, 1343 (1995).

\bibitem{Britton:1992pg}
D.~I.~Britton {\it et al.}, Phys.\ Rev.\ Lett.\  {\bf 68}, 3000
(1992); D.~I.~Britton {\it et al.}, Phys.\ Rev.\ D {\bf 46}, 885
(1992).

\bibitem{Benes:2005hn}
P.~Benes, A.~Faessler, F.~Simkovic and S.~Kovalenko, Phys.\ Rev.\
D {\bf 71}, 077901 (2005) [{\tt arXiv:hep-ph/0501295}].

\bibitem{knupeak}
D. Berghofer {\it et al.}, Proc. Intern. Conf. on Neutrino Physics
and Astrophysics (Maui, Hawaii, 1981), 67 (1981), eds. R. J.
Cence, E. Ma and A. Roberts, Vol. II (University of Hawaii,
  Honolulu, HI, 1981);
  T. Yamazaki,
  Proc. 22nd Intern. Conf, on High-energy physics (Leipzig, 1984), 262 (1984),
  eds. A. Meyer and E. Wieczorek, Vol. I (Akademie der Wiessenachaften der DDR, Leipzig, 1984).

\bibitem{Bernardi:1987ek}
G.~Bernardi {\it et al.}, Phys.\ Lett.\ B {\bf 203}, 332 (1988).

\bibitem{Badier:1986xz}
J.~Badier {\it et al.}  [NA3 Collaboration], Z.\ Phys.\ C {\bf
31}, 21 (1986).

\bibitem{Bergsma:1985is}
F.~Bergsma {\it et al.}  [CHARM Collaboration], Phys.\ Lett.\ B
{\bf 166}, 473 (1986).

\bibitem{Abreu:1996pa}
P.~Abreu {\it et al.}  [DELPHI Collaboration], Z.\ Phys.\ C {\bf
74}, 57 (1997) [Erratum-ibid.\ C {\bf 75}, 580 (1997)].

\bibitem{Adriani:1992pq}
O.~Adriani {\it et al.}  [L3 Collaboration], Phys.\ Lett.\ B {\bf
295}, 371 (1992).

\bibitem{sub}
G. B\'elanger, F. Boudjema, D. London, and H. Nadeau, Phys. Rev.
{\bf D53}, 6292 (1996) [{\tt arXiv:hep-ph/9508317}]; D. London,
{\tt arXiv:hep-ph/9907419}.

\bibitem{Kusenko:2004qc}
A.~Kusenko, S.~Pascoli and D.~Semikoz, JHEP {\bf 0511}, 028 (2005)
[{\tt arXiv:hep-ph/0405198}].

\bibitem{BEBC}
A. M. Cooper-Sarkar {\it et al.}, Phys. Lett. {\bf 160} B, 207
(1985).

\bibitem{Gallas:1994xp}
E.~Gallas {\it et al.}  [FMMF Collaboration], Phys.\ Rev.\ D {\bf
52}, 6 (1995).

\bibitem{Vaitaitis:1999wq}
A.~Vaitaitis {\it et al.}  [NuTeV Collaboration], Phys.\ Rev.\
Lett.\  {\bf 83}, 4943 (1999) [{\tt arXiv:hep-ex/9908011}].

\bibitem{Vilain:1994vg}
P.~Vilain {\it et al.}  [CHARM II Collaboration], Phys.\ Lett.\ B
{\bf 343}, 453 (1995) [Phys.\ Lett.\ B {\bf 351}, 387 (1995)].

\bibitem{Orloff:2002de}
J.~Orloff, A.~N.~Rozanov and C.~Santoni, Phys.\ Lett.\ B {\bf
550}, 8 (2002) [{\tt arXiv:hep-ph/0208075}].

\bibitem{Astier:2001ck}
P.~Astier {\it et al.}  [NOMAD Collaboration], Phys.\ Lett.\ B
{\bf 506}, 27 (2001) [{\tt arXiv:hep-ex/0101041}].

\bibitem{PDG}
PDG, S. Eidelman {\it et al.} Phys. Lett. {\bf B592}, 1 (2004).

\bibitem{delaguilaprivate}
F.~del Aguila, private communication.

\bibitem{Brooks:1999pu}
M.~L.~Brooks {\it et al.}  [MEGA Collaboration],
Phys.\ Rev.\ Lett.\  {\bf 83}, 1521 (1999)
[{\tt arXiv:hep-ex/9905013}].

\bibitem{Dohmen:1993mp}
C.~Dohmen {\it et al.}  [SINDRUM II Collaboration.],
Phys.\ Lett.\  B {\bf 317}, 631 (1993).

\bibitem{aubert}
 B. Aubert {\it et al.} [BaBar Collaboration],
Phys. Rev. Lett. {\bf 95}, 191801 (2005)[{\tt
arXiv:hep-ex/0506066}].

\bibitem{he}
Q. He {\it et al.} [CLEO Collaboration], Phys. Rev. Lett. {\bf
95}, 221802 (2005)[{\tt arXiv:hep-ex/0508031}].

\bibitem{Grossman}
Y. Grossman, Z. Ligeti and E. Nardi, Phys. Rev. {\bf D55}, 2768
(1997) [{\tt arXiv:hep-ph/9607473}].

\bibitem{Abulencia:2007rd}
A.~Abulencia {\it et al.}  [CDF Collaboration],
Phys.\ Rev.\ Lett.\  {\bf 98}, 221803 (2007)
[{\tt arXiv:hep-ex/0702051}].

\bibitem {ATLAS}
ATLAS Collaboration, Detector and Physics Performances Technical
Design Report, Vol.I, ATLAS TDR 14, CERN/LHCC 99-14, Section
2.5.9.

\bibitem{Abulencia:2005kq}
A.~Abulencia {\it et al.}  [CDF Collaboration], Phys.\ Rev.\
Lett.\  {\bf 96}, 011802 (2006)  [{\tt arXiv:hep-ex/0508051}].

\bibitem{taurec}
A.~Hektor, M.~Kadastik, M.~Muntel, M.~Raidal and L.~Rebane, Nucl.\
Phys.\  B {\bf 787}, 198 (2007)  [{\tt arXiv:0705.1495 [hep-ph]}];
P.~Fileviez Perez, T.~Han, G.-Y.~Huang, T.~Li and K.~Wang, {\tt
arXiv:0803.3450 [hep-ph]}.

\bibitem{l3}
P. Achard {\it et al.},
\journal{\plb}{517}{67}{2001} [{\tt arXiv:hep-ex/0107014}].

\bibitem{Seljak}
U. Seljak {\it et al.}, Phys. Rev. {\bf D71}, 103515 (2005) [{\tt
arXiv:astro-ph/0407372}].

\bibitem{maltoni}
F. Maltoni and T. Stelzer, JHEP 0302:027 (2003)  [{\tt
arXiv:hep-ph/0208156}].

\bibitem{Ivanov}
M. A. Ivanov and S. G. Kovalenko, Phys. Rev. {\bf D71}, 053004
(2005)  [{\tt arXiv:hep-ph/0412198}].

\bibitem{ebert}
D. Ebert, R. N. Faustov and V. O. Galkin, Phys. Lett. {\bf B635},
93 (2006) [{\tt arXiv:hep-ph/0602110}].

\bibitem{artuso}
M. Artuso {\it et al.} [CLEO Collaboration], Phys. Rev. Lett. {\bf
95}, 251801 (2005) [{\tt arXiv:hep-ex/0508057}].

\bibitem{MILC}
C. Bernard {\it et al.} [MILC Collaboration], Phys. Rev. {\bf
D66}, 094501 (2002) [{\tt arXiv:hep-lat/0206016}].

\bibitem{feldmann}
T. Feldmann, Int. J. Mod. Phys. {\bf A15}, 159 (2000) [{\tt
arXiv:hep-ph/9907491}].

\bibitem{edwards}
K. W. Edwards {\it et al.} [CLEO Collaboration], Phys. Rev. Lett.
{\bf 86}, 30 (2001) [{\tt arXiv:hep-ex/0007012}].

\bibitem{wang}
G.-L. Wang, Phys. Lett. {\bf B633}, 492 (2006) [{\tt
arXiv:math-ph/0512009}].

\end{thebibliography}
\end{document}